\documentclass{iopart}
\usepackage{epsf}
\newcommand{\rmj}{{\rm j}}
\newcommand{\rmt}{{\rm t}}

\newcommand{\ul}[1]{\underline{#1}}

\newcommand{\dul}[1]{\underline{\underline{#1}}}

\newcommand{\bea}{\begin{eqnarray}}

\newcommand{\eea}{\end{eqnarray}}

\begin{document}

\title[]{Nonstationary random acoustic and electromagnetic fields as wave diffusion processes\footnote{A summary of selected results in this paper appeared in 
\cite{arnaZurich03}. }
}

\author{L R Arnaut}
\address{National Physical Laboratory,
Time, Quantum and Electromagnetics Team,
F2--A11, MS 2F--3,
Hampton Road,
Teddington TW11 0LW,
United Kingdom.
}
\ead{luk.arnaut@npl.co.uk}


\begin{abstract}
We investigate the effects of relatively rapid variations of the boundaries of an overmoded cavity on the stochastic properties of its interior acoustic or electromagnetic field. For quasi-static variations, this field can be represented as an ideal incoherent and statistically homogeneous isotropic random scalar or vector field, respectively. A physical model is constructed showing that the field dynamics can be characterized as a generalized diffusion process. The Langevin--It\^{o} and Fokker--Planck equations are derived and their associated statistics and distributions for the complex analytic field, its magnitude and energy density are computed. 
The energy diffusion parameter is found to be proportional to the square of the ratio of the standard deviation of the source field to the characteristic time constant of the dynamic process, but is independent of the initial energy density, to first order. The energy drift vanishes in the asymptotic limit.
The time-energy probability distribution is in general not separable, as a result of nonstationarity.
A general solution of the Fokker--Planck equation is obtained in integral form, together with explicit closed-form solutions for several asymptotic cases.
The findings extend known results on statistics and distributions of quasi-stationary ideal random fields (pure diffusions), which are retrieved as special cases.
\end{abstract}



\maketitle

\tableofcontents

\newpage

\section{Introduction}

Increasingly complex environments for propagation of acoustic or electromagnetic (EM) waves call for more powerful, efficient and specialized techniques to characterize fields and associated quantities. Multiple scattering and diffraction, dynamic and complex geometries or configurations, multi-path propagation, wide-band quasi-random modulation, etc., all make deterministic analytical/computational or experimental characterization techniques increasingly less efficient. Moreover, in such applications, these conventional methods are often intrinsically inaccurate, because they typically presume single plane-wave excitation or localized idealized sources. This raises issues of analytical tractability or numerical stability, even in nonresonant ``open'' environments.
As an alternative approach, a growing trend exists toward the use of statistical methods in such cases, e.g., for characterizing random scalar and vector fields inside overmoded resonant enclosures, as test or operational environments in their own right, or as field generators or simulators of complex and multi-scattering environments.
A primary example is the mode-tuned or mode-stirred reverberation chamber (MT/MSRC) (e.g., \cite{meye1}--\cite{meye2}), which after an established career in acoustics is gaining increasing importance for application in electromagnetic compatibility (characterization of immunity, emissions, shielding, and absorption characteristics), high-intensity radio-frequency field generation, measurement of EM properties of materials, antenna characterization, or as a simulator of inhomogeneous time-varying propagation channels for radio waves in multi-terminal (MIMO) wireless communication systems \cite{bell1}--\cite{arnaRS3}. In its basic configuration, a MT/MSRC is a highly overmoded resonant cavity whose boundary or excitation conditions are perturbed dynamically -- thus generating a so-called ``stirring'' or mixing process -- either mechanically 
(e.g., through continuous rotation of a reflective paddle wheel or other diffractor that is large compared to the wavelength and exhibits an acoustically or electromagnetically reflective surface; through rotation or vibration of cavity walls; etc.), 
electronically 
(by modulating the excitation signal in phase or frequency), 
via excitation by a noise source, 
or through any combination of the above. 
More generally, such a cavity can be employed for precise modelling and measurement of spatial and temporal distributions of EM energy \cite{nye1}.
An ideal MT/MSRC generates a statistically isotropic, homogeneous, incoherent, and (in the EM case) unpolarized field, which can be represented by an isotropic angular spectrum of random plane waves.
In this way, it defines a canonical echoic EM environment that is the counterpart of unbounded (anechoic) free space.
A characteristic feature of a MT/MSRC is the extreme sensitivity of its interior field to variations in the boundary conditions, at any location, akin to wave chaos displayed by non-integrable cavities (``billiards'') with or without time-varying closed or partially opened boundaries, e.g., \cite{reve1}--\cite{fyod1}. 
As a result, mode stirring gives rise to a hybrid, i.e., amplitude-plus-frequency modulated interior field \cite{tich1}, \cite{arnaTEMCv47n4}.

Fields in the presence of rotating scattering bodies or surfaces in unbounded environments have been intensively studied for several decades, in particular for rotations of a cylindrically symmetric and axially rotating scattering object, e.g., \cite{petr1}--\cite{chua1}. More recent efforts have focused on irregular or complicated shapes inside enclosures.
Pertinently, for an overmoded cavity in which a stirring process evolves arbitrarily slowly (adiabatically) relative to the effective intrinsic relaxation time of the cavity \cite{lamb1,rich1} (that is, the weighted average of the modal decay times of all cavity modes that contribute when no stirring occurs), the overall system remains EM quasi-stationary. 
On the other hand, if the time scale of the stirring process is of the order of the intrinsic modal relaxation times (but not necessarily on a relativistic scale, and evolving still slowly compared to the period of wave oscillation), this may cause the interior field to ``slip'', i.e., to cause nonstationarity \cite{arnaTEMCv47n4}. 
In this case, the source (excitation) field and the resulting interior (cavity) field exhibit rates of fluctuation that are of comparable order of magnitude, so that the imposed perturbations are no longer tracked faithfully (i.e., on the same time scale) by the induced interior field. 
In MT/MSRCs, this situation occurs most readily for acoustic waves, because of their significantly lower velocity of propagation compared to that of EM waves. As will be shown, the nature and parameters of the cavity field then become markedly different from those in the quasi-stationary case. 

In this paper, we study quasi-monochromatic random fields that are subjected to continuously time-varying boundaries of a highly overmoded resonant cavity, generating a nonstationary random field at any interior location. The framework is that of stochastic classical wave mechanics and stochastic differential equations (SDEs).
Statistics of the field, its magnitude and energy density are obtained from their probability distribution functions (PDFs) as solutions of the associated Fokker--Planck equation (FPE). The energy density is of special importance because the incoherency of random fields makes this the fundamental EM quantity. 
The focus is on waves generated by a transmitting source (Tx) and perceived by a receiver (Rx) that are both located inside the cavity, as a canonical scenario for measurements by a field sensor, membrane, antenna, as well as radiation from, or susceptibility of an acoustic or electronic device under test. 

The paper is organized as follows. Based on a physical model developed in section \ref{sec:transients}, the SDE is derived in the present context in section \ref{sec:Langevin}, together with expressions for the drift and diffusion coefficients of the field in function of configurational and excitation parameters. These are employed 
to derive the FPEs for the magnitude and energy density of the scalar and vector complex analytic fields in section \ref{sec:Langevin2bis} which, together with section \ref{sec:systemnonstat}, forms the core of the paper. The FPEs are solved in closed form, yielding the general PDF in integral form. The asymptotic PDFs for first-order (early) time dependence are given and their physical characteristics are discussed.
The particular case of an asymptotically fully developed nonstationary process is analyzed in section \ref{sec:BEWL}. 
The results are extended in section \ref{sec:systemnonstat} to second-order systems, including the case of first-order systems responding to a nonstationary field. 
The case of time-dependent but deterministic coefficients of the SDE and FPE is treated in section \ref{sec:timedeptau}.
Finally, a summary of the main results with conclusions is given in section \ref{sec:conclusions}.

A selection of preliminary results for the case of a scalar field was summarized previously in \cite{arnaZurich03}.
The present paper follows on from a companion paper in which the nonlinearity and distortion of electromagnetic fields caused by a mode stirring process were studied from the point of view of random modulation of the interior field \cite{arnaTEMCv47n4}. 
Experimental measured results demonstrating nonstationarity in a MTRC induced by accelerated motion of cavity boundaries have been reported in \cite{arnaDEMEM012}.

\section{Transient mode-stirred fields\label{sec:transients}}
The problem at hand is the temporal transition of the local interior cavity vector field $\ul{Y}(t)$, from its initial stationary state at time $t=t_0$ to its value at a next state at $t=t_0+\Delta t$. This transition takes place under the action of an external perturbation influencing an otherwise stationary cavity field $\ul{X}(t)$ and is governed by intrinsic characteristics (modal time constants) of the cavity and by the rate of change (velocity) of the action itself. The functional relation between $\ul{Y}(t)$ and $\ul{X}(t)$ will be considered in section \ref{sec:Langevin}.

If the field were responding instantaneously to changes in the boundary conditions  -- disregarding any pure propagation effect, i.e., phase delay -- then the rate of fluctuation of $\ul{Y}(t)$ would be solely governed by that of the source (itself assumed to be stationary) at all times during the transition. 
In resonant environments, the situation is more complicated and rather intricate, particularly when coupling between modes exists, causing the transitions to become quasi-random. 

Assume $\ul{X}(t)$ and $\ul{Y}(t)$ to be randomly amplitude- and hybrid-modulated time-harmonic fields, respectively \cite{arnaTEMCv47n4}. Consider the transient field caused by an instantaneous step transition between two discrete boundary states taking place at $t=t_0$. At $t=t_{0-}$, the perceived field is in a well-defined state $\ul{Y}(t_{0-})$ whence it can be expressed as a weighted phasor sum of a finite number of vector components, viz., the amplitudes of the participating instantanous cavity eigenmodes \cite{davi1,arnaKdf} or plane-wave vector components of a realization of the angular spectrum generated at $t_0$, respectively (cf., e.g., \cite{hillv40n3,arnaPRE}):
\bea
\ul{Y}(t_0) = \sum^{\cal N}_{i=1} \ul{y}_i(t_0).
\eea
At any $t$, the local field can be represented as an analytic field with time-varying phasor $\ul{Y}(t)$ in 1-D or 3-D complex vector space for scalar or vector fields, respectively, irrespective whether or not the field is quasi-stationary \cite{arnaTEMCv47n4,arnaKdf}.

When, at $t=t_{0}$, a perturbation of the boundary conditions takes effect, each currently contributing mode or plane-wave component $\ul{y}_i(t)$ decays, on 
average\footnote{The details of the early decay of the effective field depend also on the exact instance (phase) of the $y^{(0)}_i$ at $t=t_0$. Since this phase fluctuates randomly between recurrent events, each decay process exhibits largely different early decays for each participating phasor and, hence, for the resultant phasor sum between different starting times. }, 
in accordance with its own characteristic time constant $\tau^{(0)}_i$ governed by the cavity state at $t=t_0$. These prior contributions to $\ul{Y}(t_0+\Delta t)$ will be further denoted with a superscript ``$(0)$'', as $\ul{y}^{(0)}_i(t_0+\Delta t)$. 

During the decay of each prior phasor, posterior phasors $\ul{y}^{(1)}_i$ associated with the next-state (perturbed) cavity geometry start building up, each one governed by its next-state characteristic time constant $\tau^{(1)}_i(t_0)$.
 In this process, energy is being exchanged between the participating modes owing to mode overlap in a lossy cavity \cite{soko1,soko2}, in other words $\tau^{(0)}_i \not = 0$. This coupling may cause a phasor to have its amplitude temporarily increasing before on average decaying \cite{arnaCEM11}. In other words, the decay of $\ul{y}^{(0)}_i(t)$ or even $\ul{Y}^{(0)}(t)$ is typically not monotonic. The decay and build-up processes are complicated, because the eigenmodes effectively switch during these transients (birth-death-mutation processes), while their influence is prolonged as a result of finite rise and decay times owing to continuity of fields. Thus, at any posterior instance $t_0+\Delta t$, the transition is governed by sets of developing or fading modes:
\bea
\ul{Y}(t_0+\Delta t) &= \sum^{{\cal N}^{(0)}}_{i=1} \ul{y}^{(0)}_i(t_0+\Delta t) + \sum^{{\cal N}^{(1)}}_{j=1}  \ul{y}^{(1)}_j(t_0+\Delta t)\nonumber\\
&\doteq \ul{Y}^{(0)}(t_0+\Delta t) + \ul{Y}^{(1)}(t_0+\Delta t). 
\eea
We shall refer to this form of nonstationarity as {\it stirring slip}.
It causes the phasors $\ul{y}_i(t \geq t_0)$ to decrease exponentially, on average, each with different time constants while phasors undergo a relaxed and retarded change (they locally `stretch', `shrink', and rotate) as a consequence of variations of the cavity boundary. 

\begin{figure}[htb] \begin{center} 
\begin{tabular}{c} \ 
\epsfysize=8cm 
\epsfbox{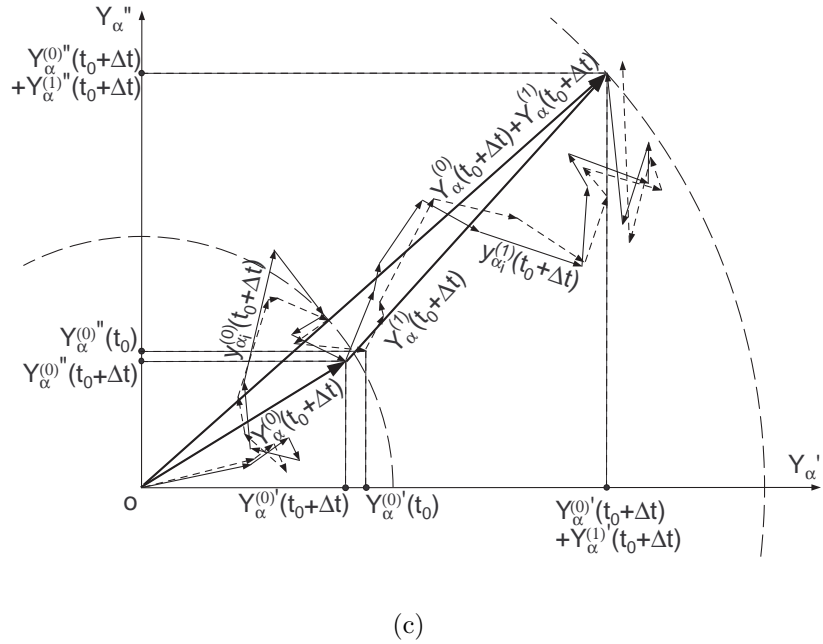}\ \\ 
(c)\\
\\
\end{tabular} \end{center}

\caption{\label{fig:randomwalk} \small
Models for transition between consecutive states of a random scalar wave in the complex plane: (a) quasi-stationary process, (b) fully developed nonstationary (BEWL) process, (c) general nonstationary process. The phasors $\ul{y}_{\alpha_i}(t_0)$ represent individual principal components (modes, spectral plane-wave components, etc.) with random aplitudes and phases. In figure (c), the dotted phasors represent the corresponding phasors of the BEWL process of figure (b), shown for reference only.}

\end{figure}

The associated phasor diagram is sketched in figure \ref{fig:randomwalk} for a scalar complex field $Y_\alpha = Y^\prime_\alpha - \rmj Y^{\prime\prime}_\alpha$. For a vector field, $\alpha$ denotes an arbitrary Cartesian component $x$, $y$ or $z$, and ${\ul{1}_\alpha}$ represents a unit vector in the direction of that component.
If the system remains quasi-stationary during the transition, ${Y}_\alpha(t_0 + \Delta t)$ is the sum ${Y}^{(1)}_\alpha (t)$ of a set of emerging posterior phasors $y^{(1)}_{\alpha_j}(t)$ and ${Y}^{(0)}_\alpha (t)$ of a set of vanishing prior phasors $y^{(0)}_{\alpha_i}(t)$ for $t_0 \leq t < t_0+\Delta t$. For general nonstationary processes, however, $Y_\alpha(t_0 + \Delta t)$ is arrived at by a more complicated dynamic superposition of a set of on-average decaying phasors $\{y^{(0)}_{\alpha_i}(t\geq t_0)\}$ and growing new phasors $\{y^{(1)}_{\alpha_j}(t\geq t_0)\}$.
In case of ``slowly'' modulated signals, the situation can formally be described by an angular spectral plane-wave expansion for harmonizable functions 
\bea
\ul{X} (t) = \int\int_\Omega \exp \left ( -\rmj \ul{k}\cdot \ul{r} \right ) {\rm d}\ul{X}(t)
\eea
in which ${\rm d}\ul{X}(t)$ is a process of nonorthogonal increments \cite{loev1}.
In the asymptotic limit of a pure Bachelier--Einstein--Wiener--L\'{e}vy (BEWL) diffusion process (random walk), the $y^{(1)}_{\alpha_j}(t>t_0)$ consist of independent increments building onto the existing prolonged $\{ y^{(0)}_{\alpha_i}(t\geq t_0) \} \equiv \{ y^{(0)}_{\alpha_i}(t_0) \}$.

Three particular regimes can be distinguished:
\begin{itemize}
\item {\it Quasi-stationary field:} In this case, ${Y}_\alpha(t_0+\Delta t)$ is statistically identical to ${Y}_\alpha(t_0)$. This can still be conceived as a special case of a random walk, viz., one that returns to the origin before each step: ${y}^{(0)}_{\alpha_i}(t_0+\Delta t) = {0}$. Hence, ${Y}^{(0)}_\alpha(t_0+\Delta t) = {0}$; cf. figure \ref{fig:randomwalk}a. In this case,
\bea
Y_\alpha(t_0+\Delta t) = Y^{(1)}_\alpha(t_0+\Delta t).
\eea
\item {\it Fully developed nonstationary field:} Here, ${Y}_\alpha(t)$ is governed by ${y}^{(0)}_{\alpha_i}(t_0+\Delta t) = {y}^{(0)}_{\alpha_i}(t_0)$ and can be represented as a traditional random walk within the complex plane, as shown in figure \ref{fig:randomwalk}b, whence
\bea
Y_\alpha(t_0+\Delta t) = Y^{(1)}_\alpha(t_0+\Delta t) + Y^{(0)}_\alpha(t_0).
\eea
In other words, $Y^{(1)}_\alpha(t_0+\Delta t)$ corresponds to the increment at $t=t_0+\Delta t$.
\item {\it Intermediate nonstationary (slipping) field:} This case is defined by $ {y}^{(0)}_{\alpha_i}(t_0+\Delta t)$ lying ``in between" $ {y}^{(0)}_{\alpha_i}(t_0)$ and ${0}$, as depicted in figure \ref{fig:randomwalk}c. Here, ${Y}^{(1)}_\alpha(t_0+\Delta t)$ ``builds onto'' a decayed version ${Y}^{(0)}_\alpha (t_0+\Delta t)$ of ${Y}^{(0)}_\alpha(t_0)$, whence
\bea
Y_\alpha(t_0+\Delta t) = Y^{(1)}_\alpha(t_0+\Delta t) + c_\alpha Y^{(0)}_\alpha (t_0)
\eea 
where $c_\alpha$ is a complex constant.
\end{itemize}
Thus, only in the quasi-stationary case does ${Y}_\alpha(t_0+\Delta t)$ not contain a contribution by ${Y}^{(0)}_\alpha (t_0+\Delta t)$.
In any case, the discretized ${Y}_\alpha(t)$ is a first-order Markov process because the transition between $Y_\alpha(t_0)$ and ${Y}_\alpha(t_0+\Delta t)$ is fully specified by $Y_\alpha(t_0)$.

\begin{figure}[htb] \begin{center} 
\begin{tabular}{c} \ 
\epsfysize=8cm 
\epsfbox{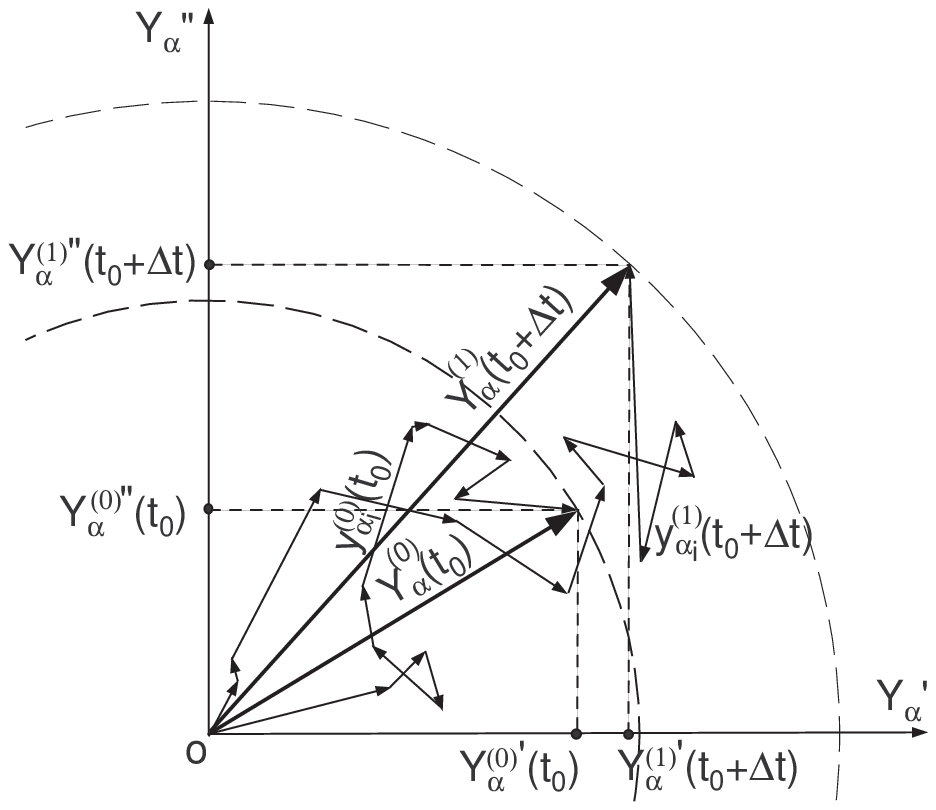}\ \\ 
(a)\\
\\
\epsfysize=8cm 
\epsfbox{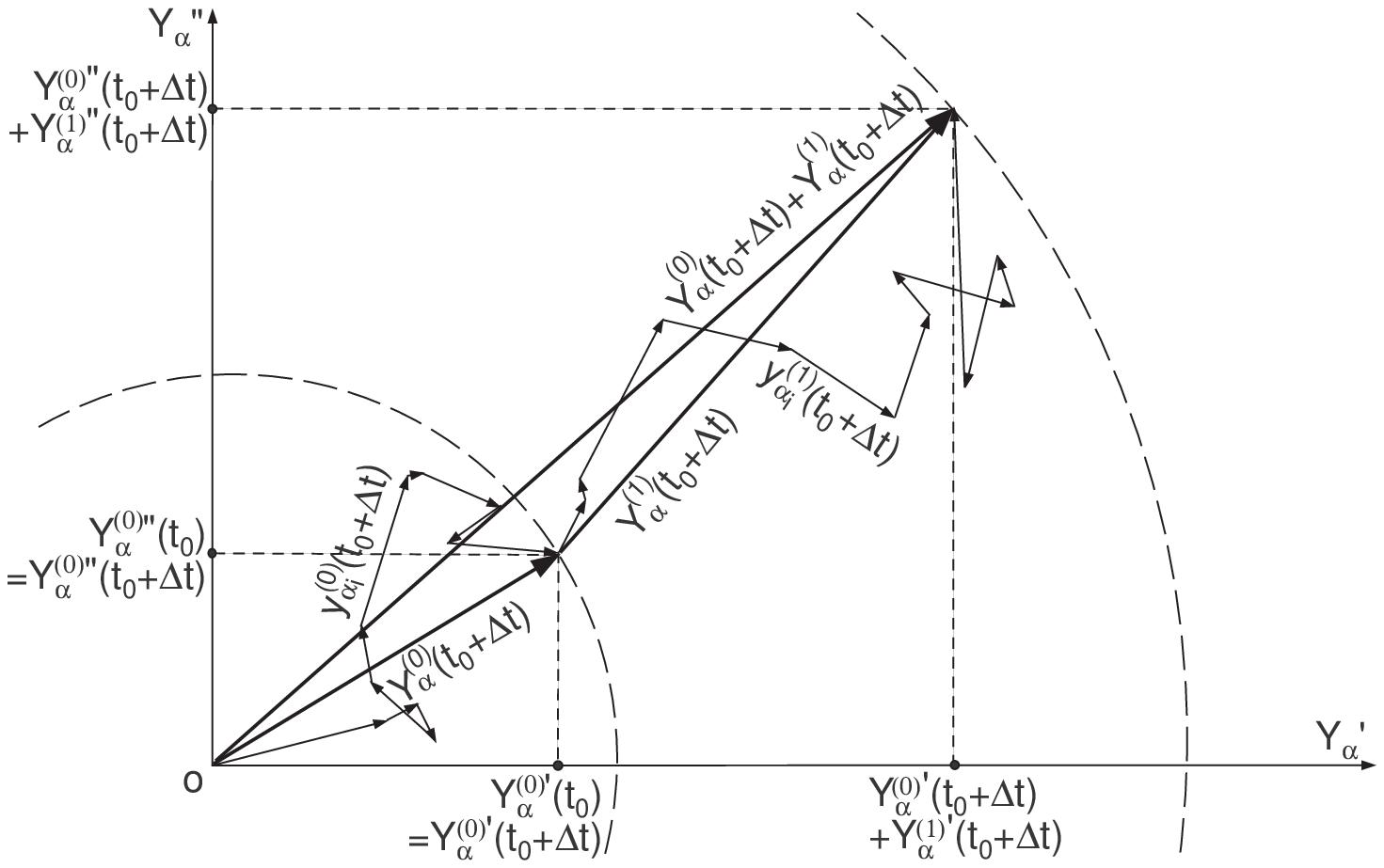}\ \\ 
(b)\\
\\
\end{tabular} \end{center}
\end{figure}

\clearpage

A description in terms of instantaneous cavity modes is only feasible insofar as the notion of instantaneous (eigen)frequency is applicable and sufficiently sharply localized. This, in itself, limits the applicability of the model to sufficiently small levels of nonstationarity \cite{arnaTEMCv47n4}.
These instantaneous modes are time-limited wavelets, rather than harmonic functions of infinite duration, enabling characterization in terms of an evolutionary spectrum \cite{prie1}.

To avoid the need for a parametric description \cite[ch. 10]{stra1}, we shall assume that the field fluctuations are independent and much more rapid than those of the state of the system and its parameters.
The latter can then be treated as constants. For the effective field, this is often a justifiable approximation. In section \ref{sec:timedeptau}, we treat the case of time constants with arbitrary but deterministic time dependence.

\section{Stochastic differential equations for real fields\label{sec:Langevin}}

In this section, we consider the relationship of an output field $Y(t)$ generated by an idealized random input field $X(t)$. Both $X(t)$ and $Y(t)$ are real modulated fields in the time domain; extension to complex analytic fields will be made in section \ref{sec:Langevin2bis}.

\subsection{Linear time-variant filtering of quasi-stationary random fields \label{sec:LTVtemp}}

Consider a real scalar quasi-stationary fluctuating acoustic or EM field $X(\ul{r},t)$ inside an enclosure, produced by a process of arbitrarily slow configurational changes (adiabatic boundary variations). This interior field $X(t)$ is generated by a given time-harmonic EM source $x_0\exp(\rmj\omega t)$ and satisfies the Helmholz or vector wave equation with quasi-statically varying boundary conditions. The rate of fluctuation of $X(t)$ is expressed by its correlation length $\tau_{\rho,X}$ calculated from the autocorrelation function (ACF) $\rho_X(t,\Delta t)$. The quasi-randomness of $X(t)$ manifests itself only on time scales well in excess of $\tau_{\rho,X}$, by definition.

The quasi-stationary field $X(t)$ serves as input to a dynamic process of realistic configurational changes, i.e., occurring with nonzero velocities of the boundaries. The corresponding output is the resulting perceived ``stirred'' field $Y(\ul{r},t)$ which is, to a certain extent, a weighted accumulation or aggregation of $X(t)$. 
The transformation from $X(\ul{r},t)$ to $Y(\ul{r},t)$ is here restricted to be linear but possibly nonuniform, i.e., $Y(\ul{r},t)$ can be obtained from $X(\ul{r},t)$ via linear time-variant filtering \cite{prie2}. 
Here, we assume that this filtering is characterized, locally, instantaneously, and in the mean, 
by a first-order process with characteristic relaxation time constant $\tau$ \cite{PRL}.
Assuming $\tau \gg \tau_{\rho,X}$, we can consider $Y(t)$ to be a first-order Markov process\footnote{This is not a fundamental limitation, because in the other case a generalized Fokker--Planck equation can still be derived, cf. e.g. \cite[eqns. (1.27) and (4.1)]{risk1}. 
} with an {\em apparently\/} white input process $X(t)$ (cf. section \ref{sec:LTV}). 
However, since $X(t)$ is a physically realizable sample-continuous noise field, the Stratonovich picture is preferred for the stochastic formulation of the sample-continuous noise field. For completeness, however, we shall also list corresponding results for the It\^{o} formalism \cite{vank1}. 

In the limit $\tau_{\rho,X} / \tau \rightarrow 0$, the integrated noise $B(\ul{r},t) \doteq \int^t X(\ul{r},t^\prime){\rm d}t^\prime$ represents a BEWL nonstationary process \cite{arnaTEMC1}, familiar from the theory of Brownian motion and random walks. 
This special case is analyzed in section \ref{sec:BEWL}.
For practical purposes, $X(t)$ and $Y(t)$ may be considered as sample-continuous random processes, e.g., by considering an ideal white noise process subjected to a `small' amount of local averaging \cite{arnaTEMC1,vanm1}. All integrated processes below are then properly defined in the mean-square sense.

We shall further be concerned with local statistical properties only, whence the dependence of $X$ and $Y$ on $\ul{r}$ will be dropped, thereby assuming that all fields are evaluated pointwise without local averaging.

\subsection{Langevin--It\^{o} equation\label{sec:LTV}}

Our first task is to derive the differential equation governing $Y(t)$ with $X(t)$ as its source term. 
To this end, we subdivide the observation interval $[t_0,t_n]$ of length ${\cal T}\doteq t_n-t_0$ into $n$ equal subintervals $[t_i,t_{i+1}]$ ($i=0,\ldots, n-1$), each of length $\delta t = t_{i+1}-t_i$, chosen such that
\bea
\tau_{\rho,X} \ll \delta t \ll {\cal T}.
\label{eq:ineqtimecst}
\eea
The input field $X(t)$ can be approximated by a series of discrete mutually independent Heaviside step functions and performs random jumps at discrete regular times $t_{i}$. We assume that the levels $X(t_i)$ are maintained between $t_i$ and $t_{i+1}$ (sample-and-hold), i.e., 
\bea
X(t_i \leq t <  t_{i+1})= X(t_{i}).
\eea
Since $\tau\not = 0$, $Y(t)$ is a smoothed weighted aggregate of $X(t)$. The fluctuations of $Y(t)$ are a result of fluctuations of 
$X(t)$ undergoing relaxed random spatial rotations and scalings of $X(t)$. Consequently, $\tau_{\rho,Y} \geq \tau_{\rho,X}$. 

\begin{figure}[htb] \begin{center} \begin{tabular}{l}
\ \epsfxsize=8cm 
\epsfbox{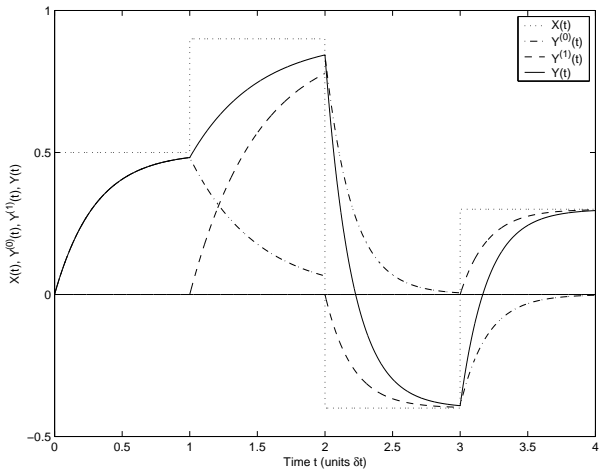}\ \\
\end{tabular}
\end{center}
{
\caption{\label{fig:nonstationary_diagram}
Discretized sample-and-hold input process $X(t)$ and resulting output process $Y(t)$. The actual continuous-time stirring process is obtained as the limit $\delta t \rightarrow 0$.}}
\end{figure}

At $t=t_{i+1}$, two changes occur simultaneously (figure \ref{fig:nonstationary_diagram}): 
(i) the previous value of the input field $X(t_i \leq t <  t_{i+1})$ ceases to exist. This can be interpreted as the ``switch-off'' of the source due to elapsing of the previous configuration. By itself, it would cause a decay in $|Y(t \geq t_{i+1})|$ to the null field (ground state); 
(ii) the next value of the source field $X(t \geq t_{i+1})$ comes into force. This can be interpreted as a ``switch-on'' of the source due to the generation of a new configuration. By itself, it would cause a transition of $Y(t \geq t_{i+1})$ from zero toward the next equilibrium value $X(t_{i+1})$ (asymptotic level if $X(t \geq t_{i+1})$ were to persist for $t\rightarrow +\infty$). 
Both effects are in competition if $|X(t_{i+1})| > |X(t)|$, while reinforcing each other if $|X(t_{i+1})| < |X(t)|$. 
The resultant field is therefore
\bea
Y(t_i \leq t < t_{i+1}) = Y(t_i) \exp \left ( - \frac{t-t_i}{\tau} \right ) + \left [ 1 - \exp \left ( - \frac{t-t_i}{\tau} \right ) \right ] X(t_i).
\label{eq:Langevinstart}
\eea
In this model, the influence on the future state at $t_{i+1}$ stretches only as far back as the end value of its immediate predecessor state at $t_i$, i.e., $Y(t)$ is a first-order Markov process. Moreover, the field does not make discontinuous jumps between consecutive states and hence $Y(t)$ is also sample-continuous.
Defining $\delta Y(t) \doteq Y(t)-Y(t_i)$, (\ref{eq:Langevinstart}) can be rewritten as the stochastic difference equation
\bea
\frac{\delta Y(t)}{\delta t} = 
\frac{\exp \left ( -\frac{\delta t}{\tau}\right ) - 1 }
     {\delta t} Y(t) 
+ \frac{1- \exp \left ( -\frac{\delta t}{\tau} \right )}
       {\delta t} X(t).
\label{eq:DE}
\eea
In the limit $\delta t \rightarrow 0$ [implying $\tau_{\rho,X} \rightarrow 0$, on account of (\ref{eq:ineqtimecst})], (\ref{eq:DE}) yields the stochastic differential equation (SDE)\footnote{If the process is irreversible \cite{weis1}, then it can be easily shown that (\ref{eq:SDE}) generalizes to $\dot{Y}(t) = - (1/\tau^{(0)}) Y(t) + (1/\tau^{(1)}) X(t)$, where 
$\tau^{(0),(1)} \doteq \lim_{t^\prime \rightarrow t_{0-},t_{0+}} \tau(t)$.}
\bea
\dot{Y}(t) = - \tau^{-1} Y(t) + \tau^{-1} X(t).
\label{eq:SDE}
\label{eq:Langevin}
\eea
This equation has been encoutered in a multitude of other applications, cf., e.g., \cite{voge1}, but note the multiplicator $1/\tau$ for the source term here.
Since $\tau$ is independent of $X(t)$ and, hence, of $Y(t)$, the SDE (\ref{eq:SDE}) is of the Langevin--It\^{o} type and has solution
\bea
Y(t) = y_0 
\exp \left ( - \frac{t-t_0}{\tau} \right ) 
+ \tau^{-1}\int^{t}_{t_0} X(t^\prime) \exp \left ( - \frac{t-t^\prime}{\tau} \right ) {\rm d}t^\prime
\label{eq:gensol}
\eea
where $y_0\doteq Y(t=t_0)$. 
The solution (\ref{eq:gensol}) expresses the fact that $Y(t)$ is an accumulation of all past values of $X(t)$ weighted by an exponential relaxation.
In certain scenarios, it may be possible to intervene manually and reset $y_0$ to zero. In general, however, $Y(t_i)$ equals the value of $Y(t)$ that was reached at the end of $[t_{i-1},t_i]$ and is hence governed by $Y(t=t_{i-1})$, $\delta t$ and $\tau$. 

\subsection{Moments, drift and diffusion coefficients}
Apart from providing information on centrality, dispersion, etc., the moments of $Y(t)$ enable determination of the coefficients of the Kramers--Moyal and Fokker--Planck equations for its probability density function (PDF) $f_Y(y,t)$. 
An ideal random, i.e., Gauss normally distributed source field $X(t)$ can be completely characterized by its mean value $\langle X(t) \rangle$ and autocovariance function, with the aid of
\bea
\langle X(t) \rangle =0
\eea
and, for an assumed wide-sense stationary $X(t)$, 
\bea
\langle X(t)X(t^\prime)\rangle 
= \sigma^2_X 
\rho_X (t-t^\prime).
\eea
In the limit of ideal white noise,
\bea
\rho_X (t-t^\prime) \rightarrow \delta (t-t^\prime).
\eea
In practice, the noise $X(t)$ is a coloured [$\rho_X(t) \not = \delta(t)$].
In such cases, the idealization to white noise may be insufficiently accurate, whence the full expressions of $\rho_X(t)$ 
must then be used. 
Several different approaches exist, in particular for colouration by an exponentially correlated $X(t)$ \cite{moss1}. Pure Markov processes associated with such an ACF are, however, not mean-square differentiable, as is physically required, because $\rho_{X}(\tau)$ is not differentiable at $\tau=0$ and therefore require at least some small degree of local averaging \cite{arnaTEMC1}, \cite{vanm1}.

From (\ref{eq:gensol}), for $t_0 \leq t, t^\prime < t_1$, 
\bea
\fl \hspace{1.5cm}
\langle Y (t) \rangle
&= 
y_0 \exp \left ( - \frac{t-t_0}{\tau}\right )
\stackrel{(\frac{t-t_0}{\tau} \rightarrow +\infty)}{ \rule[0.82mm]{2cm}{0.15mm}\hspace{-3mm}\longrightarrow} 0, 
\label{eq:mean}\\
\fl \hspace{1.5cm}
\langle Y(t) Y(t^\prime) \rangle
&= 
y^2_0 \exp \left ( - \frac{t+t^\prime-2t_0}{\tau} \right ) + 
\int^t_{t_0} 
\left [ \int^{t^\prime}_{t_0} 
\exp \left ( - \frac{t-u+t^\prime-v}{\tau} \right ) 
\frac{\sigma^2_X}{\tau^2} 
\rho_X (u-v) {\rm d}v 
\right ] 
{\rm d}u
\label{eq:YtYtprimetemp}
\nonumber\\
\\
\fl
&\rightarrow
y^2_0 \exp \left ( - \frac{t+t^\prime-2t_0}{\tau} \right ) + \frac{\sigma^2_X}{2\tau} 
\left [ \exp \left ( - \frac{|t-t^\prime|}{\tau} \right ) - \exp \left ( - \frac{t+t^\prime-2t_0}{\tau} \right ) \right ] \label{eq:cov}\\
\fl
& 
\stackrel{(\frac{t+t^\prime-2t_0}{\tau} \rightarrow +\infty )}{ \rule[0.82mm]{2cm}{0.15mm}\hspace{-3.2mm}\longrightarrow}  ~~~
\frac{\sigma^2_X}{2\tau} 
\exp \left ( - \frac{|t-t^\prime|}{\tau} \right ),
\label{eq:covar}
\eea
where (\ref{eq:cov}) was obtained by considering the limit of a delta-correlated $X(t)$.
Hence,
\bea
\sigma^2_Y (t) &\equiv& \langle Y^2(t) \rangle - \langle Y (t) \rangle^2 =
\frac{\sigma^2_X}{2\tau} \left [ 1 - \exp \left ( - \frac{2\left ( t - t_0 \right )}{\tau} \right )  \right ] 
\stackrel{\left (\frac{t-t_0}{\tau} \rightarrow +\infty\right )}{\rule[0.82mm]{2cm}{0.15mm}\hspace{-3.2mm}\longrightarrow} \frac{\sigma^2_X}{2\tau} \label{eq:var}.
\eea
Recall that, in arriving at (\ref{eq:YtYtprimetemp}), the invoked interchange of the order of ensemble averaging and integration is permitted provided that a constant $m$ exists such that $|Y(t)| < m$ and provided $\langle Y(t) \rangle$ exists for all $t$ within the integration interval \cite[section IV.5]{khol1}.
For $(t-t_0)/\tau\ll 1$, the mean value and variance evolve approximately linearly with respect to $t$:
\bea
\langle Y (t) \rangle \simeq y_0 \left ( 1 - \frac{ t - t_0 }{\tau} \right ),
~~~~~~
\sigma^2_Y \simeq \frac{\sigma^2_X}{\tau^2} \left ( t-t_0 \right ).
\eea

In practice, the stirred field $Y(t)$ is typically a narrowband hybrid amplitude-plus-phase modulated function of time, whose statistics as a result of the stirring process vary slowly relative to the fluctuations of $X(t)$ (weak nonstationarity). 
Nevertheless, the application of the analytic field concept $Y(t)$ does not restrict this process to be narrowband, for it suffices that the maximum frequency present in the spectral density does not exceed one half of the instantaneous average center frequency.

The mean rate of variation of stir states (average derivative) of the stirred field is expressed by the {drift coefficient} $D^{(1)}_Y (y,t)$, 
\bea
D^{(1)}_Y (y,t) \doteq
\lim_{\delta t \rightarrow 0} \left . \frac{\langle Y(t+\delta t)- \langle Y(t) \rangle \rangle}{1! ~\delta t} \right |_{Y=y}
= - \frac{y}{\tau} ,
\label{eq:driftsimple}
\eea
in which $y$ has a sharp, i.e., definite value at time $t$.
The mean rate of increase of the uncertainty of the state (spread of derivative) defines the field diffusion coefficient $D^{(2)}_Y (y,t)$, 
\bea
D^{(2)}_Y (y,t) 
\doteq
\lim_{\delta t \rightarrow 0} \left . \frac{\langle \left [  Y(t+\delta t) - \langle Y(t) \rangle \right ]^2 \rangle}{2! ~\delta t} \right |_{Y=y}
= \frac{\sigma^2_X}{2\tau^2}
\label{eq:diffusionsimple}
\eea
which, unlike $D^{(1)}_Y$, is independent of $y$. 
Since the source term ${X}(t)/\tau$ in (\ref{eq:Langevin}) is independent of $Y(t)$, the spurious drift $(1/2) \partial D^{(2)}(y,t) / \partial y $ is zero. 
Consequently, the It\^{o} and Stratonovich forms of the SDE (\ref{eq:Langevin}) are identical in this case \cite{vank1}. 

Two particular asymptotic modes of operation \cite{arnaTEMC1} can be retrieved as limiting cases of (\ref{eq:SDE}).
If $D^{(1)}_Y=0$ then we obtain a pure diffusion of the random field, because then $\dot{Y}(t) = \tau^{-1} X(t)$ yields the BEWL process $Y(t) = \tau^{-1} B(t)$. At the other extreme, if $\tau \rightarrow 0$ (implying no local temporal averaging, i.e., zero memory) then $\dot{Y}(t)=0$, whence the original point process is retrieved, i.e., $Y(t) = X(t)$.

\subsection{Fokker--Planck equation\label{sec:FokkerPlanck}}

On limiting the Kramers--Moyal equation for finite differences, viz.,  
\bea
\frac{\delta f_Y(y,t|y_0,t_0)}{\delta t} 
= \sum^{+\infty}_{n=1} 
\frac{(-1)^n}{n!} 
\left ( \frac{\delta}{\delta y} \right )^n 
\left [ \frac{\left \langle \left [ Y(t+\delta t) - Y(t) \right ]^n \right \rangle}{\delta t} f_Y(y,t|y_0,t_0) \right ],
\eea
to terms up to and including second order ($n\leq 2$), and upon substituting (\ref{eq:mean})--(\ref{eq:YtYtprimetemp}), we obtain
\bea
\frac{\delta f_Y(y,t|y_0,t_0)}{\delta t} 
&=&
\left [ \frac{1- \exp \left ( -\frac{\delta t}{\tau} \right )}
       {\delta t} \right ] \frac{\delta \left [ y f_Y(y,t|y_0,t_0) \right ] }{\delta y} \nonumber\\ 
&~& + \left \{ \frac{y^2_0}{2\delta t} \left [ 1 - \exp \left (-\frac{\delta t}{\tau} \right ) \right ]^2
+
\frac{\sigma^2_X}{2\tau} \left [ \frac{1- \exp \left ( -\frac{2\delta t}{\tau} \right )}
       {2\delta t} \right ] \right \} \frac{\delta^2 f_Y(y,t|y_0,t_0)}{\delta y^2}. \nonumber\\
\eea
In the limit $\delta t \rightarrow 0$, this results in the FPE 
\bea
\frac{\partial f_Y(y,t|y_0,t_0)}{\partial t} 
&=&
- \frac{\partial}{\partial y} \left [ D^{(1)}_Y(y,t) f_Y(y,t|y_0,t_0) \right ] + \frac{\partial^2}{\partial y^2} \left [ D^{(2)}_Y(y,t) f_Y(y,t|y_0,t_0)\right ] \label{eq:FPEgen}\\
&=&
\frac{1}{\tau} \frac{\partial }{\partial y} \left [ y f_Y(y,t|y_0,t_0) \right ]
+ \left ( \frac{y^2_0}{\tau} + \frac{\sigma^2_X}{2\tau^2} \right ) \frac{\partial^2 }{\partial y^2} f_Y(y,t|y_0,t_0)
\label{eq:FP0}
\eea
with specified boundary conditions $f_Y(\pm\infty,t|y_0,t_0)=0$. This PDE governs the evolution of the transition PDF (TPDF) $f_Y(\pm\infty,t|y_0,t_0)$ or the PDF $f_Y(y,t)$ whereby $f_Y(y,t=t_0)=f_Y(y_0,t_0)$, up to second order.
The deterministic part of $X(t)$ and hence of $Y(t)$ satisfies the Liouville equation, i.e., (\ref{eq:FP0}) with $\sigma^2_X=0$ and $ \tau > 0$, yielding a continuous probability current (cf. (\ref{eq:defprobcurr})].

Using standard methods \cite[section 1.11]{coff1}, the general solution of (\ref{eq:FP0}) is obtained as an Ornstein--Uhlenbeck process
\bea
f_Y(y,t|y_0,t_0) = \sqrt{ \frac{\tau} { \pi \sigma^2_X \left \{ 1 - \exp \left [ - \frac{2(t-t_0)}{\tau} \right ] \right \} }}
\exp \left [ - \frac{ \tau \left [ y - y_0 \exp \left ( - \frac{t-t_0}{\tau} \right ) \right ]^2 }
                    { \sigma^2_X \left \{ 1 - \exp \left [ - \frac{2(t-t_0)}{\tau} \right ] \right \}} 
     \right ].
\label{eq:generalt}
\eea

If, when considering a collective of boundary states, $y_0$ does not exhibit a sharp value across the ensemble but is characterized by a random variable $Y_0$ that itself exhibits the same stationary PDF, i.e., if 
\bea
f_{Y_0}(y_0,t_0) = \sqrt{\frac{\tau}{\pi \sigma^2_X}} \exp \left ( - \frac{\tau y^2_0}{\sigma^2_X} \right )
\eea
then $f_Y(y,t)$ is itself stationary throughout, with
\bea
f_Y(y,t)
&= 
C \frac{\exp \left [ \int^y_{y_0} \frac{D^{(1)}_Y (y^\prime,t)}{ D^{(2)}_Y(y^\prime,t)} {\rm d}y^\prime \right ] }
              { D^{(2)}_Y(y,t) }
&=
C \int^{+\infty}_{-\infty} f_Y(y,t|y_0,t_0) f_{Y_0}(y_0,t_0) {\rm d} y_0\nonumber\\
&=
\sqrt{\frac{\tau}{\pi\sigma^2_X}} \exp \left ( - \frac{\tau y^2}{\sigma^2_X} \right )
\label{eq:TPDFYsubensemble}
\eea
as obtained from (\ref{eq:generalt}) for $(t-t_0)/\tau \rightarrow +\infty$, 
in which $C$ is a normalization constant. This stationary solution -- sometimes referred to as the subensemble extracted from $f_Y(y,t)$ as defined by $y_0$ -- exists provided the ergodicity condition for $Y(t)$, i.e.,
\bea
{\rm l.i.m.}_{T\rightarrow +\infty} [ h_T(y) - \langle h_T(y) \rangle ] = 0
\eea
viz.,
\bea
\fl
\lim_{T\rightarrow +\infty} 
\int^T_0 \int^T_0 
\left \{ 
\int^{+\infty}_{-\infty} \int^{+\infty}_{-\infty} 
\frac{w_T \left ( y \right ) w_T \left ( y_0 \right )}{T^2} 
\left [ 
f_Y \left ( y,t|y_0,t_0 \right ) 
- f_Y \left ( y,t \right ) f_Y \left ( y_0,t_0 \right ) 
\right ] 
{\rm d}y {\rm d}y_0 
\right \} 
{\rm d}t {\rm d}t_0 = 0
\eea
is satisfied, in which $w_T[y(t)] \doteq T^{-1} \int^T_0 h[y(t)] {\rm d}t$ is an arbitrary deterministic function. In particular, the difference between the moments for boundary averaging and spatial (interior domain) averaging [i.e., for $w_T(y)=y^m$] should fall off sufficiently rapidly with increasing sample length $T$.
Thus, upon averaging over all possible values of a Gauss normal $Y_0$, the perceived variance is independent of $t$ and inversely proportional to $\tau$:
\bea
\langle \sigma^2_Y \rangle = 
\sigma^2_Y \left ( \frac{t}{\tau} \right ) = 
\frac{\sigma^2_X}{2\tau} .
\eea
Compared to $f_Y(y,t)$ for sharp initial values $y_0$, the drift now vanishes in the mean and the field variance increases by an amount $2 \sigma^2_X \exp [ - {2(t-t_0)}/{\tau} ]$, i.e., by a factor 
$
{2 \sigma^2_X} /
     \{ \exp \left [ {2(t-t_0)}/{\tau} \right ] - 1 \}
$.

\section{Application to analytic fields, energy density and magnitude\label{sec:Langevin2bis}}

\subsection{Complex analytic scalar field}
Thus far, the stochastic characterization was for real harmonic or quasi-harmonic fields $X(t)$ and $Y(t)$.
We now associate with $X(t)$ the Gabor analytic field \cite{arnaTEMCv47n4}
$X(t) - \rmj {\cal H}[X(t)]$, again defined for real $t$, where ${\cal H}[X(t)] \doteq \pi^{-1} - \hspace{-2.7mm} \int^{+\infty}_0 X(t)/(t-u){\rm d}u$ denotes the Hilbert transform of $X(t)$.
We shall further denote this complex field by $X(t) = X^\prime(t) - \rmj X^{\prime\prime}(t)$, where $X^\prime(t)$ and $X^{\prime\prime}(t)$ are the in-phase and quadrature components. Similar definitions and notations apply to the complex $Y(t)$.

If $\ul{X}(t)$ represents a vector field, its Cartesian components
$X_\alpha(t)=X^\prime_\alpha(t)-\rmj X^{\prime\prime}_\alpha (t)$ $(\alpha=x,y,z)$ are assumed to be mutually uncorrelated. 
Consider $X_\alpha(t)$ to be circular Gauss normal and delta-correlated. The output analytic field is then governed by two scalar SDEs, i.e.,
\bea
\dot{{Y}}^{\prime(\prime)}_\alpha (t) + \tau^{-1} {Y}^{\prime(\prime)}_\alpha (t) = \tau^{-1} {X}^{\prime(\prime)}_\alpha (t)
\eea
with $\langle {X}^{\prime(\prime)}_\alpha (t)\rangle = {0}$, $\langle {X}_\alpha (t) {X}_\alpha (t^\prime) \rangle =\sigma^2_{X_\alpha} \delta (t-t^\prime)$. The solution is an Ornstein-Uhlenbeck process 
\bea
{Y}^{\prime(\prime)}_\alpha (t) 
&=& {y}^{\prime(\prime)}_{\alpha_0} \exp \left ( - \frac{t-t_0}{\tau} \right ) + \tau^{-1} \int^t_{t_0} {X}^{\prime(\prime)}_\alpha \left ( t^\prime \right ) \exp \left ( - \frac{t-t^\prime}{\tau} \right ) {\rm d} t^\prime
\label{eq:gensolvec}
\eea
with the initial value ${y}^{\prime(\prime)}_{\alpha_0} \doteq {Y}^{\prime(\prime)}_{\alpha} (t=t_0)$.
The associated mean value and covariance are
\bea
\fl
\langle {Y}^{\prime(\prime)}_\alpha (t)\rangle
&=
{y}^{\prime(\prime)}_{\alpha_0} \exp \left ( - \frac{t-t_0}{\tau} \right ) ,\\
\fl
\langle {Y}^{\prime(\prime)}_\alpha (t) {Y}^{\prime(\prime)}_\alpha ( t^\prime ) \rangle
&=
{y}^{\prime(\prime)^2}_{\alpha_0} \exp \left ( - \frac{t+t^\prime-2t_0}{\tau} \right )
+
\frac{\sigma^2_{X^{\prime(\prime)}_\alpha}}{2\tau} \left [ \exp \left ( - \frac{|t - t^\prime|}{\tau} \right ) - \exp \left ( - \frac{t + t^\prime-2t_0}{\tau} \right ) \right ] .
\label{eq:covvec}
\eea
In the limit $t/\tau$, $t^\prime/\tau \rightarrow +\infty$, 
\bea
\langle Y^{\prime(\prime)}_\alpha (t) \rangle\rightarrow 0,~~~\langle {Y}^{\prime(\prime)}_\alpha (t) {Y}^{\prime(\prime)}_\alpha (t^\prime) \rangle
\rightarrow
\frac{\sigma^2_{X^{\prime(\prime)}_\alpha}}{2\tau} \exp \left ( - \frac{|t - t^\prime|}{\tau} \right ) .
\eea

\subsection{Energy density}

\subsubsection{Scalar fields or Cartesian components of a vector field \label{sec:CartEnergy}}
\paragraph{Moments\label{sec:CartMoments}}
The energy density is the physical quantity of fundamental interest, because an ideal random field is incoherent, i.e., delta-correlated with respect to time.
Consider
the normalized energy density $U_\alpha(t)={Y^\prime_\alpha}^2(t) + {Y^{\prime\prime}_\alpha}^2(t)$, after normalization by a factor relating to the constitutive properties of the stationary medium, which are not important in the statistical characterization for a deterministic medium. The stationary Cartesian density decreases for increasing $\tau$: 
\bea
\left \langle U_\alpha (t) \right \rangle 
&=& 
\left \langle {Y^\prime_\alpha}^2 (t) \right \rangle 
+ 
\left \langle {Y^{\prime\prime}_\alpha}^2 (t) \right \rangle 
= \frac{\sigma^2_{X_\alpha}}{\tau}
+ \left ( y^2_{\alpha_0} - \frac{\sigma^2_{X_\alpha}}{\tau} \right ) \exp \left [ - \frac{2(t-t_0)}{\tau} \right ] 
\rightarrow \frac{\sigma^2_{X_\alpha}}{\tau} 
\eea
where $y^2_{\alpha_0} \equiv {y^\prime_{\alpha_0} }^2  + {y^{\prime\prime}_{\alpha_0} }^2 = u_{\alpha_0}$ and $\sigma^2_{X_\alpha} \doteq \sigma^2_{X^\prime_\alpha} = \sigma^2_{X^{\prime\prime}_\alpha} $. 
On account of the Isserlis moment theorem for a circular Gauss normal  $X_\alpha(t)$, it follows that $\left \langle {Y^{\prime(\prime)}_\alpha}^4 \right \rangle = 3 \left \langle {Y^{\prime(\prime)}_\alpha}^2 \right \rangle^2$, whence
\bea
\sigma^2_{U_\alpha} (t) = 2 \left \langle U_\alpha(t) \right \rangle^2 - 4 \left \langle {Y^\prime_\alpha}^2(t) \right \rangle \left \langle {Y^{\prime\prime}_\alpha}^2(t) \right \rangle .
\eea

\paragraph{SDE\label{sec:CartSDE}}

From 
\bea
\left \langle U_\alpha(t+\delta t)-U_\alpha(t) \right \rangle
&=&
\left \langle {{Y}^{\prime}_\alpha      }^2(t+\delta t)-{{Y}^{\prime      }_\alpha}^2(t) \right \rangle +
\left \langle {{Y}^{\prime\prime}_\alpha}^2(t+\delta t)-{{Y}^{\prime\prime}_\alpha}^2(t) \right \rangle \nonumber\\
&=&
\left ( 
\frac{\sigma^2_{X_\alpha}}{2 \tau} 
- y^{\prime^2}_{\alpha_0}
\right ) 
\exp \left [ - \frac{2 (t-t_0)}{\tau} \right ] \left [ 1 - \exp \left ( - \frac{2\delta t}{\tau} \right ) \right ]\nonumber\\
&~&+
\left ( 
\frac{\sigma^2_{X_\alpha}}{2 \tau} 
- y^{\prime\prime^2}_{\alpha_0}
\right ) 
\exp \left [ - \frac{2 (t-t_0)}{\tau} \right ] \left [ 1 - \exp \left ( - \frac{2\delta t}{\tau} \right ) \right ]
\label{eq:firstdiffu}
\eea
it follows that the energy drift coefficient is constant with respect to $u_\alpha$:
\bea
D^{(1)}_{U_\alpha} (u_{\alpha}, t) = \lim_{\delta t\rightarrow 0} \left. \frac{\left \langle U_\alpha(t+\delta t)-U_\alpha(t) \right \rangle
}{\delta t}\right|_{U_\alpha=u_\alpha}
 = 
\frac{2}{\tau}
\left ( \frac{\sigma^2_{X_\alpha}}{\tau} - u_{\alpha_0} \right ) \exp \left [ -\frac{2 (t-t_0)}{\tau} \right ]
\label{eq:driftUCart}
\eea
where $u_{\alpha_0} \doteq U_\alpha(t_0) \equiv y^2_{\alpha_0}$.
The energy diffusion coefficient is calculated in \ref{app:diffcffCart} as
\bea
D^{(2)}_{U_\alpha} (u_\alpha,t) 
&=& \lim_{\delta t\rightarrow 0} \left. \frac{\left \langle \left [ U_\alpha(t+\delta t) - U_\alpha(t) \right ]^2 \right \rangle}{2\delta t} \right|_{U_\alpha=u_\alpha}
= \frac{2 \sigma^2_{X_\alpha}}{\tau^2} u_\alpha .
\label{eq:diffusionUCart_copy}
\eea
The SDE for $U_\alpha(t)$ can be obtained from (\ref{eq:driftUCart})--(\ref{eq:diffusionUCart_copy}) and is nonlinear. Although the transformation from the complex analytic field to the real energy density is a multivariate one, the SDE can in this particular case be uniquely determined via inversion \cite[section 3.4.1]{risk1} because $Y^\prime_\alpha(t)$ and $Y^{\prime\prime}_\alpha(t)$ are orthogonal, whence the diffusion dyadic $\dul{D}^{(2)}_{Y^\prime_\alpha Y^{\prime\prime}_\alpha}(y^\prime_\alpha,y^{\prime\prime}_\alpha,t)$ is diagonal.
In the It\^{o} formulation, the SDE is 
\bea
\dot{U}_\alpha(t) 
&=& 
D^{(1)}_{U_\alpha} (U_\alpha, t) + \frac{\sqrt{2 D^{(2)}_{U_\alpha}(U_\alpha, t) }}{\sigma_{X_\alpha}} \dot{B}_\alpha(t) \label{eq:LangevUCartOUIto_gen}\\ 
&=& 
\frac{2}{\tau} 
\left \{
\left (
\frac{\sigma^2_{X_\alpha}}{\tau} 
- u_{\alpha_0} 
\right ) 
\exp \left [ - \frac{2 (t-t_0)}{\tau} \right ] 
+ 
\sqrt{U_\alpha(t)} \dot{B}_\alpha(t)\label{eq:LangevUCartOUIto}
\right \}
\eea
in which we have written $\dot{B}_\alpha(t)$ for $X_\alpha(t)$, 
whereas in the Stratonovich formulation,
\bea
\dot{U}_\alpha(t) 
&=& \left ( D^{(1)}_{U_\alpha}(U_\alpha, t) - \frac{1}{2} \frac{\partial D^{(2)}_{U_\alpha}(U_\alpha, t) }{\partial u_\alpha}\right ) + \frac{\sqrt{2 D^{(2)}_{U_\alpha}(U_\alpha, t) }}{\sigma_{X_\alpha}} \dot{B}_\alpha(t) \label{eq:LangevUCartOUStrat_gen}
\\ 
&=& 
\frac{2}{\tau} 
\left \{
\left (
\frac{\sigma^2_{X_\alpha}}{\tau} 
- u_{\alpha_0} \right ) \exp \left [ - \frac{2 (t-t_0)}{\tau} \right ] 
- \frac{\sigma^2_{X_\alpha}}{2\tau}
+ 
\sqrt{U_\alpha(t)} \dot{B}_\alpha(t)
\right \} ,
\label{eq:LangevUCartOUStrat}
\eea
involving a multiplicative nonlinear white noise source process that is no longer of the Langevin--It\^{o} type, which can be traced to the fact that $D^{(2)}_{U_\alpha}$ depends on $u_\alpha(t)$. Since the transformation from $Y_\alpha(t)$ to $U_\alpha(t)$ is nonlinear, the Stratonovich formulation contains the additional spurious drift term $-\sigma_{X_\alpha}^2 / \tau^2 $.
The SDEs (\ref{eq:LangevUCartOUIto}) and (\ref{eq:LangevUCartOUStrat}) can easily be integrated numerically, for example using the Cauchy--Euler method, to generate sample functions.

\paragraph{FPE\label{sec:CartFPE}}
Following (\ref{eq:FPEgen}), the associated FPE for $f_{U_\alpha}(u_\alpha,t|u_{\alpha_0},t_0)$ is
\bea
\frac{\partial}{\partial t}                           f_{U_\alpha}\left ( u_\alpha,t|u_{\alpha_0},t_0 \right ) 
&=&
- \frac{2}{\tau} \left ( \frac{\sigma^2_{X_\alpha}}{\tau} - u_{\alpha_0} \right ) \exp \left [ - \frac{2 \left ( t - t_0 \right )}{\tau} \right ]
\frac{\partial}{\partial u_\alpha}                    f_{U_\alpha}\left ( u_\alpha,t|u_{\alpha_0},t_0 \right ) \nonumber\\
&~&+
\frac{2\sigma^2_{X_\alpha}}{\tau^2} 
\frac{\partial^2}{\partial u^2_\alpha} \left [ u_\alpha f_{U_\alpha}\left ( u_\alpha,t|u_{\alpha_0},t_0 \right ) \right ] 
\label{eq:FPenergystirCart}
\eea
with specified initial value $U_\alpha(t_0) \doteq u_{\alpha_0}$, i.e., initial condition
\bea
f_{U_\alpha} \left ( u_\alpha, t_0 \right ) = \delta \left ( u_\alpha - u_{\alpha_0} \right )
\label{eq:IC_FPenergystirCart}
\eea
and boundary conditions
\bea
f_{U_\alpha}(0,t|u_{\alpha_0},t_0)= f_{U_\alpha}(+\infty,t|u_{\alpha_0},t_0)= 0.
\label{eq:BC_FPenergystirCart}
\eea
In (\ref{eq:FPenergystirCart}), the product of the field variate with its PDF occurs through its second derivative, rather than its first derivative in (\ref{eq:FP0}).
The sought $f_{U_\alpha}(u_\alpha,t)$ can be obtained via an integral transformation of (\ref{eq:FPenergystirCart})
(cf. \ref{app:solvingFPECart}) and can be expressed as the Laplace inversion formula
\bea
f_{U_\alpha}(u_\alpha,t) &=& 
\left ( \rmj 2 \pi \right )^{-1} 
\int^{\gamma+\rmj\infty}_{\gamma-\rmj\infty} 
{\cal F}_{S} \left ( s,t \right )
 \exp \left ( u_\alpha s \right )
{\rm d}s 
\label{eq:FPEgeneralsolutionCartEnergy}
\eea
for $\gamma > 0$, where
\bea
\fl \hspace{1cm}
{\cal F}_{S} \left ( s,t \right )
&=
\exp \left [ - \frac{\tau^2 u_{\alpha_0} }{2 \sigma^2_{X_\alpha} \left ( t - t_0 + \frac{\tau^2}{2 \sigma^2_{X_\alpha} s} \right )} 
\right ]
\nonumber\\
\fl
&~~~~~~~~
\times
\exp \left \{
-
\int^{t - t_0 + \frac{\tau^2}{2 \sigma^2_{X_\alpha} s} }_{{\frac{\tau^2}{2 \sigma^2_{X_\alpha} s} }}
\frac{1}{t^{\prime\prime}} \left ( 1 - \frac{u_{\alpha_0}\tau}{\sigma^2_{X_\alpha}} \right ) 
\exp \left [ - \frac{2}{\tau} \left ( t - t_0 - t^{\prime\prime} \right ) - \frac{\tau}{\sigma^2_{X_\alpha} s} \right ] {\rm d} t^{\prime\prime} 
\right \}
\nonumber\\
\fl
&~~~+
\int^{t - t_0 + \frac{\tau^2}{2 \sigma^2_{X_\alpha} s} }_{{\frac{\tau^2}{2 \sigma^2_{X_\alpha} s} }}
\left \{ -1 + \left ( 1- \frac{u_{\alpha_0}\tau}{\sigma^2_{X_\alpha}} \right ) \exp \left [ - \frac{2}{\tau} \left ( t - t_0 - t^{\prime\prime} \right ) - \frac{\tau}{\sigma^2_{X_\alpha} s} \right ] 
\right \} 
\frac{2 \sigma^2_{X_\alpha} f_{U_\alpha}(0+,t)}{\tau^2}
\nonumber\\
\fl
&~~~~~~~~
\times
\exp \left \{
\int^{t - t_0 + \frac{\tau^2}{2 \sigma^2_{X_\alpha} s} }_{t^{\prime\prime}}
\frac{1}{t^{\prime\prime\prime}} \left ( 1 - \frac{u_{\alpha_0}\tau}{\sigma^2_{X_\alpha}} \right ) 
\exp \left [ -\frac{2}{\tau} \left ( t - t_0 - t^{\prime\prime\prime} \right ) - \frac{\tau}{\sigma^2_{X_\alpha} s} \right ] 
{\rm d} t^{\prime\prime\prime} 
\right \}
{\rm d} t^{\prime\prime} 
\nonumber\\
\fl
&~~~
\times
\exp \left \{
-
\int^{t - t_0 + \frac{\tau^2}{2 \sigma^2_{X_\alpha} s} }_{{\frac{\tau^2}{2 \sigma^2_{X_\alpha} s} }}
\frac{1}{t^{\prime\prime}} \left ( 1 - \frac{u_{\alpha_0}\tau}{\sigma^2_{X_\alpha}} \right ) 
\exp \left [ - \frac{2}{\tau} \left ( t - t_0 - t^{\prime\prime} \right ) - \frac{\tau}{\sigma^2_{X_\alpha} s} \right ] {\rm d} t^{\prime\prime} 
\right \}.
\label{eq:PDEfinalLTCart}
\eea
If the imposed boundary condition is $f_{U_\alpha}(0+,t) =0$ for all $t$, then ${\cal F}_S(s,t)$ is limited to the first term in (\ref{eq:PDEfinalLTCart}) only. If, in addition, $f_{U_\alpha}(u_\alpha,t_0)=0$ for all $u_\alpha$, then
\bea
{\cal F}_S(s,t) &=& 
\exp \left ( - \exp \left [ - \frac{2}{\tau} \left ( t - t_0 \right ) - \frac{\tau}{\sigma^2_{X_\alpha} s} \right ]
\left \{
{\rm Ei} \left [ \frac{2 }{\tau} \left ( t - t_0 \right ) + \frac{\tau}{\sigma^2_{X_\alpha} s} \right ]
-
{\rm Ei} \left [ \frac{\tau}{\sigma^2_{X_\alpha} s} \right ]
\right \}
\right )
\nonumber\\
&=&
\frac{
\sum^{+\infty}_{m=0} 
\left \{ 
\exp \left [ - \frac{2}{\tau} \left ( t - t_0 \right ) - \frac{\tau}{\sigma^2_{X_\alpha} s} \right ]
{\rm E}_1 \left [ - \frac{2}{\tau} \left ( t - t_0 \right ) - \frac{\tau}{\sigma^2_{X_\alpha} s} \right ]
\right \}^m
/ m! 
}
{
\exp \left [ - \frac{2}{\tau} \left ( t - t_0 \right ) \right ]~~
\sum^{+\infty}_{n=0} 
\left [ 
\exp \left ( - \frac{\tau}{\sigma^2_{X_\alpha} s} \right )
{\rm E}_1 \left ( - \frac{\tau}{\sigma^2_{X_\alpha} s} \right )
\right ]^n
/ n!
}
\label{eq:PDEfinalLTsimplified}
\eea
where
$
{\rm Ei} \left ( z < 0 \right ) \doteq - \int^{+\infty}_{-z} [\exp (-z^\prime)/z^\prime] {\rm d}z^\prime
$ is the exponential integral function \cite[eqn. (8.211.1)]{grad1} and
$
{\rm E}_1 \left ( z \right ) \doteq \int^{+\infty}_z [ \exp (-z^\prime)/z^\prime ] {\rm d}z^\prime = \exp \left ( - z \right ) \sum^{+\infty}_{n=0} {(-1)^n n!}/{z^{n+1}} \equiv - {\rm Ei} \left ( - z \right )
$. 
With the aid of the residue theorem, (\ref{eq:FPEgeneralsolutionCartEnergy})--(\ref{eq:PDEfinalLTCart}) can also be expressed as (cf. \ref{app:solvingFPECart})
\bea
f_{U_\alpha}\left ( u_\alpha, t \right ) = {\rm Res}_{s=0} \left [ {\cal F}_S(s,t) \exp \left ( u_\alpha s \right ) \right ] 
\label{eq:residue}
\eea
corresponding to the coefficient of $s^{-1}$ in the Laurent series expansion of ${\cal F}_S(s,t) \exp \left ( u_\alpha s \right )$, i.e.,
\bea
\fl \hspace{1cm}
{\cal F}_S(s,t) \exp \left ( u_\alpha s \right )
&=
\sum^{+\infty}_{\ell=0} \left \{
u_\alpha s - \exp \left [ - \frac{2\left(t-t_0\right )}{\tau} \right ]
\sum^{+\infty}_{m=0} \left ( - \frac{\tau}{\sigma^2_{X_\alpha} s} \right )^m \left [ {\rm ln} 
\left ( 1 + \frac{2\left(t-t_0\right ) \sigma^2_{X_\alpha} s}{\tau^2}  \right )
\right. \right.\nonumber\\ 
\fl
&~\left. \left.~~~~~~~~~~~~~~~~~~~~~~~~~~~~~~~
+ \sum^{+\infty}_{n=1} \frac{\left ( \frac{\tau}{\sigma^2_{X_\alpha} s} \right )^n}{n!\thinspace n }
\sum^n_{p=0} \left ( \begin{array}{c}n\\p\end{array} \right ) \left [ \frac{2\left (t-t_0\right )}{\tau} \right ]^p \right ] \right \}^{\ell} / \ell !
\label{eq:residue_expan}
\eea
which contains infinite numbers of both positive and negative powers of $s$.

\paragraph{Limit PDFs\label{sec:limitPDFCartEnergy}}
In the limit $(t-t_0)/\tau \rightarrow 0$, the exponential time dependence in (\ref{eq:FPenergystirCart}), (\ref{eq:FPenergystirCart_copy}), (\ref{eq:PDEtrfCart_copy}) and (\ref{eq:FPEtoODE}) disappears. 
The limit PDF can be obtained in closed form using a transformation of variables in the FPE (cf. \ref{app:limitPDFCartEnergy}) yielding
\bea
f_{U_\alpha} \left ( u_\alpha,t \right ) 
&=& 
\frac{\tau^2 }
     {2 \left ( t - t_0 \right ) \sigma^2_{X_\alpha} 
     }
\left ( \frac{u_{\alpha}}{u_{\alpha_0}}
\right )^{\nu }
\exp \left [ - \frac{ \tau^2 \left ( u_\alpha + u_{\alpha_0} \right )}
                    { 2 \left ( t - t_0 \right ) \sigma^2_{X_\alpha} }                                  
     \right ] 
I_{2\nu}
 \left [ \frac{ \tau^2 \sqrt{u_\alpha \thinspace u_{\alpha_0} }}{\left ( t - t_0 \right ) \sigma^2_{X_\alpha} } 
    \right ]
\label{eq:condprobenergyCart_smallt_largetau}
\eea
valid for $(t-t_0)/\tau \rightarrow 0$,
which involves in general a Bessel function of negative fractional order $2\nu$, with $\nu$ given by (\ref{eq:defnu}):
\bea
\nu \doteq -\frac{u_{\alpha_0}\tau}{2\sigma^2_{X_{\alpha}}} \leq 0 .
\label{eq:defnu_copy}
\eea 
[Recall that $I_{2\nu}(\cdot) \not = I_{-2\nu}(\cdot)$ unless $2 \nu$ is an integer number.] Observe that (\ref{eq:condprobenergyCart_smallt_largetau}) is in general not separable with respect to $u_\alpha$ and $t$, unlike for quasi-stationary fields.

As a special case, if $u_{\alpha_0} \tau / \sigma^2_{X_\alpha} \ll 1$ in addition, then $\mu=1/2$, $\nu=0$, 
whence (\ref{eq:condprobenergyCart_smallt_largetau}) reduces to a PDF for a diffusing and drifting energy density characterized by a Nakagami--Rice distribution with PDF
\bea
f_{U_\alpha} \left ( u_\alpha,t \right ) 
=
\frac{\tau^2 }
     {2 \left ( t - t_0 \right ) \sigma^2_{X_\alpha} 
     }
\exp \left [ - \frac{ \tau^2 \left ( u_\alpha + u_{\alpha_0} \right )}
                    { 2 \left ( t - t_0 \right ) \sigma^2_{X_\alpha} }                                  
     \right ] 
I_0 \left [ \frac{ \tau^2 \sqrt{u_\alpha \thinspace u_{\alpha_0} }}{\left ( t - t_0 \right ) \sigma^2_{X_\alpha} } 
    \right ]
\label{eq:condprobenergyCart_smallt_smalltau}
\eea
whose variance increases linearly with time. The order (type) of the PDF has asymptotically increased to zero.
This PDF constitutes a special case of the general solution (\ref{eq:FPEgeneralsolutionCartEnergy}). For arbitrary $t$, it corresponds to a noncentral $\chi^2_2$ PDF including a constant bias $u_{\alpha_0}$. Note that for $u_{\alpha_0}=0$, (\ref{eq:condprobenergyCart_smallt_smalltau}) reduces to a negative exponential ($\chi^2_2$) PDF for which $f_{U_\alpha}(0+,t)\not = 0$, i.e., it does not strictly satisfy the prescribed boundary condition at $u_{\alpha_0}=0$. Therefore, the latter PDF represents a singular case. 

At the other extreme, if $(t-t_0)/\tau \rightarrow + \infty$ (limit of long time or vanishingly short memory) then
(\ref{eq:FPenergystirCart}) becomes independent of $u_{\alpha_0}$, viz.,
\bea
\frac{\partial f_{U_\alpha}}{\partial t} = 
2 \frac{\partial f_{U_\alpha}}{\partial u_\alpha} 
+
u_\alpha 
\frac{\partial^2 f_{U_\alpha}}{\partial u_\alpha^2} .
\label{eq:FPE_Ualphalarget}
\eea
This FPE coincides with (\ref{eq:FPE_Ualphasmallt}) if we choose 
$\mu=0$, $\nu=-1/2$ in (\ref{eq:defnu}). 
An asymptotically stationary solution exists, because the coefficients of the FPE (\ref{eq:FPenergystirCart}) become asymptotically independent of time.
Hence, from (\ref{eq:condprobenergyCart_smallt_largetau}), the PDF for $(t-t_0)/\tau \rightarrow + \infty$ approaches
\bea
f_{U_\alpha} \left ( u_\alpha,t \right ) 
&=& 
\frac{\tau^2 }
     {2 \left ( t - t_0 \right ) \sigma^2_{X_\alpha} 
     }
\sqrt{\frac{u_{\alpha_0}}{u_\alpha}} 
\exp \left [ - \frac{ \tau^2 \left ( u_\alpha + u_{\alpha_0} \right )}
                    { 2 \left ( t - t_0 \right ) \sigma^2_{X_\alpha} }                                  
     \right ] 
I_1 \left [ \frac{ \tau^2 \sqrt{u_\alpha \thinspace u_{\alpha_0} }}{\left ( t - t_0 \right ) \sigma^2_{X_\alpha} } 
    \right ] .
\label{eq:condprobenergyCart_larget}
\eea

Comparing (\ref{eq:FPE_Ualphasmallt}) with (\ref{eq:FPE_Ualphalarget}), and (\ref{eq:condprobenergyCart_smallt_largetau}) with (\ref{eq:condprobenergyCart_larget}), it is found that the change in drift
gives rise to a different functional dependence of $f_{U_\alpha}(u_\alpha,t) $ on $u_\alpha$.
Thus, as time progresses, not only do the parameters and moments of the distribution evolve, in a continuous manner, but also the nature of the PDF itself. Both features are discussed in more detail below. In the special case when $\nu=-1/2$, i.e., $u_{\alpha_0}=\sigma^2_{X_\alpha}/\tau$, the initial and final asymptotic PDFs coincide.

At the other extreme, the asymptotically stationary state ($t/\tau \rightarrow +\infty$) corresponds to $Y_\alpha(t)$ performing a random walk in the complex plane with full
return to its reference state $(u_{\alpha_0},t_0)$ after each completed step (complete instantaneous relaxation). The underlying PDF of the diffusing $Y_\alpha(t)$ is then of course identical to the one for $X_\alpha(t)$, viz., a circular Gauss normal PDF with time-independent parameters. Therefore,
(\ref{eq:condprobenergyCart_smallt_smalltau}) reduces to a central $\chi^2_2$ PDF, as expected: 
\bea
f_{U_\alpha}\left ( u_\alpha,t \right ) = \frac{\tau}{\sigma^2_{X_\alpha}} \exp \left ( -\frac{\tau u_\alpha}{\sigma^2_{X_\alpha}} \right ) .
\eea

In general, the probability current for $f_{U_\alpha}(u_\alpha,t)$, i.e., 
\bea
\jmath_{U_\alpha} (u_\alpha,t)
&\doteq& D^{(1)}_{U_\alpha}(u_\alpha,t)
 f_{U_\alpha}(u_\alpha,t)
 - \frac{1}{2} \frac{\partial }{\partial u_\alpha} \left [ D^{(2)}_{U_\alpha}(u_\alpha,t)
 f_{U_\alpha}(u_\alpha,t)
 \right ] \label{eq:defprobcurr}\\
&=& \frac{\sigma^2_{X_\alpha}}{\tau^2} \left ( f_{U_\alpha} - u_\alpha \frac{\partial f_{U_\alpha}}{\partial u_\alpha} \right ) - \frac{2 u_{\alpha_0}}{\tau} f_{U_\alpha}(u_\alpha,t)
\eea
is the sum of a convection current $\dot{u}_\alpha f_{U_\alpha}(u_\alpha,t)$ plus a diffusion current $\partial f_{U_\alpha} (u_\alpha,t) / \partial u_\alpha$ of probability as the PDF evolves in time. Expressions for $\jmath_{U_\alpha}$ follow with the aid of the above expressions for $f_{U_\alpha}$, but are in general rather cumbersome. For the special case $u_{\alpha_0} =0$, however,
\bea
\jmath_{U_\alpha} = \frac{1}{2(t-t_0)} 
\exp \left ( - \frac{\tau^2 u_\alpha}{2(t-t_0) \sigma^2_{X_\alpha}} \right )
\left [ 1 + \frac {\tau^2 u_\alpha}{2(t-t_0) \sigma^2_{X_\alpha}} \right ],
\eea
which vanishes in the limit $(t-t_0)/\tau\rightarrow +\infty$.

\paragraph{First-order time-dependence of PDF: asymptotic PDF\label{sec:firstorderCart}}
Having investigated the solution $f_{U_\alpha}(u_\alpha, t)$ of (\ref{eq:FPenergystirCart}) for asymptotically small and large times, we now consider the dynamics of the PDF as a function of time. As mentioned before, the PDF for general $t$ is obtained by inversion of (\ref{eq:PDEfinalLTCart}), or (\ref{eq:PDEfinalLTsimplified}) in particular. 
With the boundary condition $f_{U_\alpha}(0+,t)=0$, the kernel
(\ref{eq:PDEfinalLTCart}) reads
\bea
\fl \hspace{1cm}
{\cal F}_S \left ( s, t \right )
&=
\exp \left [ - \frac{ u_{\alpha_0}s}{1+\frac{2(t-t_0) \sigma^2_{X_\alpha} s}{\tau^2} } \right ] \nonumber\\
\fl
&~~~\times
\exp \left \{ - \int^{t-t_0 + \frac{\tau^2}{2\sigma^2_{X_\alpha} s}}_{\frac{\tau^2}{2\sigma^2_{X_\alpha} s}} \frac{\left ( 1 - \frac{u_{\alpha_0} \tau}{\sigma^2_{X_\alpha}} \right )}{t^{\prime\prime}} \exp \left [ - \frac{2}{\tau} \left ( t-t_0 + \frac{\tau^2}{2\sigma^2_{X_\alpha} s} - t^{\prime\prime} \right ) \right ]                                                     {\rm d}t^{\prime\prime} \right \}.
\label{eq:Laplacetrfspecial_temp}
\eea
For $0 < (t-t_0)/\tau \ll 1$, on defining $t^{\prime\prime\prime} \doteq t^{\prime\prime} - \tau^2 / (2\sigma^2_{X_\alpha} s)$ and with\footnote{Note that this must be valid for all values of $s$ when integrating the inverse Laplace transformation (\ref{eq:FPEgeneralsolutionCartEnergy}). In particular, for the part of the Bromwich contour parallel to the imaginary axis in the complex plane, this requires an appropriate limit process for the principal value of the integral when $\Im[ s ] \rightarrow 0$. The contribution of the essential isolated singular point $s=0$ in the inversion integral can be neglected if it is assumed that $\sigma^2_{X_\alpha} / \tau$ and the positive $\Re[ s]\equiv \gamma$ in (\ref{eq:FPEgeneralsolutionCartEnergy}) are both sufficiently large whence 
$ ( \sigma^2_{X_\alpha} / \tau) \gamma \gg 1$ and that $2 t^{\prime\prime\prime}/\tau \ll 1$ for $t^{\prime\prime\prime} \in[0,t-t_0]$.} 
$t^{\prime\prime\prime}/ \tau \ll 1$, the exponential function in the kernel can be expanded to first order in $(t-t_0)/\tau$ with reference to $t^{\prime\prime\prime}=0$, yielding
\bea
\fl \hspace{1cm}
{\cal F}_S \left ( s, t \right )
&=
\exp \left \{ - \frac{ u_{\alpha_0}s}{1+\frac{2(t-t_0) \sigma^2_{X_\alpha} s}{\tau^2} }
-
\left ( 1 - \frac{u_{\alpha_0} \tau}{\sigma^2_{X_\alpha}} \right )
\exp \left [ - \frac{2\left ( t - t_0 \right ) }{\tau} \right ] 
\int^{t-t_0}_{0} \frac{1+ \frac{2 }{\tau} t^{\prime\prime\prime}
     }
     {t^{\prime\prime\prime} + \frac{\tau^2}{2\sigma^2_{X_\alpha} s}
     } 
{\rm d}t^{\prime\prime\prime} 
\right \}
\label{eq:Laplacetrfspecial_temp2}
\nonumber\\
\fl
&=
\exp \left ( - \frac{ u_{\alpha_0}s}{1+ s t^\prime} \right ) \left ( 1 + s t^\prime \right )^{- \left ( 1 - \frac{u_{\alpha_0}\tau}{\sigma^2_{X_\alpha}} \right ) \left ( 1 - \frac{\tau}{2\sigma^2_{X_\alpha}s} \right ) \exp \left [ - \frac{2 (t-t_0) }{\tau} \right ]}\nonumber\\
\fl
&~~~
\times
\exp \left \{ - \left ( 1 - \frac{u_{\alpha_0}\tau}{\sigma^2_{X_\alpha}} \right ) \exp \left [ - \frac{2(t-t_0)}{\tau} \right ] \frac{2(t-t_0)}{\tau} \right \}
\label{eq:Laplacetrfspecial_temp3a}\\
&\simeq
\left [ 1 - \frac{2(t-t_0)}{\tau} \left ( 1- \frac{u_{\alpha_0}\tau}{\sigma^2_{X_\alpha}} \right ) \right ]
\exp \left ( - \frac{ u_{\alpha_0}s}{1+ s t^\prime} \right ) 
\left ( 1 + s t^\prime \right )^\varpi
\label{eq:Laplacetrfspecial_temp3}
\eea
in which $t^\prime\doteq  2 \sigma^2_{X_\alpha} (t-t_0)/\tau^2$ [cf. (\ref{eq:deftprime})] and 
\bea
\varpi \doteq  {- 1 + \frac{u_{\alpha_0}\tau}{\sigma^2_{X_\alpha}} + \frac{2(t-t_0)}{\tau}  - \frac{2 (t-t_0) u_{\alpha_0} }{\sigma^2_{X_\alpha}} }.
\eea
Except for an additional $s$-independent scaling factor and the fact that the exponent of $(1+st^\prime)$ is now time-dependent, (\ref{eq:Laplacetrfspecial_temp3}) is of the same form as (\ref{eq:Laplacetrfspecial}). Consequently, as an extension of (\ref{eq:condprobenergyCart_smallt_largetau}), we now obtain
\bea
f_{U_\alpha} \left ( u_\alpha,t \right ) 
\simeq 
\frac{\tau^2 }
     {2 \left ( t - t_0 \right ) \sigma^2_{X_\alpha} 
     }
\left ( \frac{u_{\alpha}}{u_{\alpha_0}}
\right )^{\nu(t) }
\exp \left [ - \frac{ \tau^2 \left ( u_\alpha + u_{\alpha_0} \right )}
                    { 2 \left ( t - t_0 \right ) \sigma^2_{X_\alpha} }                                  
     \right ] 
I_{2\nu(t)}
 \left [ \frac{ \tau^2 \sqrt{u_\alpha \thinspace u_{\alpha_0} }}{\left ( t - t_0 \right ) \sigma^2_{X_\alpha} } 
    \right ],
\label{eq:condprobenergyCart_lint}
\eea
valid for $(t-t_0)/\tau \ll 1$, in which the time-dependent parameter
\bea
\nu(t) = - \frac{u_{\alpha_0} \tau}{2 \sigma^2_{X_\alpha}} + \frac{(t-t_0) u_{\alpha_0}}{\sigma^2_{X_\alpha}} - \frac{t-t_0}{\tau} 
\eea
exhibits a linear dependence on $(t-t_0)/\tau$. Again, $\nu(t)$ becomes constant with respect to time when 
\bea
u_{\alpha_0} = \frac{\sigma^2_{X_\alpha}}{\tau} .
\label{eq:condnut}
\eea
This corresponds to $\nu$ in (\ref{eq:defnu_copy}) taking on the value $-1/2$ and, hence, (\ref{eq:condprobenergyCart_lint}) like (\ref{eq:condprobenergyCart_smallt_largetau}) to coincide with the limit PDF (\ref{eq:condprobenergyCart_larget}) at all (early) times, as indicated before. 

\begin{figure}[t] \begin{center} 
\begin{tabular}{c} \ 
\epsfxsize=10cm 
\epsfbox{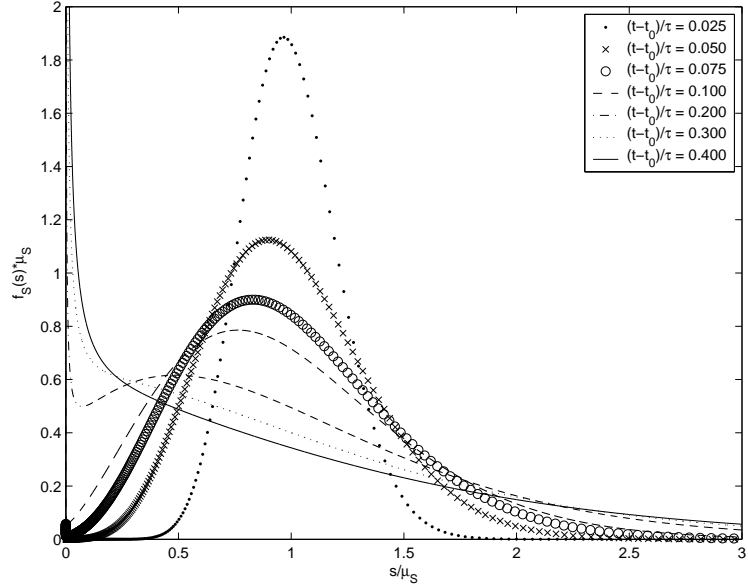}\
\end{tabular} \end{center}
\caption{\label{fig:Idistr_nonstat_S_PDF_p1} \small
Evolution of short-time asymptotic probability density functions of the energy density for scalar field ($p=1$).}
\end{figure}

\begin{figure}[t] \begin{center} 
\begin{tabular}{c} \ 
\epsfxsize=10cm 
\epsfbox{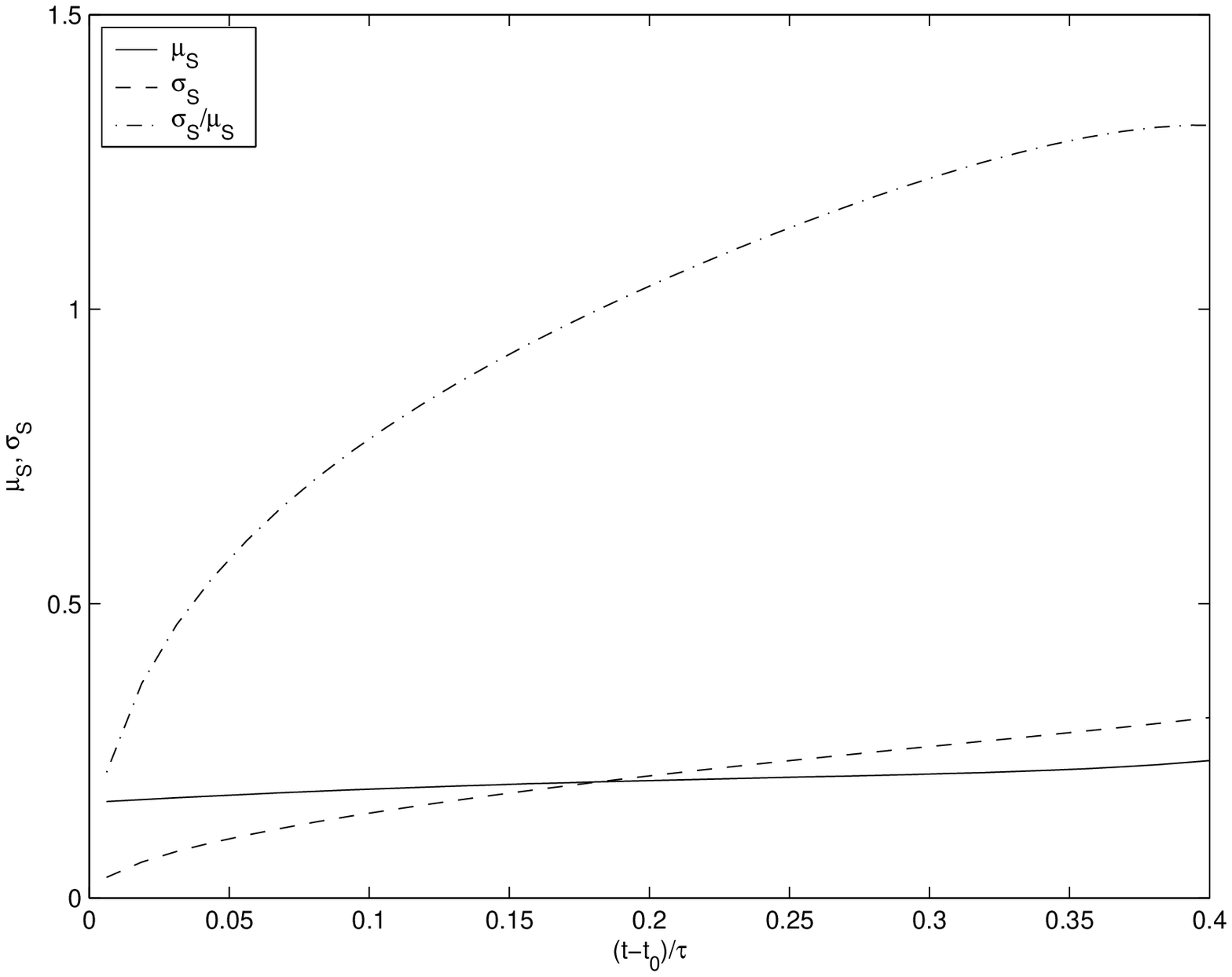}\
\end{tabular} \end{center}
\caption{\label{fig:Idistr_nonstat_S_musigma_p1} \small
Evolution of instantaneous mean value, standard deviation and coefficient of variation of the energy density for scalar field ($p=1$).
}
\end{figure}

Figure \ref{fig:Idistr_nonstat_S_PDF_p1} shows the asymptotic PDF (\ref{eq:condprobenergyCart_lint}) for selected values of $(t-t_0)/\tau \ll 1$ with $t_0=0$, $\tau=0.3$ s, $\sigma_{X_\alpha}=0.3$ V/m.
The corresponding dependencies of the mean, standard deviation, and coefficient of variation are shown in figure \ref{fig:Idistr_nonstat_S_musigma_p1}.

When $\tau / \sigma^2_{X_\alpha} \not \ll 1$, the term $\tau / ( \sigma^2_{X_\alpha} s)$ in the exponent of $(1+s t^\prime)$ in (\ref{eq:Laplacetrfspecial_temp3a}) must be accounted for. On application of contour integration in the complex $s$-plane [cf. \ref{sec:Mellin} with the definitions (\ref{eq:defaCart})-- (\ref{eq:defcCart})], $f_{U_\alpha}(u_\alpha,t)$ can be expressed as (\ref{eq:finalsolMellinF}), i.e.,
\bea
f_{U_\alpha} (u_\alpha,t) 
&=& 
\frac{1}{\pi} \int^{+\infty}_{\frac{\tau^2}{2(t-t_0)\sigma^2_{X_\alpha}}} 
\left [ \frac{2(t-t_0)\sigma^2_{X_\alpha}}{\tau^2} x - 1 \right ]^{ - \left ( 1 - \frac{u_{\alpha_0}\tau}{\sigma^2_{X_\alpha}} \right ) \left [ 1 - \frac{2(t-t_0)}{\tau} + \frac{\tau}{\sigma^2_{X_\alpha} x} \right ] }\nonumber\\
&~&~~~~ \times
\sin \left \{ \pi \left ( 1 - \frac{u_{\alpha_0}\tau}{\sigma^2_{X_\alpha}} \right ) \left [ 1 - \frac{2(t-t_0)}{\tau} + \frac{\tau}{\sigma^2_{X_\alpha} x} \right ] \right \}
\exp \left ( - u_\alpha x \right )
{\rm d} x.
\label{eq:finalsolMellinFCart}
\eea

\paragraph{TPDF for random initial value (extraction of subensemble)}
For a $\chi^2_2$-distributed random initial energy density $U_{\alpha_0}$ exhibiting the same statistics as $U_\alpha$ itself, i.e.,
\bea
f_{U_{\alpha_0}} (u_{\alpha_0}) = \frac{\tau}{\sigma^2_{X_\alpha}} \exp \left ( - \frac{\tau u_{\alpha_0}}{\sigma^2_{X_\alpha}} \right ),
\eea
the PDF $f_{U_\alpha}(u_\alpha,t)$ coincides with the asymptotic $\chi^2_2$-distributed TPDF:
\bea
f_{U_\alpha}(u_\alpha,t)
&=
\int^{+\infty}_{0} f_{U_\alpha}(u_\alpha,t|u_{\alpha_0},t_0) f_{U_{\alpha_0}}(u_{\alpha_0}) {\rm d} u_{\alpha_0}\nonumber\\
&=
\frac{\tau}
     {\sigma^2_X \left \{ 1 + \exp \left [ - \frac{t-t_0}{\tau} \right ] - \exp \left [ - \frac{2(t-t_0)}{\tau} \right ] \right \}} 
\nonumber\\&~\times 
\exp 
\left [ 
- \frac{\tau u_\alpha}{\sigma^2_X \left \{ 1 + \exp \left [ - \frac{t-t_0}{\tau} \right ] - \exp \left [ - \frac{2(t-t_0)}{\tau} \right ] \right \}}
\right ] \nonumber\\
&\stackrel{\left ( t-t_0 \right ) \ll \tau}{ \rule[0.82mm]{1cm}{0.15mm}\hspace{-3mm}\longrightarrow}  ~~~
\frac{\tau}{\sigma^2_X \left (1 + \frac{t-t_0}{\tau} \right )}
\exp \left [ - \frac{\tau u_\alpha}{\sigma^2_X \left ( 1+ \frac{t-t_0}{\tau} \right )} \right ]
\label{eq:TPDFUalphasubensemble}
\eea
which, unlike (\ref{eq:TPDFYsubensemble}), remains time-dependent.

For more general random $U_{\alpha_0}$, the issue of estimation of its PDF arises. 
This can be achieved in a systematic manner using a Bayesian scheme, i.e., starting from an assumed prior PDF $f_{U_{\alpha_0}}(u_{\alpha_0})$ and adding information of the observed $f_{U_\alpha}(u_\alpha,t|u_{\alpha_0},t_0)$ to yield an improved posterior estimate:
\bea
f_{U_{\alpha_0}}(u_{\alpha_0}|u_\alpha)
=C
f_{U_{\alpha}}(u_\alpha|u_{\alpha_0})
f_{U_{\alpha_0}}(u_{\alpha_0})
\eea
where $C$ is a normalization constant. This expression can be used in an iterative scheme to update the posterior distribution. For the prior distribution, one can use a known or a chosen ad hoc (Gauss normal or other) distribution centered around a classical (e.g., maximum likelihood) estimate of the mean value.

\subsubsection{Vector field\label{sec:Total}}
In the following, we shall use the subscript $\rmt$ to explicitly denote quantities associated with the total (i.e., vector) field, consistent with the use of the subscript $\alpha$ before. For consistency, results will be expressed in terms of statistics relating to ${X_\rmt}$ rather than ${X_\alpha}$, in order to yield self-sufficient distributions and statistics.

For the energy density of the mixed total field $Y_\rmt(t)$, we readily find in analogy with sections \ref{sec:CartMoments} and \ref{sec:CartSDE} that
\bea
\left \langle U_\rmt (t) \right \rangle 
&=& 
\left \langle {Y^\prime_\rmt}^2 (t) \right \rangle 
+ 
\left \langle {Y^{\prime\prime}_\rmt}^2 (t) \right \rangle 
\rightarrow \frac{\sigma^2_{X_\rmt}}{\tau} \equiv \frac{3 \sigma^2_{X_\alpha}}{\tau},\\
D^{(1)}_{U_\rmt}\left ( u_\rmt,t \right )
&=&
\frac{2}{\tau} \left ( \frac{\sigma^2_{X_\rmt} }{\tau} - u_{\rmt_0} \right ) \exp \left [ - \frac{2 \left ( t - t_0 \right )}{\tau} \right ] \nonumber\\ 
&=&
\frac{6}{\tau} \left ( \frac{\sigma^2_{X_\alpha} }{\tau} - u_{\alpha_0} \right ) \exp \left [ - \frac{2 \left ( t - t_0 \right )}{\tau} \right ]
=
3 D^{(1)}_{U_\alpha}(u_\alpha,t),
\label{eq:driftUtot}
\\
D^{(2)}_{U_\rmt}\left ( u_\rmt,t \right )
&=&
\frac{2 \sigma^2_{X_\rmt}}{3\tau^2} u_{\rmt} 
=
3 D^{(2)}_{U_\alpha}(u_\alpha,t),
\label{eq:diffusionUtot}
\eea
where $\sigma^2_{X_\rmt} \doteq \sigma^2_{X_{x}} + \sigma^2_{X_{y}} + \sigma^2_{X_{z}} = 3 \sigma^2_{X_\alpha}$ and
$U_\rmt=U_x+U_y+U_z=3U_\alpha$.
The associated SDE follows as
\bea
\dot{U}_\rmt (t) = 
\frac{2}{\tau}
\left \{
\left ( 
\frac{\sigma^2_{X_\rmt}}{\tau} - u_{\rmt_0} \right ) \exp \left [ - \frac{2 \left ( t - t_0 \right )}{\tau} \right ] + \sqrt{\frac{U_\rmt(t)}{3}} \dot{B}_\rmt(t)
\right \}
\eea
in the It\^{o} formulation, or
\bea
\dot{U}_\rmt (t) = 
\frac{2}{\tau}
\left \{
\left ( 
\frac{\sigma^2_{X_\rmt}}{\tau} - u_{\rmt_0} \right ) \exp \left [ - \frac{2 \left ( t - t_0 \right )}{\tau} \right ] - \frac{\sigma^2_{X_\rmt}}{6\tau} + \sqrt{\frac{U_\rmt(t)}{3}} \dot{B}_\rmt(t)
\right \}
\eea 
in the Stratonovich formulation.
The corresponding FPE is
\bea
\frac{\partial}{\partial t}                           f_{U_\rmt}\left ( u_\rmt,t|u_{\rmt_0},t_0 \right ) 
&=&
- \frac{2}{\tau} \left ( \frac{\sigma^2_{X_\rmt} }{\tau} - u_{\rmt_0} \right ) \exp \left [ - \frac{2 \left ( t - t_0 \right )}{\tau} \right ]
\frac{\partial}{\partial u_\rmt}                    f_{U_\rmt}\left ( u_\rmt,t|u_{\rmt_0},t_0 \right ) \nonumber\\
&~&+
\frac{2\sigma^2_{X_\rmt}}{3 \tau^2} 
\frac{\partial^2}{\partial u^2_\rmt} \left [ u_\rmt f_{U_\rmt}\left ( u_\rmt,t|u_{\rmt_0},t_0 \right ) \right ] 
\label{eq:FPenergystirTot}
\eea
with $u_{\rmt_0} \doteq U_\rmt(t_0) $, i.e.,
\bea
f_{U_\rmt}(u_\rmt,t_0) = \delta ( u_\rmt - u_{\rmt_0} )
\label{eq:IC_FPenergystirTot}
\eea 
and 
\bea
f_{U_\rmt} (0, t|u_{\rmt_0},t_0) = f_{U_\rmt}(+\infty,t|u_{\rmt_0},t_0) = 0.
\label{eq:BC_FPenergystirTot}
\eea
Its transformation to a first-order ODE and its solution via Laplace transformation are obtained in a similarly way as in \ref{app:solvingFPECart}, but some subtle changes in the coefficients arise. Details are given in \ref{app:solvingFPEVecEnergy}
and lead to the general solution again being given by the inversion formula (\ref{eq:FPEgeneralsolutionCartEnergy}) but now with
\bea
\fl\hspace{1cm}
{\cal F}_{S} \left ( s,t \right )
&=
\exp \left [ - \frac{3\tau^2 u_{\rmt_0} }{2 \sigma^2_{X_\rmt} \left ( t - t_0 + \frac{3\tau^2}{2 \sigma^2_{X_\rmt} s} \right )} 
\right ]
\nonumber\\
\fl
&~~~~~~~~
\times
\exp \left \{
-
\int^{t - t_0 + \frac{3\tau^2}{2 \sigma^2_{X_\rmt} s} }_{{\frac{3\tau^2}{2 \sigma^2_{X_\rmt} s} }}
\frac{3}{t^{\prime\prime}} \left ( 1 - \frac{u_{\rmt_0}\tau}{\sigma^2_{X_\rmt}} \right ) 
\exp \left [ - \frac{2}{\tau} \left ( t - t_0 - t^{\prime\prime} \right ) - \frac{3\tau}{\sigma^2_{X_\rmt} s} \right ] {\rm d} t^{\prime\prime} 
\right \}
\nonumber\\
\fl
&~~~+
\int^{t - t_0 + \frac{3\tau^2}{2 \sigma^2_{X_\rmt} s} }_{{\frac{3\tau^2}{2 \sigma^2_{X_\rmt} s} }}
\left \{
-1 + 3 \left ( 1- \frac{u_{\rmt_0}\tau}{\sigma^2_{X_\rmt}} \right ) \exp \left [ - \frac{2}{\tau} \left ( t - t_0 - t^{\prime\prime} \right ) - \frac{3\tau}{\sigma^2_{X_\rmt} s} \right ] \right \} 
\frac{2 \sigma^2_{X_\rmt} f_{U_\rmt}(0+,t)}{3\tau^2}
\nonumber\\
&~~~~~~~~
\times
\exp \left \{
+
\int^{t - t_0 + \frac{3\tau^2}{2 \sigma^2_{X_\rmt} s} }_{t^{\prime\prime}}
\frac{3}{t^{\prime\prime\prime}} \left ( 1 - \frac{u_{\rmt_0}\tau}{\sigma^2_{X_\rmt}} \right ) 
\exp \left [ -\frac{2}{\tau} \left ( t - t_0 - t^{\prime\prime\prime} \right ) - \frac{3\tau}{\sigma^2_{X_\rmt} s} \right ] 
{\rm d} t^{\prime\prime\prime} 
\right \}
{\rm d} t^{\prime\prime} 
\nonumber\\
&~~~
\times
\exp \left \{
-
\int^{t - t_0 + \frac{3\tau^2}{2 \sigma^2_{X_\rmt} s} }_{{\frac{3\tau^2}{2 \sigma^2_{X_\rmt} s} }}
\frac{3}{t^{\prime\prime}} \left ( 1 - \frac{u_{\rmt_0}\tau}{\sigma^2_{X_\rmt}} \right ) 
\exp \left [ - \frac{2}{\tau} \left ( t - t_0 - t^{\prime\prime} \right ) - \frac{3\tau}{\sigma^2_{X_\rmt} s} \right ] {\rm d} t^{\prime\prime} 
\right \}.
\label{eq:PDEfinalLTTot}
\eea
If $f_{U_\rmt}(0,t)=0$ then ${\cal F}_S(s,t)$ is limited to the first term in (\ref{eq:PDEfinalLTTot}) only. If, in addition, $f_{U_\rmt}(u_\rmt,t_0)=0$ then
\bea
\fl\hspace{1cm}
{\cal F}_S(s,t) &=& 
\exp \left \{ - 3 \exp \left [ - \frac{2}{\tau} \left ( t - t_0 \right ) - \frac{3\tau}{\sigma^2_{X_\rmt} s} \right ]
\left [
{\rm Ei} \left [ \frac{2 }{\tau} \left ( t - t_0 \right ) + \frac{3\tau}{\sigma^2_{X_\rmt} s} \right ]
-
{\rm Ei} \left ( \frac{3\tau}{\sigma^2_{X_\rmt} s} \right )
\right ]
\right \}.
\label{eq:PDEfinalLTTot_zero}
\eea
In either case, $f_{U_\rmt}(u_\rmt,t)$ is obtained via inverse transformation of (\ref{eq:PDEfinalLTTot}), or (\ref{eq:PDEfinalLTTot_zero}) in particular.

As a special case, we consider again the early-time or large-relaxation limit $(t-t_0)/\tau \rightarrow 0$. Upon scaling $t$ as
\bea
t^\prime \doteq \frac{2 \sigma^2_{X_\rmt} }{3\tau^2} (t-t_0) 
\label{eq:deftprimeTot}
\eea
the FPE (\ref{eq:FPenergystirTot}) becomes
\bea
\frac{\partial}{\partial t^\prime}                           f_{U_\rmt}\left ( u_\rmt,t^\prime \right ) 
&=&
\left ( -1 + \frac{3 u_{\rmt_0} \tau}{\sigma^2_{X_\rmt}} \right ) \frac{\partial}{\partial u_\rmt}                    f_{U_\rmt}\left ( u_\rmt,t^\prime \right ) 
+
\frac{\partial^2}{\partial u^2_\rmt} \left [ u_\rmt f_{U_\rmt}\left ( u_\rmt,t^\prime \right ) \right ]
\label{eq:FPenergystirTot_smallt}
\eea
whence (\ref{eq:PDEfinalLTTot}) simplifies to
\bea
{\cal F}_S \left ( s, t^\prime \right )
&=& 
\exp \left ( - \frac{u_{\rmt_0}s}{1+s t^\prime} \right ) 
\exp \left [ - 3 \left ( 1-\frac{u_{\rmt_0}\tau}{\sigma^2_{X_\rmt}} \right )\int^{s^{-1}+t^\prime}_{s^{-1}} \frac{{\rm d}t^{\prime\prime} }{t^{\prime\prime}} 
                                                    \right ]\nonumber\\
&=&
\exp \left ( - \frac{u_{\rmt_0}s }{1+s t^\prime} \right )
\left ( 1 + s t^\prime \right )^{-3 + \frac{3 u_{\rmt_0} \tau}{\sigma^2_{X_\rmt}} } .
\label{eq:LaplacetrfspecialTot}
\eea
Except for the exponent of the second factor, (\ref{eq:LaplacetrfspecialTot}) is identical to the result for the scalar case (\ref{eq:Laplacetrfspecial}). Therefore, we can again use (\ref{eq:GRAD}) with $\alpha \doteq {t^\prime}^{-1} + s$, $\beta \doteq \sqrt{u_{\rmt_0}}/t^\prime$, $\nu \doteq \mu-(1/2)$, but now defining
\bea
\nu
\doteq 
1-\frac{3u_{\rmt_0}\tau}{2\sigma^2_{X_\rmt}} .
\label{eq:defnuTot}
\eea
With this choice of parameters, the limit PDF of $U_\rmt$ for $(t-t_0)/\tau \rightarrow 0$ is
\bea
f_{U_\rmt} \left ( u_\rmt,t \right ) 
&=& 
\frac{3\tau^2}
     {2 \left ( t - t_0 \right )\sigma^2_{X_\rmt} 
     }
\left ( \frac{u_\rmt}{u_{\rmt_0}} \right )^\nu
\exp \left [ - \frac{ 3 \tau^2 \left ( u_\rmt + u_{\rmt_0} \right ) }
                    { 2 \left ( t - t_0 \right ) \sigma^2_{X_\rmt} }                                  
     \right ] 
I_{2\nu} \left [ \frac{3 \tau^2 \sqrt{u_\rmt \thinspace u_{\rmt_0} }}{\left ( t - t_0 \right ) \sigma^2_{X_\rmt} } 
    \right ] .
\label{eq:condprobenergyTot_smallt}
\eea
Although this result is formally identical with (\ref{eq:condprobenergyCart_smallt_largetau}), recall the different definition of $\nu$ in (\ref{eq:defnuTot}) compared to (\ref{eq:defnu}).
If, in addition, $u_{\rmt_0} \tau / \sigma^2_{X_\rmt} \ll 1$ then $\mu=3/2$, $\nu=1$ whence (\ref{eq:condprobenergyTot_smallt}) specializes to
\bea
f_{U_\rmt} \left ( u_\rmt,t \right ) 
=
\frac{3 \tau^2}
     {2 \left ( t - t_0 \right ) \sigma^2_{X_\rmt} 
     }
\left ( \frac{u_\rmt}{u_{\rmt_0}} \right )
\exp \left [ - \frac{ 3 \tau^2 \left ( u_\rmt + u_{\rmt_0} \right )}
                    { 2 \left ( t - t_0 \right )\sigma^2_{X_\rmt} }                                  
     \right ] 
I_2 \left [ \frac{3 \tau^2 \sqrt{u_\rmt \thinspace u_{\rmt_0} }}{ \left ( t - t_0 \right )\sigma^2_{X_\rmt} } 
    \right ] ,
\label{eq:condprobenergyTot_smallt_smalltau}
\eea
which is to be compared with (\ref{eq:condprobenergyCart_smallt_smalltau}). 
By contrast, the limit $(t-t_0)/\tau \rightarrow +\infty$ yields the {\em same} limit PDF as the one for $U_\alpha$, i.e., (\ref{eq:condprobenergyCart_larget}), after replacing $u_{\alpha_0}$ with $u_{\rmt_0}$, viz.,
\bea
f_{U_\rmt} \left ( u_\rmt,t \right ) 
=
\frac{3\tau^2 }
     {2 \left ( t - t_0 \right ) \sigma^2_{X_\rmt} 
     }
\sqrt{\frac{u_{\rmt_0}}{u_\rmt}} 
\exp \left [ - \frac{ 3 \tau^2 \left ( u_\rmt + u_{\rmt_0} \right )}
                    { 2 \left ( t - t_0 \right ) \sigma^2_{X_\rmt} }                                  
     \right ] 
I_1 \left [ \frac{ 3 \tau^2 \sqrt{u_\rmt \thinspace u_{\rmt_0} }}{\left ( t - t_0 \right ) \sigma^2_{X_\rmt} } 
    \right ].
\label{eq:condprobenergyTot_larget}
\eea

\begin{figure}[t] \begin{center} 
\begin{tabular}{c} \ 
\epsfxsize=10cm 
\epsfbox{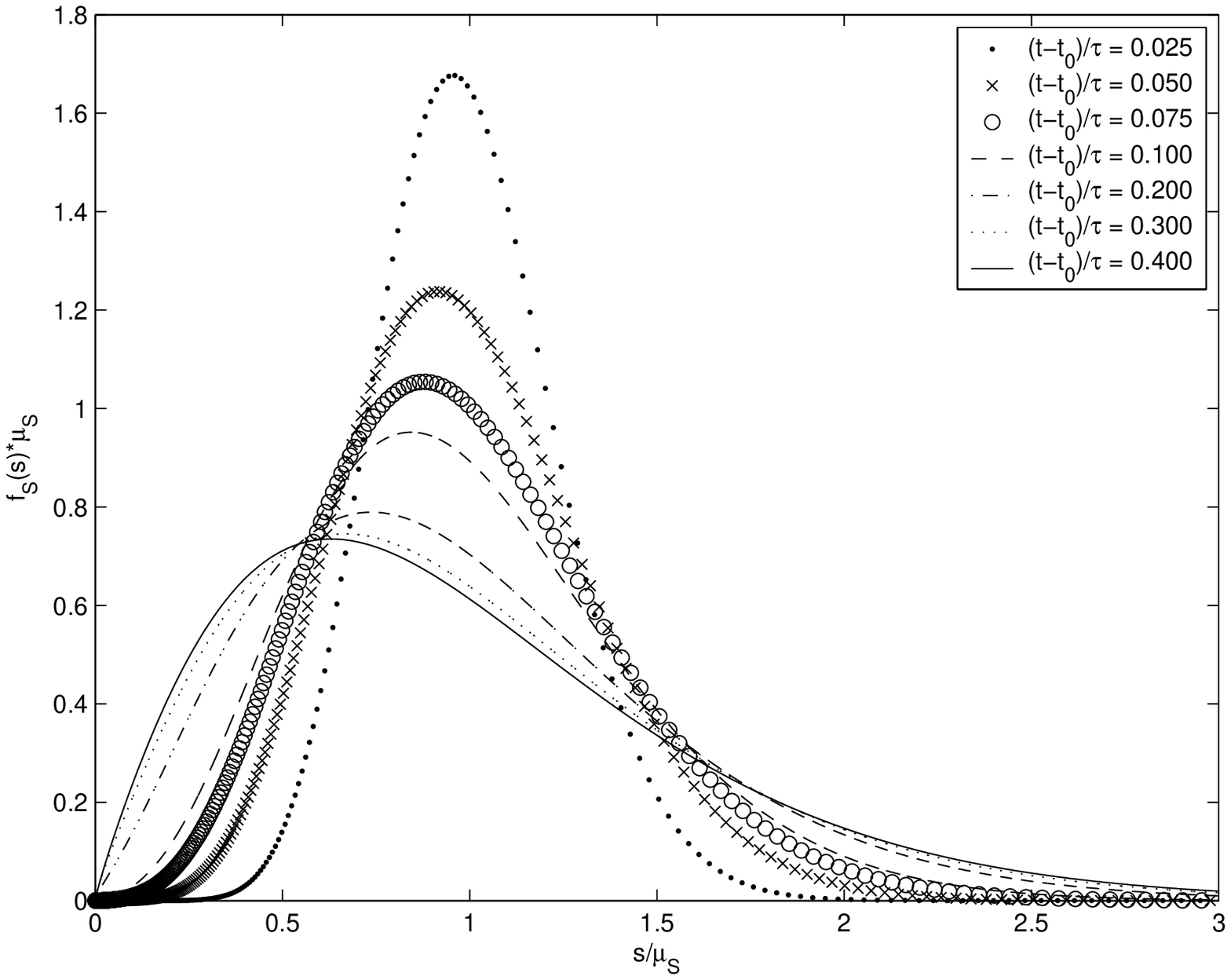}\ 
\end{tabular} \end{center}
\caption{\label{fig:Idistr_nonstat_S_PDF_p3} \small
Evolution of short-time asymptotic probability density functions of the energy density for vector field ($p=3$).}
\end{figure}

\begin{figure}[t] \begin{center} 
\begin{tabular}{c} \ 
\epsfxsize=10cm 
\epsfbox{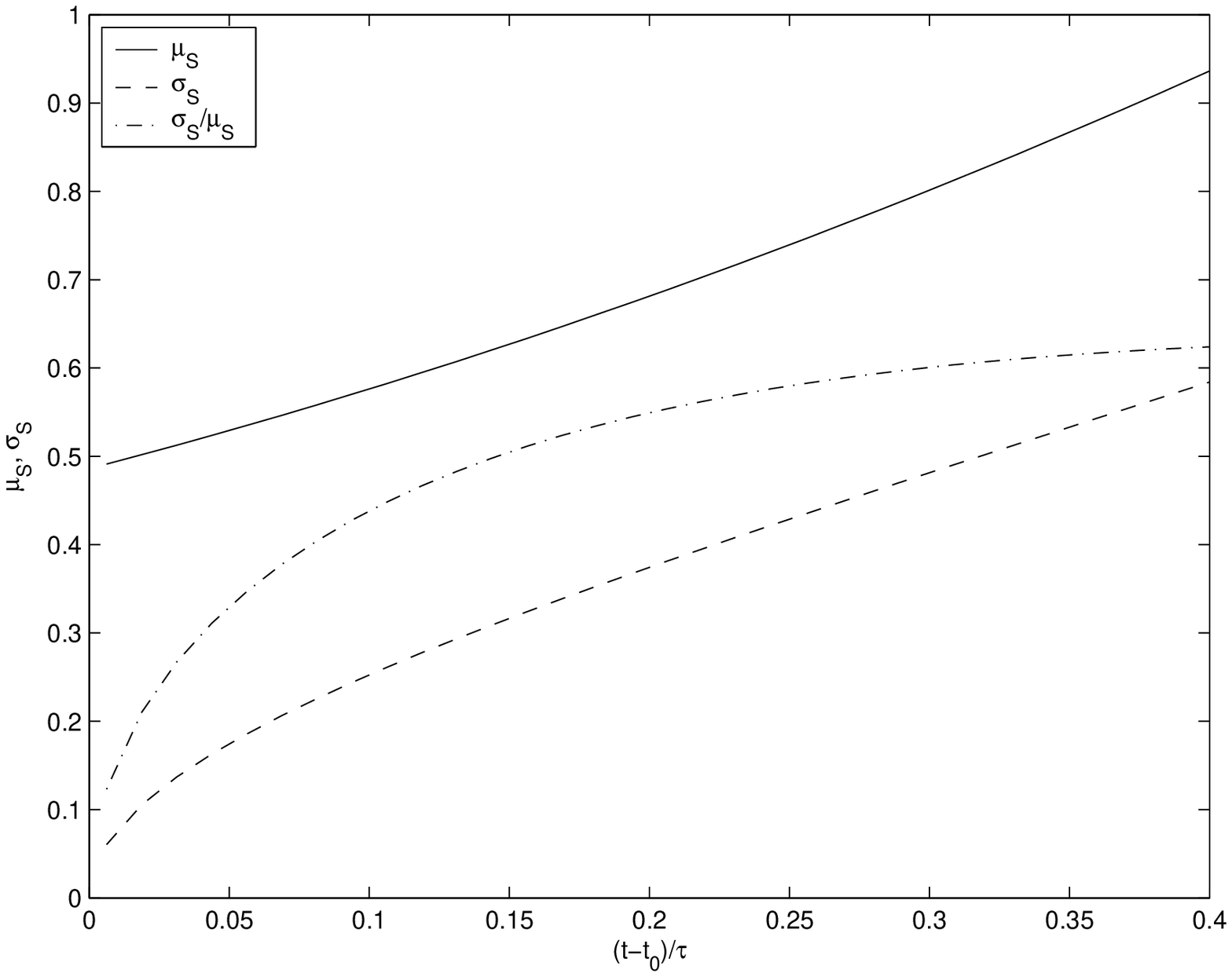}\
\end{tabular} \end{center}
\caption{\label{fig:Idistr_nonstat_S_musigma_p3} \small
Evolution of instantaneous mean value, standard deviation and coefficient of variation of the energy density for vector field ($p=3$).
}
\end{figure}

\paragraph{First-order time-dependence of PDF\label{sec:firstorderTot}}
For $(t-t_0)/\tau \ll 1$, the evolution of the PDF for the total energy density can be analyzed in a manner analogous to that for the Cartesian energy density in section \ref{sec:firstorderCart}. As mentioned before, the PDF for general $t$ is obtained by inversion of (\ref{eq:PDEfinalLTCart}), or (\ref{eq:PDEfinalLTsimplified}) in particular. 
Instead of (\ref{eq:Laplacetrfspecial_temp})--(\ref{eq:Laplacetrfspecial_temp3}), we now have with $f_{U_\rmt}(0+,t)=0$ that
\bea
\fl \hspace{1.5cm}
{\cal F}_S \left ( s, t \right )
&=
\exp \left [ - \frac{ u_{\rmt_0}s}{1+\frac{2(t-t_0) \sigma^2_{X_\rmt} s}{3\tau^2} } \right ] \nonumber\\
\fl 
&~~~\times
\exp \left \{ - \int^{t-t_0 + \frac{3\tau^2}{2\sigma^2_{X_\rmt} s}}_{\frac{3\tau^2}{2\sigma^2_{X_\rmt} s}} ~~~~\frac{3 \left ( 1 - \frac{u_{\rmt_0} \tau}{\sigma^2_{X_\rmt}} \right )}{t^{\prime\prime}} \exp \left [ - \frac{2}{\tau} \left ( t-t_0 + \frac{3\tau^2}{2\sigma^2_{X_\rmt} s} - t^{\prime\prime} \right ) \right ]                                                     {\rm d}t^{\prime\prime} \right \}
\label{eq:Laplacetrfspecial_temp_tot} \\
\fl
&=
\exp \left ( - \frac{ u_{\rmt_0}s}{1+ s t^\prime} \right ) \left ( 1 + s t^\prime \right )^{- 3 \left ( 1 - \frac{3 u_{\rmt_0}\tau}{\sigma^2_{X_\rmt}} \right ) \left ( 1 - \frac{3\tau}{2\sigma^2_{X_\rmt}s} \right ) \exp \left [ - \frac{2 (t-t_0) }{\tau} \right ]}\nonumber\\
\fl
&~~~\times 
\exp \left \{ - \left ( 1 - \frac{3 u_{\rmt_0}\tau}{\sigma^2_{X_\rmt}} \right ) \exp \left [ - \frac{2(t-t_0)}{\tau} \right ] \frac{2(t-t_0)}{\tau} \right \}
\label{eq:Laplacetrfspecial_temp3a_tot}\\
\fl
&\simeq
\left [ 1 - \frac{2(t-t_0)}{\tau} \left ( 1 - \frac{3 u_{\rmt_0} \tau}{\sigma^2_{X_\rmt}} \right ) \right ]
\exp \left ( - \frac{u_{\rmt_0}s}{1+ s t^\prime} \right ) 
\left ( 1 + s t^\prime \right )^\varpi 
\label{eq:Laplacetrfspecial_temp3_tot}
\eea
in which $t^\prime$ is again defined by (\ref{eq:deftprimeTot}) and
\bea
\varpi\doteq {- 3 \left [ 1 - \frac{3 u_{\rmt_0}\tau}{\sigma^2_{X_\rmt}} - \frac{2(t-t_0)}{\tau} + \frac{2 (t-t_0) u_{\rmt_0} }{\sigma^2_{X_\rmt}} \right ] } 
.
\eea
The resulting PDF shows again a time-dependent order, i.e., (\ref{eq:condprobenergyTot_smallt}) generalizes to
\bea
f_{U_\rmt} \left ( u_\rmt,t \right ) 
&=& 
\frac{3\tau^2}
     {2 \left ( t - t_0 \right )\sigma^2_{X_\rmt} 
     }
\left ( \frac{u_\rmt}{u_{\rmt_0}} \right )^{\nu(t)}
\exp \left [ - \frac{ 3 \tau^2 \left ( u_\rmt + u_{\rmt_0} \right ) }
                    { 2 \left ( t - t_0 \right ) \sigma^2_{X_\rmt} }                                  
     \right ] 
I_{2\nu(t)} \left [ \frac{3 \tau^2 \sqrt{u_\rmt \thinspace u_{\rmt_0} }}{\left ( t - t_0 \right ) \sigma^2_{X_\rmt} } 
    \right ]
\label{eq:condprobenergyTot_smallnonzerot}
\eea
where
\bea
\nu(t) \doteq 1 - \frac{3 u_{\rmt_0} \tau}{2 \sigma^2_{X_\rmt}} + \frac{3(t-t_0) u_{\rmt_0}}{\sigma^2_{X_\rmt}} - \frac{t-t_0}{\tau} .
\label{eq:defnuTotfirstorder}
\eea
If $\tau / \sigma^2_{X_\rmt} \not \ll 1$, then (\ref{eq:condprobenergyTot_smallnonzerot}) generalizes to [cf. (\ref{eq:finalsolMellinF}) with the definitions (\ref{eq:defaTot})--(\ref{eq:defcTot})], 
\bea
f_{U_\rmt} (u_\rmt,t) 
&=& 
\frac{1}{\pi} \int^{+\infty}_{\frac{3\tau^2}{2(t-t_0)\sigma^2_{X_\rmt}}} 
\left [ \frac{2(t-t_0)\sigma^2_{X_\rmt}}{3 \tau^2} x - 1 \right ]^{ - 3 \left ( 1 - \frac{3 u_{\rmt_0}\tau}{\sigma^2_{X_\rmt}} \right ) \left [ 1 - \frac{2(t-t_0)}{\tau} + \frac{3\tau}{\sigma^2_{X_\rmt} x} \right ] }\nonumber\\
&~&~~~~ \times
\sin \left \{ \pi \left ( 1 - \frac{3 u_{\rmt_0}\tau}{\sigma^2_{X_\rmt}} \right ) \left [ 1 - \frac{2(t-t_0)}{\tau} + \frac{3 \tau}{\sigma^2_{X_\rmt} x} \right ] \right \}
\exp \left ( - u_\rmt x \right )
{\rm d} x .
\label{eq:finalsolMellinFTot}
\eea
Figure \ref{fig:Idistr_nonstat_S_PDF_p3} shows the asymptotic PDF (\ref{eq:condprobenergyTot_smallnonzerot}) for selected values of $(t-t_0)/\tau \ll 1$ with $t_0=0$, $\tau=0.3$ s, $\sigma_{X_\alpha}=0.3$ V/m, $u_{\rmt_0} = 0.9 \sigma^2_{X_\rmt}$.
The corresponding dependencies of the mean, standard deviation, and coefficient of variation are shown in figure \ref{fig:Idistr_nonstat_S_musigma_p3}.

\subsubsection{Generalized field}

On occasion, it is expedient to handle energy densities with more than three degrees of freedom, for example when collating data for $\ell$ statistically independent locations or for differently generated realizations. There are then in general $2p \doteq 2m\ell$ degrees of freedom, where $m=1$ or $3$ for single-point Cartesian or vector fields, respectively.
The analysis and results of section \ref{sec:Total} can be repeated mutatus mutandis. Denoting $U_p(t=t_0) \doteq u_{p_0}$, the Stratonovich SDE and the FPE are now
\bea
\dot{U}_p (t) = 
\frac{2}{\tau}
\left \{
\left ( 
\frac{\sigma^2_{X_p}}{\tau} - u_{p_0} \right ) \exp \left [ - \frac{2 \left ( t - t_0 \right )}{\tau} \right ] - \frac{\sigma^2_{X_p}}{2p\tau} + \sqrt{\frac{U_{p}(t)}{p}} \dot{B}_p(t)
\right \}
\eea
and
\bea
\frac{\partial}{\partial t}                           f_{U_{p}}\left ( u_{p},t|u_{{p}_0},t_0 \right ) 
&=&
- \frac{2}{\tau} \left ( \frac{\sigma^2_{X_{p}} }{\tau} - u_{p_0} \right ) \exp \left [ - \frac{2 \left ( t - t_0 \right )}{\tau} \right ]
\frac{\partial}{\partial u_{p}}                    f_{U_{p}}\left ( u_{p},t|u_{{p}_0},t_0 \right ) \nonumber\\
&~&+
\frac{2\sigma^2_{X_{p}}}{p \tau^2} 
\frac{\partial^2}{\partial u^2_{p}} \left [ u_{p} f_{U_{p}}\left ( u_{p},t|u_{{p}_0},t_0 \right ) \right ] 
\label{eq:FPenergystirGen}
\eea
respectively, with 
$f_{U_p}(u_p,t_0) = \delta(u_p-u_{p_0})$ and with the boundary values
$f_{U_p} (0, t|u_{p_0},t_0) = f_{U_p}(+\infty,t|u_{p_0},t_0) = 0$, and the $p$-dimensional circular Gauss normal white noise process $B_p$ defined by
\bea
\dot{B}_p(t) 
\doteq \frac{ \sum^{p}_{n=1} B_{n} (t) \dot{B}_{n}(t) } {||B_p(t)||} 
= \frac{ \sum^{p}_{n=1} \left [ B^\prime_{n} (t) \dot{B}^\prime_{n} (t) + B^{\prime\prime}_{n} (t) \dot{B}^{\prime\prime}_{n} (t) \right ] } { \sqrt{ \sum^p_{n=1} \left [ B^\prime_{n} (t) \right ]^2 + \left [ B^{\prime\prime}_{n} (t) \right ]^2 } } 
\eea
in which $B_n\equiv B^\prime_{n}- \rmj B^{\prime\prime}_{n}$.
Corresponding to (\ref{eq:condprobenergyCart_lint}), we have the solution
\bea
f_{U_p} \left ( u_p,t \right ) 
&=& 
\frac{p\tau^2}
     {2 \left ( t - t_0 \right )\sigma^2_{X_p} 
     }
\left ( \frac{u_p}{u_{p_0}} \right )^{\nu(t)}
\exp \left [ - \frac{ p \tau^2 \left ( u_p + u_{p_0} \right ) }
                    { 2 \left ( t - t_0 \right ) \sigma^2_{X_p} }    
     \right ] 
I_{2\nu(t)} \left [ \frac{p \tau^2 \sqrt{u_p \thinspace u_{p_0} }}{\left ( t - t_0 \right ) \sigma^2_{X_p} } 
    \right ]
\label{eq:condprobenergyGen_smallnonzerot}
\eea
now with
\bea
\nu(t) &\doteq& \frac{p-1}{2} - \frac{p\thinspace u_{p_0} \tau}{2 \sigma^2_{X_p}} - \left ( \frac{1}{\tau}  - \frac{p u_{p_0} \tau }{\sigma^2_{X_p}} \right ) \frac{t-t_0}{\tau} .
\label{eq:nu_condprobenergyGen_smallnonzerot}
\eea

\subsubsection{Extraction of subensemble}
An ideal random initial energy density $U_0 = |Y_0|^2$ exhibits a $\chi^2_2$-distribution with standard deviation $2\sigma^2_X$. Performing an additional averaging, the $u_{p_0}$ in the above expressions is to be replaced by 
$\langle U_{p_0} \rangle = 2 \sigma^2_{X_p}/\tau \equiv p \sigma^2_{X_\alpha}/\tau$. 
This coincides with (\ref{eq:condnut}), whence the type of $f_{U_\alpha}$ (although not its statistics) then becomes again time-independent. Hence we retrieve (\ref{eq:condprobenergyCart_larget}) with $u_{\alpha_0}$ replaced as indicated. For the total or generalized field, however, the order of the Bessel function in $f_{U_p}$ remains time-dependent and the PDF does not approach (\ref{eq:condprobenergyTot_larget}) when $(t-t_0)/\tau \rightarrow +\infty$, except if 
$\sigma^2_{X_p}/(p\tau) \equiv \sigma^2_{X_\alpha}/\tau$ in which case $\nu(t) = (p/2)-1$.

\subsection{Field magnitude (envelope)}
The instantaneous field magnitude (amplitude) $A(t)$ of an analytic process $Y(t)$ is defined by the envelope $\sqrt{ {Y^\prime}^2(t)+ {Y^{\prime\prime}}^2(t)}$ and can be obtained from $U(t)$ via the variate transformation $A[U(t)] = \sqrt{U(t)}$.

\subsubsection{Scalar fields or Cartesian components}

From (\ref{eq:driftUCart}) and (\ref{eq:diffusionUCart}), we obtain the drift and diffusion coefficients of $A_\alpha(t)$ as
\bea
D^{(1)}_{A_\alpha} (a_{\alpha},t)
&=& \frac{\partial A_\alpha}{\partial t}
+ \frac{\partial A_\alpha}{\partial U_\alpha} D^{(1)}_{U_\alpha} (a_\alpha,t)
+ \frac{\partial^2 A_\alpha}{\partial U^2_\alpha} D^{(2)}_{U_\alpha} (a_\alpha,t) \label{eq:trfvardriftCartA}\\
&=&
\left ( \tau a_\alpha \right )^{-1}  \left \{ \left ( \frac{\sigma^2_{X_\alpha}}{\tau}  - a^2_{\alpha_0} \right ) \exp \left [ - \frac{2 (t-t_0)}{\tau} \right ] - \frac{\sigma^2_{X_\alpha}}{2\tau} \right \}
\label{eq:driftACart}
\\
D^{(2)}_{A_\alpha} (a_{\alpha},t)
&=& 
\left ( \frac{\partial A_\alpha}{\partial U_\alpha} \right )^2 D^{(2)}_{U_\alpha} 
=
\frac{\sigma^2_{X_\alpha}}{2\tau^2}
\label{eq:trfvardiffusionCartA}
\eea
in which $a_{\alpha} \equiv | y_{\alpha} | \doteq \sqrt{u_{\alpha}} $.
The SDE for $A_\alpha(t)$ is thus quasi-linear. In the It\^{o} formalism, this equation is obtained on application of the It\^{o} differentiation formula \cite{okse1}:
\bea
{\rm d}{A}_\alpha(t) 
&=& \frac{\partial A_\alpha}{\partial t} {\rm d}t 
+ \frac{\partial A_\alpha}{\partial U_\alpha} {\rm d}{U}_\alpha 
+ \frac{1}{2} \frac{\partial^2 A_\alpha}{\partial U^2_\alpha} \left ( {\rm d}{U}_\alpha \right )^2
=\frac{{\rm d}{U}_\alpha(t)}{2 A_\alpha(t)} - \frac{({\rm d} U_\alpha)^2}{8 A^3_\alpha(t)},
\eea 
in which (\ref{eq:LangevUCartOUIto}) is substituted for $\dot{U}_\alpha$ and with $({\rm d}U_\alpha)^2 = (4 U_\alpha \sigma^2_{X_\alpha}/\tau^2) {\rm d}t$, yielding
\bea
\dot{A}_\alpha(t) 
&=& 
\left [ \tau A_\alpha(t) \right ]^{-1} 
\left \{
\left (
\frac{\sigma^2_{X_\alpha}}{\tau} 
- a^2_{\alpha_0} 
\right ) \exp \left [ - \frac{2 (t-t_0)}{\tau} \right ] 
- \frac{\sigma^2_{X_\alpha}}{2\tau}
\right \}
+ 
\frac{\dot{B}_\alpha(t)}{\tau} .
\label{eq:LangevACartOUIto}
\eea
This SDE can also be obtained directly as $\dot{A}_\alpha(t) = D^{(1)}_{A_\alpha} + \left ( \sqrt{2 D^{(2)}_{A_\alpha}} / \sigma_{X_\alpha} \right ) \dot{B}_\alpha(t)$. 
A form of (\ref{eq:LangevACartOUIto}) solely involving amplitude (rather than field) statistics can be obtained by replacing $\sigma_{X_\alpha}$ with $\sigma_{A_\alpha}/\sqrt{2-(\pi/2)}$ for a Rayleigh distributed $A_\alpha$ \cite{arnaTEMC1}. 
In the Stratonovich formulation, the last term in (\ref{eq:trfvardriftCartA}) is omitted. Substitution  of (\ref{eq:LangevUCartOUStrat}) into $\dot{A}_\alpha(t) = {\dot{U}_\alpha(t)} / [{2 A_\alpha(t)}]$  
then yields again (\ref{eq:LangevACartOUIto}). Thus, unlike for $U_\alpha(t)$, no spurious drift of $A_\alpha(t)$ occurs, as is also apparent from the fact that $D^{(2)}_{A_\alpha}$ is independent of $a_\alpha$.

For the special case of a process dominated by relaxation ($\tau \rightarrow +\infty$), (\ref{eq:LangevACartOUIto}) reduces to a result that has been obtained previously using a different approach \cite[eqn. (4.4.41b)]{gard1}.

Using (\ref{eq:trfvardriftCartA}) and (\ref{eq:trfvardiffusionCartA}), the FPE for $A_\alpha(t)$ is obtained as
\bea
\fl \hspace{1.5cm}
\frac{\partial}{\partial t}                           f_{A_\alpha}\left ( a_\alpha,t|\thinspace a_{\alpha_0},t_0 \right )
&=&
\frac{\partial}{\partial a_\alpha}     
\left [ 
\left \{ \left ( \frac{a_{\alpha}}{\tau} - \frac{\sigma^2_{X_\alpha}}{\tau^2 a_\alpha} \right ) \exp \left [ - \frac{2 (t-t_0)}{\tau} \right ] + \frac{\sigma^2_{X_\alpha}}{2\tau^2 a_\alpha } \right \}
f_{A_\alpha}\left ( a_\alpha,t|\thinspace a_{\alpha_0},t_0 \right ) 
\right ] \nonumber\\
\fl
&~&+ \frac{\sigma^2_{X_\alpha}}{2 \tau^2}
\frac{\partial^2}{\partial a^2_\alpha} f_{A_\alpha}\left ( a_\alpha,t|\thinspace a_{\alpha_0},t_0 \right ) .
\label{eq:FPEOUACart}
\eea
As an alternative to solving (\ref{eq:FPEOUACart}), the (T)PDF for $f_{A_\alpha}$ can be determined more straightforwardly from variate transformation of the corresponding distribution for $U_\alpha$. For example, 
the limit distribution of $A_\alpha(t)$ when $(t-t_0)/\tau$ and $a^2_{\alpha_0} \tau / \sigma^2_{X_\alpha} \ll 1$ is obtained from (\ref{eq:condprobenergyCart_smallt_smalltau}) as
\bea
\fl
f_{A_\alpha}\left ( a_\alpha,t | a_{\alpha_0},t_0\right ) = 
\frac{ \tau^2 a_\alpha}{\left ( t - t_0 \right )\sigma^2_{X_\alpha} } 
\exp \left [ - \frac{\tau^2 \left ( a^2_\alpha+a^2_{\alpha_0} \right ) }{2 \left ( t-t_0 \right )\sigma^2_{X_\alpha} } \right ]
I_0 \left [ \frac{ \tau^2 a_\alpha a_{\alpha_0} }{ \left ( t-t_0 \right )\sigma^2_{X_\alpha} } \right ],~~(a_\alpha>0,t>t_0)
\label{eq:temp1} .
\label{eq:condprobmag}
\eea
This distribution has also been encountered in a related problem of one-dimensional power dissipation in electronic circuits subjected to external illumination by random fields \cite{arnaRadioSci}.

\subsubsection{Vector field}
For $A_\rmt(t)$ we obtain, using (\ref{eq:driftUtot}), (\ref{eq:diffusionUtot}) and the definitions (\ref{eq:trfvardriftCartA}) and (\ref{eq:trfvardiffusionCartA}) applied to $A_\rmt(t)$,
\bea
D^{(1)}_{A_\rmt} (a_\rmt, t)
&=&
\left [ \tau a_\rmt(t) \right ]^{-1}  \left \{ \left ( \frac{\sigma^2_{X_\rmt}}{\tau}  - a^2_\rmt \right ) \exp \left [ - \frac{2 (t-t_0)}{\tau} \right ] - \frac{\sigma^2_{X_\rmt}}{6\tau} \right \}
\label{eq:driftATot}
\\
D^{(2)}_{A_\rmt} (a_\rmt, t)
&=&
\frac{\sigma^2_{X_\rmt}}{6\tau^2}.
\eea
The corresponding SDE is
\bea
\dot{A}_\rmt (t) &=&
\left [ \tau A_\rmt(t) \right ]^{-1}  \left \{ \left ( \frac{\sigma^2_{X_\rmt}}{\tau}  - a^2_{\rmt_0} \right ) \exp \left [ - \frac{2 (t-t_0)}{\tau} \right ] - \frac{\sigma^2_{X_\rmt}}{6\tau} \right \}
+ \frac{\dot{B}_\rmt(t) }{\sqrt{3}\tau} 
\label{eq:LangevinOUAt}
\eea
and the FPE reads
\bea
\fl \hspace{1.5cm}
\frac{\partial}{\partial t}                           f_{A_\rmt}\left ( a_\rmt,t|\thinspace a_{\rmt_0},t_0 \right )
&=&
\frac{1}{\tau}
\left \{ \left ( a^2_{\rmt_0} - \frac{\sigma^2_{X_\rmt}}{\tau} \right ) \exp \left [ - \frac{2 (t-t_0)}{\tau} \right ] + \frac{\sigma^2_{X_\rmt}}{6\tau} \right \}
\frac{\partial}{\partial a_\rmt}     \left [ \frac{ f_{A_\rmt}\left ( a_\rmt,t|\thinspace a_{\rmt_0},t_0 \right ) }{a_\rmt} \right ] \nonumber\\
\fl
&~&~~~+ \frac{\sigma^2_{X_\rmt}}{6 \tau^2}
\frac{\partial^2}{\partial a^2_\rmt} f_{A_\rmt}\left ( a_\rmt,t|\thinspace a_{\rmt_0},t_0 \right ) 
\label{eq:FPEOUAt}
\eea
whose short-term fundamental solution, assuming an unbiased underlying field, is given by
\bea
f_{A_\rmt} \left ( a_\rmt,t \right ) = 
\frac{3a^3_\rmt}{\left ( t - t_0 \right ) a^2_{\rmt_0}\sigma^2_{X_\rmt} } 
\exp \left [ - 
\frac{3\tau^2(a^2_\rmt + a^2_{\rmt_0})}{2 \left ( t-t_0 \right ) \sigma^2_{X_\rmt} } \right ] 
I_2 \left [ \frac{ 3 \tau^2 a_\rmt a_{\rmt_0}}{(t-t_0) \sigma^2_{X_\rmt} } \right ] .
\label{eq:condprobamplitudeTot}
\eea

\subsubsection{Generalized field}
For a generalized field with $2p$ degrees of freedom, the SDE and FPE follow by straightforward extension as
\bea
\dot{A}_p (t) &=&
\left [ \tau A_p(t) \right ]^{-1}  \left \{ \left ( \frac{\sigma^2_{X_p}}{\tau}  - a^2_{p_0} \right ) \exp \left [ - \frac{2 (t-t_0)}{\tau} \right ] - \frac{\sigma^2_{X_p}}{2p \tau} \right \}
+ \frac{\dot{B}_p(t) }{\sqrt{p}\thinspace\tau} 
\label{eq:LangevinOUAp}
\eea
and
\bea
\fl 
\frac{\partial}{\partial t}
f_{A_p}\left ( a_p,t|\thinspace a_{p_0},t_0 \right )
&=&
\frac{1}{\tau}
\left \{ \left ( a^2_{p_0} - \frac{\sigma^2_{X_p}}{\tau} \right ) \exp \left [ - \frac{2 (t-t_0)}{\tau} \right ] + \frac{\sigma^2_{X_p}}{2p\tau} \right \}\nonumber\\
\fl
&\times&
\frac{\partial}{\partial a_p}     \left [ \frac{ f_{A_p}\left ( a_p,t|\thinspace a_{p_0},t_0 \right ) }{a_p} \right ] 
+ \frac{\sigma^2_{X_p}}{2p \tau^2}
\frac{\partial^2}{\partial a^2_p} f_{A_p}\left ( a_p,t|\thinspace a_{p_0},t_0 \right ) 
\label{eq:FPEOUAp}
\eea
respectively, with short-term fundamental solution, again assuming an unbiased underlying field, given by
\bea
f_{A_{p}} \left ( a_{p},t \right ) = 
\frac{p \thinspace a_{p_0} }{\left ( t - t_0 \right ) \sigma^2_{X_p} } \left ( \frac{a_p
}{ a_{p_0} } \right )^p \exp \left [ - 
\frac{p \tau^2 \left ( a^2_{p} + a^2_{{p}_0} \right ) }{2 \left ( t-t_0 \right ) \sigma^2_{X_p}} \right ] 
I_{p-1} \left [ \frac{p \tau^2 a_{p} a_{{p}_0}}{(t-t_0) \sigma^2_{X_p} } \right ] .
\label{eq:condprobamplitudep}
\eea


\section{Example: BEWL diffusion process 
\label{sec:BEWL}}
The above results for the dissipative Ornstein--Uhlenbeck diffusion can be specialized to the case of a Bachelier--Einstein--Wiener--L\'{e}vy (BEWL) process $B_p(t)$, i.e., a pure diffusion, corresponding to the limit $t/\tau \rightarrow 0$ in the above (cf. figure \ref{fig:randomwalk}b). In this case takes, the field requires an arbitrarily long time to relax to its asymptotic equilibrium state. 
Here, we formulate the results immedately in terms of the generalized field, from which local scalar ($p=1$) and vector ($p=3$) fields follow as particular cases.

\subsection{Energy density}
\subsubsection{SDE}
For a BEWL process, we have that
\bea
\left \langle B_p(t) \right \rangle = 0,~~~~
\left \langle B_p(t-t_0)B_p(t^\prime-t_0) \right \rangle = \frac{\sigma^2_{X_p}}{2 \tau^2} \left ( t+t^\prime -2 t_0 -|t-t^\prime| \right ).
\eea 
Following an analysis similar to that in section \ref{sec:CartEnergy}, we obtain
\bea
\left \langle U_p(t+\delta t)-U_p(t) \right \rangle = 
\frac{2}{\tau} \left ( \frac{\sigma^2_{X_p}}{\tau} - u_{p_0} \right ) \delta t ,
\eea
\bea
\left \langle [ U_p(t+\delta t)-U_p(t) ]^2 \right \rangle = 
\frac{4\sigma^2_{X_p}}{\tau^2} \thinspace U_p (t) ~ \delta t 
+ 4 p (1+p) (\delta t)^2 + \Or \left [ (\delta t)^3 \right ].
\eea 
Combined with (\ref{eq:LangevUCartOUIto_gen}) and (\ref{eq:LangevUCartOUStrat_gen}) applied to a generalized field, the SDEs for $U_p$ in the It\^{o} and Stratonovich formulations become, respectively,
\bea
\dot{U}_p(t) 
&=& 
\frac{2}{\tau} \left ( \frac{\sigma^2_{X_p}}{\tau} - u_{p_0} + \sqrt{ \frac{U_p(t)}{p}} \dot{B}_p(t) \right ) ,
\eea
\bea
\dot{U}_p(t) 
&=& 
\frac{2}{\tau} 
\left [ \left ( 1- \frac{1}{2p} \right ) \frac{\sigma^2_{X_p}}{\tau} - u_{p_0} + \sqrt{ \frac{U_p(t)}{p} } \dot{B}_p(t) 
\right ].
\eea

\subsubsection{FPE}
Following (\ref{eq:FPEgen}), the associated FPE for $f_{U_p}(u_p,t|u_{p_0},t_0)$ is
\bea
\fl \hspace{1.5cm}
\frac{\partial}{\partial t}                          f_{U_p}\left ( u_p,t|u_{p_0},t_0 \right ) 
&=
- \frac{2}{\tau} \left ( \frac{\sigma^2_{X_p}}{\tau} - u_{p_0} \right )
\frac{\partial}{\partial u_p}     \left [               f_{U_p}\left ( u_p,t|u_{p_0},t_0 \right ) \right ] \nonumber\\
&~~~
+
\frac{2 \sigma^2_{X_p}}{p\tau^2}
\frac{\partial^2}{\partial u^2_p} \left [ u_p f_{U_p}\left ( u_p,t|u_{p_0},t_0 \right ) \right ] 
\label{eq:FPenergy1}
\eea
now with the boundary conditions $f_{U_{p}}(0,t|u_{p_0},t_0)=f_{U_p}(+\infty,t|u_{p_0},t_0)=0$. 

With the former condition, it is verified that the probability current (\ref{eq:defprobcurr}), which on account of current continuity can also be expressed as
\bea
\jmath_{U_p} ( u_p, t ) = 
\int^{u_p} 
\left \{
\frac{2}{\tau} \left ( \frac{\sigma^2_{X_p}}{\tau} - u_{p_0} \right ) f_{U_p}\left ( u^\prime_p \right ) - \frac{2 \sigma^2_{X_p}}{\tau^2}
\frac{\partial}{\partial u^\prime_p} \left [ u^\prime_p f_{U_p} \left ( u^\prime_p \right ) \right ] 
\right \}
{\rm d} u^\prime_p,
\eea
vanishes at $u_p = 0$ for all $t$. This may be expected on physical grounds, on account of the positivity of energy [$f_{U_p}(u_p < 0,t) = 0$]. In general, however, $\jmath_{U_p}(u_p\geq 0,t) \not = 0$ except for the trivial case $u_{p_0} = 0$.

\subsection{Field magnitude}

Through the variate transformation for the field magnitude $A_p(t) \propto \sqrt{U_p(t)}$, we obtain the drift and diffusion coefficients as
\bea
D^{(1)}_{A_p} (a_p,t)
=
\left ( 1-\frac{1}{2p} \right ) \frac{\sigma^2_{X_p}}{\tau^2 a_p(t)}
,
\label{eq:driftmagBEWL}
\eea
\bea
D^{(2)}_{A_p} (a_p,t) 
=
\frac{\sigma^2_{X_p}}{2 p \tau^2} 
\label{eq:diffusionmagBEWL}
\eea
whence the SDE in the It\^{o} formulation is
\bea
\dot{A}_p(t) 
=
D^{(1)}_{A_p} + \frac{\sqrt{2 D^{(2)}_{A_p}}}{\sigma_{X_p}} \dot{B}_p(t) 
=
\left ( 1-\frac{1}{2p} \right ) \frac{\sigma^2_{X_p}}{\tau^2 A_p(t)}
+ \frac{\dot{B}_p(t) }{\sqrt{p} \thinspace \tau} .
\label{eq:LangevinACartBEWLIto}
\eea
Alternatively, (\ref{eq:LangevinACartBEWLIto}) may be obtained from the It\^{o} formula. 
The SDE in the Stratonovich formulation is identical, because $D^{(2)}_{A_p}$ is independent of $a_p$.
The corresponding FPE follows again by substitution of (\ref{eq:driftmagBEWL}) and (\ref{eq:diffusionmagBEWL}).

\section{Response of first- and second-order systems to interior field\label{sec:systemnonstat}}
In certain practical applications, the goal is to determine how a (for sake of simplicity) first-order system with characteristic time constant $\tau$ responds to a nonstationary cavity field.
The analysis below complements a similar analysis for nonlinearity and distortion given in section IV.B of \cite{arnaTEMCv47n4}. To this end, in section \ref{sec:secondorderRx} we first determine the response of a second-order system to an ideal white noise source field (quasi-stationary interior field), by converting a single second-order SDE into a system of two coupled first-order SDEs with well-separated time scales. This approach is akin to a widely used technique for analyzing a coloured noise process \cite{moss1}. The result is then used in section \ref{sec:nonstatinterior} to find the response of a first-order system to a nonstationary effective interior field. Throughout this section, we assume a scalar field for simplicity, thereby omitting its subscript $\alpha$ for brevity.

\subsection{Quasi-stationary cavity field and second-order receiver system\label{sec:secondorderRx}}

The results of sections \ref{sec:Langevin} and \ref{sec:Langevin2bis} for a simple first-order Rx can be extended to higher-order systems. For a second-order system represented by an RLC equivalent circuit, the response is now governed by
\bea
\ddot{Y}(t) + 2 \zeta \dot{Y}(t) + \omega^2_o Y(t) = v(t) + {\cal T}^{-2} \dot{B}(t) \label{eq:2ndorderSDE}
\eea
where $v(t)$ is a deterministic excitation (e.g., direct illumination or biasing field), $\omega^{2}_o = (LC)^{-1}$ and $2\zeta = \omega_o/Q_o$, in which $Q_o = \omega_o RC$ or $(\omega_o RC)^{-1}$ for a parallel or series RLC circuit model, respectively. 
The mean, variance, and covariance of $Y(t)$ and $\dot{Y}(t)$ can be derived by converting the single second-order SDE (\ref{eq:2ndorderSDE}) into a system of two coupled first-order SDEs, as shown in \ref{app:secondordersyst}, yielding
\bea
{\mu}_{Y}(t) 
&=&
\left ( 2 \sqrt{\zeta^2-\omega^2_o} \right )^{-1} 
\nonumber\\ &~&
\times 
\left \{ 
\left ( \zeta + \sqrt{\zeta^2-\omega^2_o} \right ) \exp \left [ - \left ( \zeta - \sqrt{\zeta^2-\omega^2_o} \right ) t \right ]  y_0
\right . 
\nonumber\\ &~&
~~~\left.
-
\left ( \zeta - \sqrt{\zeta^2-\omega^2_o} \right ) \exp \left [ - \left ( \zeta + \sqrt{\zeta^2-\omega^2_o} \right ) t \right ] y_0 
\right .
\nonumber\\ &~&
~~~\left . 
+ 
\exp \left [ - \left ( \zeta - \sqrt{\zeta^2-\omega^2_o} \right ) t \right ] \dot{y}_0 
-
\exp \left [ - \left ( \zeta + \sqrt{\zeta^2-\omega^2_o} \right ) t \right ] \dot{y}_0 
\right \} 
,
\label{eq:musecondorderRx}
\eea
\bea
{\mu}_{\dot{Y}}(t) 
&=&
\left ( 2 \sqrt{\zeta^2-\omega^2_o} \right )^{-1} 
\nonumber\\ &~&
\times 
\left \{ 
\omega^2_o \exp \left [ - \left ( \zeta + \sqrt{\zeta^2-\omega^2_o} \right ) t \right ] y_0
-
\omega^2_o \exp \left [ - \left ( \zeta - \sqrt{\zeta^2-\omega^2_o} \right ) t \right ] y_0 
\right .
\nonumber\\ &~& 
~~~\left . 
+ 
\left ( \zeta + \sqrt{\zeta^2-\omega^2_o} \right ) 
\exp \left [ - \left ( \zeta + \sqrt{\zeta^2-\omega^2_o} \right ) t \right ] \dot{y}_0 
\right .
\nonumber\\ &~&
~~~\left . 
-
\left ( \zeta - \sqrt{\zeta^2-\omega^2_o} \right ) 
\exp \left [ - \left ( \zeta - \sqrt{\zeta^2-\omega^2_o} \right ) t \right ] \dot{y}_0 
\right \} 
,
\label{eq:mudotsecondorderRx}
\eea
\bea
{\sigma}^2_{Y}(t) &=&
\frac{\sigma^2_X }{8 {\cal T}^4 \left ( \zeta^2 - \omega^2_o \right )}
\left [
\frac{
1 - \exp \left [ - 2 \left ( \zeta + \sqrt{\zeta^2-\omega^2_o} \right ) t \right ] }{\zeta + \sqrt{\zeta^2-\omega^2_o}} \right . 
\nonumber\\
&~& 
+
\left. 
\frac{1 -
\exp \left [ - 2 \left ( \zeta - \sqrt{\zeta^2-\omega^2_o} \right ) t \right ] }{\zeta - \sqrt{\zeta^2-\omega^2_o}}
- \frac{2}{\zeta} \left [
1 - \exp \left ( - 2 \zeta t \right ) \right ]
\right ],
\label{eq:sigmasqsecondorderRx}
\\
{\sigma}^2_{\dot{Y}}(t) &=&
\frac{\sigma^2_X }{8 {\cal T}^4 \left ( \zeta^2 - \omega^2_o \right )}
\left [
\left ( \zeta + \sqrt{\zeta^2-\omega^2_o} \right ) 
\left \{
1 - \exp \left [ - 2 \left ( \zeta + \sqrt{\zeta^2-\omega^2_o} \right ) t \right ] 
\right \}
\right.
\nonumber\\
&~&
+
\left. 
\left ( \zeta - \sqrt{\zeta^2-\omega^2_o} \right )  
\left \{ 1 - \exp \left [ - 2 \left ( \zeta - \sqrt{\zeta^2-\omega^2_o} \right ) t \right ] \right \} \right. \nonumber\\
&~&
\left.
- \frac{2 \omega^2_o}{\zeta}
\left [ 1 - \exp \left ( - 2 \zeta t \right ) \right ]
\right ],
\label{eq:sigmasqdotsecondorderRx}
\\
{\sigma}_{Y\dot{Y}}(t) &=&
\frac{\sigma^2_X }{8 {\cal T}^4 \left ( \zeta^2 - \omega^2_o \right )}
\left [
\left \{
1 - \exp \left [ - 2 \left ( \zeta + \sqrt{\zeta^2-\omega^2_o} \right ) t \right ] 
\right \}
\right . 
\nonumber\\
&~&
+
\left. 
\left \{ 1 - \exp \left [ - 2 \left ( \zeta - \sqrt{\zeta^2-\omega^2_o} \right ) t \right ] \right \}
- 2
\left [ 1 - \exp \left ( - 2 \zeta t \right ) \right ]
\right ].
\label{eq:sigmasqmixsecondorderRx}
\eea
Note that the mean value of the transient (although not the variance) now depends not only on $y_0$ but also on $\dot{y}_0$, i.e., the initial velocity of the mixed field. Thus, the result will in general depend on whether the reset occurs at the beginning of the stirring process (reset from ``stand-still''), or during the process (``flying-start'' reset). As a generalization, for a third- or higher-order Rx a dependence on the initial field acceleration $\ddot{y}_0$ or higher-order time-derivatives occurs, respectively.

From (\ref{eq:Langevin2D}), the FPE for the joint PDF $f_{Y,\dot{Y}}(y,\dot{y})$ with $v(t)=0$ follows as 
\bea
\frac{\partial f_{Y,\dot{Y}} (y,\dot{y},t) }
     {\partial t} 
&=& 
- \dot{y} \frac{\partial f_{Y,\dot{Y}}(y,\dot{y},t) }{\partial y} + \omega^2_o \thinspace y \frac{\partial f_{Y,\dot{Y}}(y,\dot{y},t) }{\partial \dot{y}} 
\nonumber\\ 
&~& + 
2\zeta \frac{\partial }{\partial \dot{y}} \left [ \dot{y} f_{Y,\dot{Y}}(y,\dot{y},t) \right ]
+ \frac{\sigma^2_X}{2 {\cal T}^4} \frac{\partial^2 f_{Y,\dot{Y}}(y,\dot{y},t) }{\partial \dot{y}^2}.
\label{eq:FPEjoint}
\eea
The stationary joint PDF $f_{Y,\dot{Y}}(y,\dot{y},+\infty)$ is obtained from (\ref{eq:FPEjoint}) as
\bea
\fl \hspace{1.5cm}
f_{Y,\dot{Y}}(y,\dot{y},+\infty) 
= f_{Y}(y,+\infty) f_{\dot{Y}}(\dot{y},+\infty)
= C \exp \left ( - \frac{2\zeta            {\cal T}^4}{\sigma^2_X} \dot{y}^2 \right ) 
    \exp \left ( - \frac{2\zeta \omega^2_o {\cal T}^4}{\sigma^2_X}     {y}^2 \right ) 
\label{eq:FPEjointstatsol}
\eea
i.e., $f_{Y,\dot{Y}}$ factorizes into two Gauss normal marginal PDFs for $Y$ and $\dot{Y}$.
The corresponding FPE for the marginal PDF $f_Y(y,t)$ is well known to be (cf. eq. (I.4.245) in \cite{stra1}) 
\bea
\fl \hspace{1.5cm}
\frac{\partial f_{Y} (y,t) }
     {\partial t} 
&=
\frac{\omega^2_o}{2\zeta} \left [ 1 + \left ( \frac{\omega_o}{2\zeta} \right )^2 \right ]
\frac{\partial}{\partial y} \left [ y f_Y(y,t) \right ]
+
\frac{\sigma^2_X}{2\left ( 2 \zeta {\cal T}^2 \right )^2}
\left [ 1 + \left ( \frac{\omega_o}{2\zeta} \right )^2 \right ]
\frac{\partial^2 f_{Y}(y,t) }{\partial {y}^2} .
\label{eq:FPEmarg}
\eea
The stationary marginal PDF $f_Y(y,+\infty)$ can also be directly obtained from (\ref{eq:FPEmarg}) and is similar to the corresponding PDF  
for a first-order process, viz.,
\bea
f_Y(y,+\infty) = \sqrt{\frac{2\zeta \omega^2_o {\cal T}^4 }{{\pi} \thinspace \sigma^2_X}} \exp \left ( - \frac{ 2\zeta \omega^2_o {\cal T}^4}{\sigma^2_X} y^2\right ),
\label{eq:FPEmargstatsolY}
\eea
whence $\mu_Y(t\rightarrow +\infty) =0$, $\sigma_Y (t\rightarrow +\infty) = \sqrt{Q_o/(2\omega^3_o{\cal T}^4 )} \thinspace \sigma_X $. As may be expected, lowering the $Q$-factor of the Rx is seen to result in a reduced variability of the perceived field. This effect, which must be accounted for, is irrespective of any loading effect that insertion of the Rx may impose on the stationary cavity field which excites it.
The marginal PDF $f_{\dot{Y}}(\dot{y},+\infty)$ follows from (\ref{eq:FPEjointstatsol}) and (\ref{eq:FPEmargstatsolY}) as
\bea
f_{\dot{Y}}(\dot{y},+\infty) = \sqrt{\frac{2\zeta {\cal T}^4 }{{\pi} \thinspace \sigma^2_X}} \exp \left ( - \frac{ 2\zeta {\cal T}^4}{\sigma^2_X} \dot{y}^2\right ),
\label{eq:FPEmargstatsoldotY}
\eea
whence $\mu_{\dot{Y}} (t\rightarrow +\infty) =0$, $\sigma_{\dot{Y}} (t\rightarrow +\infty) = \sqrt{Q_o/(2\omega_o{\cal T}^4 )} \thinspace \sigma_X $. The latter distribution is relevant to the characterization of the rate of fluctuation, whose sample maximum value is important in applications \cite{arnaTEMCv47n4}.
The distributions of the energy density and field magnitude are qualitatively the same as for the first-order Rx.

\subsection{Nonstationary cavity field with first-order Rx\label{sec:nonstatinterior}}
The previous analysis presumed $X(t)$ to be quasi-stationary, i.e., to represent an input field that is sufficiently slowly varying relative to the intrinsic time constant $\tau_{\rm ch}$ of the mechanically static propagation environment (cavity).
If the mode-stirring process is sufficiently rapid (e.g., for electronic stirring), however, then the field presented as input to a detector or test device with its own characteristic time constant $\tau$ may itself become nonstationary. In effect, the nonstationary cavity field $Y(t)$ now becomes the new input field $X(t)$ for this device. This complicates an analytical treatment. A time-domain numerical calculation then seems the most appropriate approach. 
Alternatively, a complex frequency $\sigma+\rmj\omega$ may be introduced to incorporate transients or nonstationarity into this $X(t)$ \cite{arnaTEMCv47n4}. In order to gain further insight, however, we shall here pursue an analytical solution under the assumption that $X(t)$ is {\em short-term\/} stationary (i.e., quasi-stationary within a sufficiently narrow time interval).
Thus, in a certain sense, the cavity now ``generates'' its own transfer function for the stirring process as a result of the finite response time $\tau_{\rm ch} $ of the chamber (cavity) which is long  in real time compared to the scale of fluctuation $\tau_{\rho,X}$ of the stirring process. 

Specifically, in this case, 
\bea
\left \{
\begin{array}{c}
\dot{X}(t) + \tau^{-1}_{\rm ch} X(t) =  \tau^{-1}_{\rm ch} \dot{\cal B}(t) \\ 
\dot{Y}(t) + \tau^{-1}          Y(t) =  \tau^{-1}                     X(t) , 
\end{array}
\right .
\label{eq:LangevinChamber}
\eea
thereby assuming that
\bea
\tau_{\rho,X} \ll \tau_{\rm ch} \ll \tau \ll 1 .
\label{ineq:chamber}
\eea
Each one of the first two inequalities in (\ref{ineq:chamber}) enables either one of the equations in (\ref{eq:LangevinChamber}) still to be treated as a SDE driven by a white noise process (i.e., a source function that is fluctuating very rapidly relative to the response function).
In (\ref{eq:LangevinChamber}), $X(t) = \dot{B}(t)$ is the actual and possibly nonstationary cavity field, as would be perceived by an idealized instantaneously responding dot sensor or component [$\tau=0$, i.e., $Y(t)=X(t)$], while ${\cal B}(t)$ is defined as the time-integral of a now fictitious {\it quasi-stationary} cavity field $Z(t)$.
Equation (\ref{eq:LangevinChamber}) holds in the mean, when the higher-order fluctuations of the field decay are neglected, and can be written as the single second-order SDE
\bea
\ddot{Y}(t) + \frac{ \tau + \tau_{\rm ch} }{\tau \tau_{\rm ch} } \dot{Y}(t) + \left ( {\tau \tau_{\rm ch}} \right )^{-1} Y(t) = \left ( { \tau \tau_{\rm ch}} \right )^{-1} \dot{\cal B}(t)  .
\eea

From (\ref{eq:FPEjoint}) and (\ref{eq:FPEmarg}) with $2\zeta = \tau^{-1} + \tau^{-1}_{\rm ch}$ and $\omega^2_o = {\cal T}^{-2} = (\tau \tau_{\rm  ch})^{-1}$, the joint 2-D PDF $f_{Y,\dot{Y}}(y,\dot{y})$ and marginal 1-D PDF $f_Y(y,t)$ with the initial condition $f_Y (y_0,t_0 ) = \delta(y-y_0)$ have associated FPEs 
\bea
\frac{\partial f_{Y,\dot{Y}} (y,\dot{y},t) }
     {\partial t} 
&=& 
- \dot{y} \frac{\partial f_{Y,\dot{Y}}(y,\dot{y},t) }{\partial y} + \left ( \tau \tau_{\rm ch} \right )^{-1} y \frac{\partial f_{Y,\dot{Y}}(y,\dot{y},t) }{\partial \dot{y}} 
\label{eq:FPEjoint2}
\nonumber\\ 
&~& + 
\frac{\tau+\tau_{\rm ch}}{\tau \tau_{\rm ch}} \frac{\partial }{\partial \dot{y}} \left [ \dot{y} f_{Y,\dot{Y}}(y,\dot{y},t) \right ]
+ \frac{\sigma^2_X}{2 \left ( \tau \tau_{\rm ch} \right )^2} \frac{\partial^2 f_{Y,\dot{Y}}(y,\dot{y},t) }{\partial \dot{y}^2}\\
\frac{\partial f_Y(y,t)}{\partial t}
&=&
\frac{1}{\tau} \frac{\partial }{\partial y} \left [ y f_Y(y,t) \right ]
+
\frac{\sigma^2_Z}{2\tau^2}\left ( 1- \frac{\tau_{\rm ch}}{\tau} \right )
\frac{\partial^2 f_Y(y,t)}{\partial y^2} ,
\label{eq:FPEmarg2}
\eea
where the latter equation is valid to first order in the ratio $(\tau_{\rm ch}/\tau)$. 
The stationary joint PDF follows from (\ref{eq:FPEjoint2}) as
\bea
f_{Y,\dot{Y}}(y,\dot{y},+\infty) &=& 
C \exp 
\left [ -
\frac{\tau \tau_{\rm ch} \left ( \tau + \tau_{\rm ch} \right )}{\sigma^2_Z} \dot{y}^2 \right ]
\exp
\left [ -
\frac{\left ( \tau + \tau_{\rm ch} \right )}{\sigma^2_Z} \left ( y^2 - y^2_0 \right ) \right ] .
\label{eq:FPEjointstatsol2}
\eea
The stationary marginal PDFs follow from (\ref{eq:FPEmarg2})--(\ref{eq:FPEjointstatsol2}) as
\bea
f_Y(y,+\infty) &=& C \exp \left [ - \frac{2}{\sigma^2_Z} \int^y_{y_0} \left ( 1 + \frac{\tau_{\rm ch}}{\tau} \right ) \frac{y^\prime}{\tau} {\rm d}y^\prime + \left ( 1 + \frac{\tau_{\rm ch}}{\tau} \right ) \right ]\nonumber\\
&=&
\sqrt{ \frac{\tau+\tau_{\rm ch}}{\pi\sigma^2_Z}} \exp \left [ - \frac{ (\tau+\tau_{\rm ch}) }{\sigma^2_Z} y^2 \right ] ,
\label{eq:nonstat_stat}\\
f_{\dot{Y}}(\dot{y},+\infty)
&=&
\sqrt{ \frac{ \tau\tau_{\rm ch} \left ( \tau+\tau_{\rm ch} \right )}{\pi\sigma^2_Z}} \exp \left [ - \frac{ \tau\tau_{\rm ch} \left ( \tau+\tau_{\rm ch} \right ) }{\sigma^2_Z} \dot{y}^2 \right ] 
\label{eq:nonstat_statdotY}\\
&\simeq&
\sqrt{ \frac{ \tau^2\tau_{\rm ch}}{\pi\sigma^2_Z}} \exp \left ( - \frac{ \tau^2\tau_{\rm ch}}{\sigma^2_Z} \dot{y}^2 \right )
\eea
where we made use of the fact that $\tau_{\rm ch} \ll \tau$.

The general nonstationary solution can be obtained as a series expansion, but its expression is more cumbersome. We limit ourselves here to calculation of the time dependence of the first- and second-order moments.
Following from (\ref{eq:musecondorder}) and (\ref{eq:sigmasecondorder}), we obtain
\bea
\mu_Y(t) &=& 
\frac{ \exp \left ( -\frac{t}{\tau} \right ) - \frac{\tau}{\tau_{\rm ch}} \exp \left ( -\frac{t}{\tau_{\rm ch}} \right )}{ 1 - \frac{\tau_{\rm ch}}{\tau} } y_0
+
\frac{ \exp \left ( -\frac{t}{\tau} \right ) - \exp \left ( - \frac{t}{\tau_{\rm ch}} \right )}{ 1 - \frac{\tau_{\rm ch}}{\tau} } \tau_{\rm ch}
\dot{y}_0,\\
\mu_{\dot{Y}}(t) &=& 
-
\frac{ \exp \left ( -\frac{t}{\tau} \right ) - \exp \left ( -\frac{t}{\tau_{\rm ch}} \right )}{ 1 - \frac{\tau}{\tau_{\rm ch}} } \frac{y_0}{\tau}
-
\frac{ \frac{\tau_{\rm ch}}{\tau} \exp \left ( -\frac{t}{\tau} \right ) + \exp \left ( - \frac{t}{\tau_{\rm ch}} \right )}{ 1 - \frac{\tau_{\rm ch}}{\tau} }
\dot{y}_0,
\eea
\bea
\sigma^2_Y(t) &=&  
\frac{\sigma^2_Z}{\left ( \tau - \tau_{\rm ch} \right )^2}
\left [ 
\frac{ 1 - \exp \left ( - \frac{2t}{\tau} \right )}{2/\tau}
+
\frac{ 1 - \exp \left ( - \frac{2t}{\tau_{\rm ch}} \right )}{2/\tau_{\rm ch}}
\right . \nonumber\\
&~& ~~~~~~~~~~~~~~~
\left.
-
\frac{ 2 \left \{ 1 - \exp \left [ - \left ( \tau^{-1} + \tau^{-1}_{\rm ch} \right ) t \right ] \right \} }{ \tau^{-1} + \tau^{-1}_{\rm ch} }
\right ],\\
\sigma^2_{\dot{Y}}(t) &=&  
\frac{\sigma^2_Z}{\left ( \tau - \tau_{\rm ch} \right )^2}
\left [ 
\frac{ 1 - \exp \left ( - \frac{2t}{\tau} \right )}{2\tau}
+
\frac{ 1 - \exp \left ( - \frac{2t}{\tau_{\rm ch}} \right )}{2\tau_{\rm ch}}
\right . \nonumber\\
&~& ~~~~~~~~~~~~~~~
\left.
-
\frac{ 2 \left \{ 1 - \exp \left [ - \left ( \tau^{-1} + \tau^{-1}_{\rm ch} \right ) t \right ] \right \} }{ \tau + \tau_{\rm ch} }
\right ],\\
\sigma_{Y\dot{Y}}(t) &=&  
\frac{\sigma^2_Z}{2\left ( \tau - \tau_{\rm ch} \right )^2}
\left [ 
\exp \left ( - \frac{t}{\tau} \right ) - \exp \left ( - \frac{t}{\tau_{\rm ch}} \right ) 
\right ]^2 .
\eea
Since $\tau_{\rm ch} \ll \tau$, the Rx experiences only a second-order asymptotic effect, viz.,
\bea
\mu_Y (t \rightarrow + \infty) &\simeq& \left ( y_0 +  \tau_{\rm ch} \dot{y}_0 \right ) \exp \left (- \frac{t}{\tau} \right ) \rightarrow 0 ,
\label{eq:nonstatmeanY}\\
\mu_{\dot{Y}} (t \rightarrow + \infty) & \simeq& - \frac{y_0 +  \tau_{\rm ch} \dot{y}_0 }{\tau} \exp \left ( - \frac{t}{\tau} \right ) \rightarrow 0 ,
\label{eq:nonstatmeanYdot}
\eea
\bea
\sigma^2_Y (t\rightarrow +\infty) \rightarrow 
\frac{\sigma^2_Z}{{2\left ( \tau + \tau_{\rm ch} \right )}} \simeq
\frac{\sigma^2_Z}{{2\tau}} \left ( 1 - \frac{\tau_{\rm ch}}{\tau} \right ) ,
\label{eq:nonstatvarY}
\eea
\bea
\sigma^2_{ \dot{Y}} (t\rightarrow +\infty) \rightarrow 
\frac{\sigma^2_Z}{{2\tau^2 \tau_{\rm ch}} \left ( 1 - \frac{\tau_{\rm ch}}{\tau} \right )^2 } \simeq 
\frac{\sigma^2_Z}{{2\tau^2 \tau_{\rm ch}}} \left ( 1 + 2 \frac{\tau_{\rm ch}}{\tau} \right ) ,
\label{eq:nonstatvarYdot}
\eea
\bea
\sigma_{ Y\dot{Y}} (t\rightarrow +\infty) \rightarrow 0 ,
\eea
with (\ref{eq:nonstatmeanY}) and (\ref{eq:nonstatvarY}) leading to the PDF (\ref{eq:nonstat_stat}). The variances (\ref{eq:nonstatvarY}) and (\ref{eq:nonstatvarYdot}) indicate that nonstationary mode stirring yields a lower root-mean-square (rms) {\em magnitude\/} of fluctuation $\sqrt{\langle Y^2(t) \rangle}$ but a larger rms {\em rate\/} of fluctuation $\sqrt{\langle \dot{Y}^2(t) \rangle}$ compared to (quasi-)stationary mixing [i.e., stirring or tuning ($\tau_{\rm ch}/\tau \rightarrow 0$)]. In the limit $\tau_{\rm ch} \rightarrow +\infty$, we retrieve the characteristics of a Markov process, i.e., $\sigma_{\dot{Y}} \rightarrow +\infty$, as may be expected. Furthermore, (\ref{eq:nonstatmeanY}) and (\ref{eq:nonstatmeanYdot}) show that no asymptotic bias occurs ($\mu_Y$, $\mu_{\dot{Y}} \rightarrow 0$), although the initial bias (for $t/\tau \rightarrow 0$) now increases when $\tau_{\rm ch} \dot{y}_0 \not = 0$.

\section{SDE and FPE with time-dependent deterministic coefficients 
\label{sec:timedeptau}}
\begin{figure}[t] \begin{center} 
\begin{tabular}{c} \ 
\epsfxsize=8cm 
\epsfbox{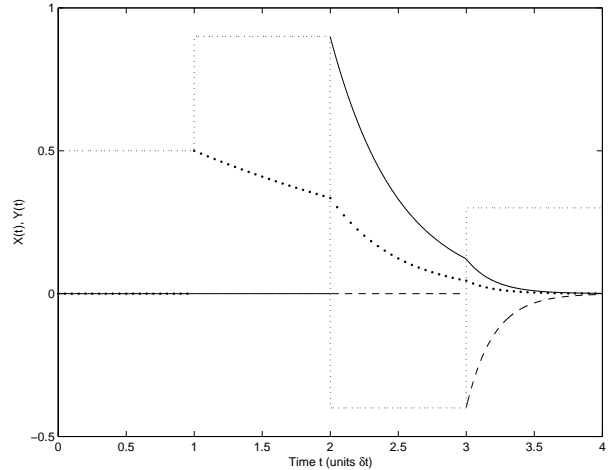}\ \\ 
\\
\end{tabular} \end{center}
\caption{\label{fig:nonstationarydiagramextended} \small
Time dependence of the amplitude of the discretized nonstationary real-valued stirring process with state-dependent decay constant $\tau(t)$. Dotted, solid and dashed curves represent the contribution to $Y(t)$ when having reached steady-state at the starting time, i.e., $ Y(t_0=i\delta t) = X(i\delta t)$, $i=1,2,3$. These contributions evolve with different time constants governed by the subinterval [$t_{i},t_{i+1}]$.}
\end{figure}

In the above, we assumed that the decay constant $\tau$ was independent of the stir state at time $t$. For example, in the case of a nonstationary response of an electronic circuit immersed in a stationary mode-tuned field $X(t)$, $\tau$ is entirely determined by the circuit itself and is then indeed independent of $t$. 
By contrast, when analyzing a mode-stirred cavity field resulting from nonstationary mechanical or electronic stirring, the chamber's decay constant $\tau = Q/\omega\propto\omega^{-1/2}$ is governed by the particular set of participating eigenmodes at each stir state. These modes and, hence, $\tau$ then depend on $t$ (cf. figure \ref{fig:nonstationarydiagramextended}). 
In a first approximation for this scenario, we may assume $\tau$ (being an average value taken across a typically large number of modes within a frequency band that is narrow relative to the center frequency) to be varying relatively slowly compared to the local $X(t)$ and hence for $\tau$ to be a quasi-deterministic but time-varying function of $t$ (Bachelier process). 
Repeating the foregoing analysis of a discretized stirring process, under the often plausible assumption that the decay of $|Y(t_{0-})|$ toward zero is governed by the same characteristic time as the competing growth from zero to $|Y(t_{0-}+\Delta t)|$, and subsequently taking the limit $\delta t \rightarrow 0$,
the SDE (\ref{eq:Langevin}) becomes
\bea
\dot{Y}(t) = - \tau^{-1} (t) Y(t) + \tau^{-1} (t) X(t)
\label{eq:SDEtimedep}
\label{eq:Langevintimedep}
\eea
now with the solution 
\bea
Y(t) = y_0 
\exp \left [ - \int^t_{t_0} \frac{{\rm d}t^\prime}{\tau(t^\prime)} \right ]
+ \int^{t}_{t_0} \frac{X(t^\prime)}{\tau(t^\prime)} \exp \left [ - \int^t_{t^\prime} \frac{{\rm d}t^{\prime\prime}}{\tau(t^{\prime\prime})} \right ] {\rm d}t^\prime ,
\label{eq:gensoltimedep}
\eea
which can be integrated for specified deterministic $\tau(t)$.
For example, assume that the {\em rate\/} of change of $\tau$ is independent of $t$ and denoted by $\nu=\dot{\tau}$, and that $\tau(t)$ can be linearized when taking the limit $\delta t\rightarrow 0$, so that we can write
\bea
\tau(t) = 
\left ( t - t_1 \right ) \nu_{12} + \tau_1 
\eea
for $t_1 \leq t < t_2$, whence
\bea
\int^{t_2}_{t_1} \frac{{\rm d}t}{\tau(t)} = \frac{1}{\nu_{12}}
{\rm ln} \left ( 1 + \frac{\nu_{12} (t_2 - t_1)}{\tau_1} \right ) \simeq \frac{t_2-t_1}{\tau_1}.
\eea
Then (\ref{eq:gensoltimedep}) becomes
\bea
Y(t) 
&=& 
y_0 \left [ \frac{\tau \left ( t_0 \right )}{\tau \left ( t \right )} \right ]^{{1}/{\nu_{0t}}}
+ \int^{t}_{t_0} \frac{X(t^\prime)}{\left ( t^\prime - t_0 \right ) \nu_{0t} + \tau \left ( t_0 \right )} 
\left [ \frac{\tau \left ( t^\prime \right )}{\tau \left ( t \right )} \right ]^{{1}/{\nu_{t^\prime t}}} {\rm d}t^\prime.
\label{eq:gensoltimedepfinal}
\eea

For given $\tau(t)$, the drift and diffusion coefficients can be explicitly calculated. 
In general, the resulting FPE is still of the form
\bea
\frac{\partial f_{U_p}(u_p,t)}{\partial t} 
=
g(t) \frac{\partial f_{U_p}(u_p,t)}{\partial u_p} 
+
h(t) u_p \frac{\partial^2 f_{U_p}(u_p,t)}{\partial u_p^2},
\label{eq:FPEtimedep}
\eea
which can again be solved by Laplace transformation and conversion to a first-order ODE, by proceeding as in \ref{app:solvingFPECart}. Instead of (\ref{eq:trfvarPDE}), the appropriate transformations of variables are now
\bea 
s^\prime \doteq s+ w(s,t),~~~~~t^\prime \doteq t+ u(s,t)
\eea
where we define
\bea
w(s,t) &\doteq & g(t_0) s^2 t - s^2 \int^t_{t_0} g(t^\prime) {\rm d}t^\prime\\
u(s,t) &\doteq & \left [ g(t) s \right ]^{-1},
\eea
with which the coefficient of $\partial {\cal F}_{S^\prime}(s^\prime,t^\prime)/\partial s^\prime$ in (\ref{eq:FPEtoODE}) is constant with respect to $t^\prime$ and the coefficient  of $\partial {\cal F}_{S^\prime}(s^\prime,t)/\partial t^\prime$ vanishes again.

In the case when the time dependence of $\tau(t)$ is weak, we can immediately state the FPE in its first-order approximation, i.e., (\ref{eq:FPEtimedep}) with
\bea
h(t) &=& \frac{2 D^{(2)}_{U_p}(u_p,t)}{u_p} - D^{(1)}_{U_p}(u_p,t) \simeq \frac{4 \sigma^2_{X_p}}{p~\tau^2(t)} - \frac{2}{\tau(t)} \left ( \frac{\sigma^2_{X_p}}{\tau(t)} - u_{p_0} \right )
\exp \left [ - \frac{2(t-t_0)}{\tau(t)} \right ]\\
g(t) &=& \frac{D^{(2)}_{U_p}(u_p,t)}{u_p} \simeq \frac{2\sigma^2_{X_p}}{p~\tau^2(t)}.
\eea

\section{Conclusions\label{sec:conclusions}}
Nonstationarity of scalar (acoustic) and vector (electromagnetic) random classical fields in complex propagation environments has been modelled as a diffusion process and analyzed statistically based on a description in terms of stochastic differential equations. It was shown that closed-form PDF solutions of the FPE for the energy density can be derived if the underlying field is ideal Gauss normal, without recourse to a typical polynomial expansion of the solution for the associated Sturm--Liouville problem. It was found that, in the general case, PDFs of the time-varying energy density are not separable with respect to the energy density and time variables. Limiting and asymptotic forms of these PDFs in the short- and long-time limits have been obtained. The main results are contained in (\ref{eq:LangevUCartOUStrat}), (\ref{eq:FPenergystirGen})--(\ref{eq:condprobenergyGen_smallnonzerot}), (\ref{eq:LangevinOUAp})--(\ref{eq:condprobamplitudep}), (\ref{eq:musecondorderRx})--(\ref{eq:sigmasqmixsecondorderRx}), (\ref{eq:FPEmargstatsolY})--(\ref{eq:FPEmargstatsoldotY}). The energy distribution of a generalized field was found to exhibit a Bessel $I$ distribution with time-varying order and statistics. The results are, in particular, important for the characterization of sample statistics for nonstationary fields, whose random fluctuations are governed by dynamics that exhibit significant departures from those for stationary systems \cite{lead1}.

\ack
This work was supported in part by the Electrical Programme and the Software Support for Metrology Programme of the U.K. Department of Trade and Industry National Measurement System Policy Unit.
I wish to thank W. Weiglhofer (deceased), G. Borzdov, G. Kristensson, L. Jonsson, T. Mansfield, and L. Wright for discussions and comments.

\clearpage

\setcounter{section}{1}

\appendix

\section{~~~~~~~~~~~~~~~Diffusion coefficient for scalar or Cartesian enery density\label{app:diffcffCart}}

In this Appendix, we derive the expression (\ref{eq:diffusionUCart_copy}) for the diffusion coefficient $D^{(2)}_{U_\alpha}(u_\alpha,t)$ for $U_\alpha(t) \doteq {Y^{\prime}_\alpha}^2(t) + {Y^{\prime\prime}_\alpha}^2(t)$ when $Y_\alpha$ is circular Gauss normal at any $t$.
To this end, we expand $\langle [U_\alpha(t+\delta t)-U_\alpha(t)]^2 \rangle$ with respect to $Y^{\prime(\prime)}_\alpha(t)$.
Making use of the fact that the process $Y_\alpha$ has independent increments $Y_\alpha(t+\delta t)-Y_\alpha(t)$ that are conditional to $Y_\alpha(t)$ and, hence, to $U_\alpha(t)$, we obtain after straightfowarded calculation
\bea
\fl \hspace{1.5cm}
\left \langle \left [ U_\alpha(t+\delta t)-U_\alpha(t) \right ]^2 \right \rangle
&=
\left \langle \left [ Y^{\prime}_\alpha       (t+\delta t) - Y^{\prime}_\alpha       (t) \right ]^4 \right \rangle
+
\left \langle \left [ Y^{\prime\prime}_\alpha (t+\delta t) - Y^{\prime\prime}_\alpha (t) \right ]^4 \right \rangle \nonumber\\
\fl
&~ +
2 \left \langle \left [ Y^{\prime}_\alpha       (t+\delta t)- Y^{\prime}_\alpha       (t) \right ]^2 \right \rangle 
         \left \langle \left [ Y^{\prime\prime}_\alpha (t+\delta t)- Y^{\prime\prime}_\alpha (t) \right ]^2 \right \rangle \nonumber\\
\fl
&~ +
4 \left [ {Y^{\prime}_\alpha}^2       (t) \left \langle \left [ Y^{\prime}_\alpha       (t+\delta t)- Y^{\prime}_\alpha       (t) \right ]^2 \right \rangle
        + {Y^{\prime\prime}_\alpha}^2 (t) \left \langle \left [ Y^{\prime\prime}_\alpha (t+\delta t)- Y^{\prime\prime}_\alpha (t) \right ]^2 \right \rangle
  \right ]\nonumber\\ 
\fl
&~ +
8 ~Y^{\prime}_\alpha (t) Y^{\prime\prime}_\alpha (t) ~
          \left \langle \left [ Y^{\prime}_\alpha       (t+\delta t)- Y^{\prime}_\alpha       (t)  \right ] \left [ Y^{\prime\prime}_\alpha (t+\delta t)- Y^{\prime\prime}_\alpha (t) \right ] \right \rangle .
\label{eq:seconddiffu}
\eea
The first pair of terms in (\ref{eq:seconddiffu}) involves higher-order moments that are in general difficult to express in terms of $Y^{\prime(\prime)}_\alpha(t)$, unless the ${Y}_\alpha(t)$ are circular Gauss normal. In this case, using the Isserlis moment theorem, we have that
\bea
\left \langle \left [ Y^{\prime(\prime)}_\alpha (t+\delta t) - Y^{\prime(\prime)}_\alpha (t) \right ]^4 \right \rangle
=
3 \left \langle \left [ Y^{\prime(\prime)}_\alpha (t+\delta t) - Y^{\prime(\prime)}_\alpha (t) \right ]^2 \right \rangle^2 
.
\eea
For the third and fourth terms in (\ref{eq:seconddiffu}),
\bea
\fl 
\left \langle \left [ Y^{\prime(\prime)}_\alpha (t+\delta t) - Y^{\prime(\prime)}_\alpha (t) \right ]^2 \right \rangle
&=
\left \langle \left [ Y^{\prime(\prime)}_\alpha (t+\delta t) \right ]^2 \right \rangle +
\left \langle \left [ Y^{\prime(\prime)}_\alpha (t  ) \right ]^2 \right \rangle - 2
\left \langle Y^{\prime(\prime)}_\alpha (t+\delta t) Y^{\prime(\prime)}_\alpha (t  ) \right \rangle\nonumber\\
&=
\frac{\sigma^2_{X_\alpha}}{\tau} \left [ 1 - \exp \left ( - \frac{\delta t}{\tau} \right ) \right ]
+
\left ( \frac{\sigma^2_{X_\alpha}}{2\tau} - y^{\prime(\prime)^2}_{\alpha_0} \right )
\exp \left [ - \frac{2 (t-t_0)}{\tau} \right ]\nonumber\\
&~~~
\times
\left [ -1 + 2 \exp \left ( - \frac{\delta t}{\tau} \right ) - \exp \left ( - \frac{2 \delta t}{\tau} \right ) \right ] .
\label{eq:seconddiffutemp} 
\eea
In the limit $\delta t\rightarrow 0$, to first order in $\delta t$, the expression (\ref{eq:seconddiffutemp}) reduces to $(\sigma^2_{X_\alpha}/\tau^2)\delta t$ which is independent of $y^{\prime(\prime)}_{\alpha_0}$ to this order.
Finally, the last term in (\ref{eq:seconddiffu}) vanishes, because ${Y}^\prime_\alpha(t)$ and ${Y}^{\prime\prime}_\alpha(t)$ are uncorrelated. 
Thus, provided ${Y}^\prime_\alpha(t)$ and ${Y}^{\prime\prime}_\alpha(t)$ are identically distributed, 
(\ref{eq:seconddiffu}) becomes
\bea
\fl 
\left \langle \left [ U_\alpha(t+\delta t)-U_\alpha(t) \right ]^2 \right \rangle
&=
8 \left \langle \left [ Y^{\prime}_\alpha (t+\delta t) - Y^{\prime}_\alpha (t) \right ]^2 \right \rangle^2
+ 4 \left \langle \left [ Y^{\prime}_\alpha (t+\delta t) - Y^{\prime}_\alpha (t) \right ]^2 \right \rangle U_\alpha(t) \\
&=
\frac{4\sigma^2_{X_{\alpha}}}{\tau^2}
U_{\alpha}(t) \delta t +
\Or [(\delta t)^2]
\label{eq:seconddiffuresult}
\eea
whence the diffusion coefficient for the Cartesian energy density is
\bea
D^{(2)}_{U_\alpha} (u_\alpha,t) 
&= \lim_{\delta t\rightarrow 0} \left. \frac{\left \langle \left [ U_\alpha(t+\delta t) - U_\alpha(t) \right ]^2 \right \rangle}{2\delta t} \right|_{U_\alpha=u_\alpha}
= \frac{2 \sigma^2_{X_\alpha}}{\tau^2} u_\alpha .
\label{eq:diffusionUCart}
\eea
This coefficient is independent of $u_{\alpha_0}$ because the dependence of (\ref{eq:seconddiffutemp}) on $\delta t$ is only of second order.

\section{~~~~~~~~~~~~~~~Solution of FPE for Cartesian energy density \label{app:solvingFPECart}}
In this Appendix, we derive the general solution (\ref{eq:FPEgeneralsolutionCartEnergy})--(\ref{eq:PDEfinalLTCart}) of the FPE-based boundary value problem (\ref{eq:FPenergystirCart})--(\ref{eq:BC_FPenergystirCart}), i.e.,
\bea
\frac{\partial}{\partial t}                           f_{U_\alpha}\left ( u_\alpha,t|u_{\alpha_0},t_0 \right ) 
&=&
- \frac{2}{\tau} \left ( \frac{\sigma^2_{X_\alpha}}{\tau} - u_{\alpha_0} \right ) \exp \left [ - \frac{2 \left ( t - t_0 \right )}{\tau} \right ]
\frac{\partial}{\partial u_\alpha}                    f_{U_\alpha}\left ( u_\alpha,t|u_{\alpha_0},t_0 \right ) \nonumber\\
&~&+
\frac{2\sigma^2_{X_\alpha}}{\tau^2} 
\frac{\partial^2}{\partial u^2_\alpha} \left [ u_\alpha f_{U_\alpha}\left ( u_\alpha,t|u_{\alpha_0},t_0 \right ) \right ] 
\label{eq:FPenergystirCart_copy}
\eea 
for the Cartesian energy density $U_\alpha$ with initial value
$
U_\alpha(t_0) \doteq u_{\alpha_0}
$, i.e.,  initial condition
\bea
f_{U_\alpha} \left ( u_\alpha, t_0 \right ) = \delta \left ( u_\alpha - u_{\alpha_0} \right )
\label{eq:IC_FPenergystirCart_copy}
\eea
and boundary conditions
\bea
f_{U_\alpha}(0,t|u_{\alpha_0},t_0)= f_{U_\alpha}(+\infty,t|u_{\alpha_0},t_0)= 0.
\label{eq:BC_FPenergystirCart_copy}
\eea
Since $U_\alpha$ takes positive values only, we perform a Laplace
transformation of the stochastic PDE with respect to $U_\alpha$, viz., 
\bea
{\cal F}_S(s,t) \doteq \int^{+\infty}_0 f_{U_\alpha}(u_\alpha,t) \exp \left ( - u_\alpha s \right ) {\rm d} u_\alpha,~~~~~ \Re\left [ s \right ] > \gamma_0.
\eea
Here, $\gamma_0$ denotes the abscissa of convergence for $f_{U_\alpha}(u_\alpha,t)$, with $\gamma_0 < +\infty$ because $f_{U_\alpha}(u_\alpha,t)$ is of exponential order owing to the upper boundary condition in (\ref{eq:BC_FPenergystirCart_copy}).
The transformation of (\ref{eq:FPenergystirCart_copy}) is
\bea
&~&\frac{\partial {\cal F}_S(s,t)}{\partial s} 
+ 
\frac{\tau^2}{2 \sigma^2_{X_\alpha} s^2} \frac{\partial {\cal F}_S(s,t)}{\partial t} 
+
\frac{1}{s} 
\left ( 1 - \frac{u_{\alpha_0} \tau}{\sigma^2_{X_\alpha}} \right ) \exp \left [ - \frac{2 \left ( t - t_0 \right )}{\tau} \right ] 
{\cal F}_S(s,t) \nonumber\\
&~&~~~~~~~~~~~~~~~~~~~~~~~~~~~ 
= \left \{ - 1 + \left ( 1 - \frac{u_{\alpha_0}\tau}{\sigma^2_{X_\alpha}} \right ) 
\exp \left [ - \frac{2 \left ( t - t_0 \right )}{\tau} \right ] 
\right \}
\frac{f_{U_\alpha}(0+,t) }{s^2} 
\label{eq:PDEtrfCart_copy}
\eea
now subject to the corresponding initial condition
\bea
{\cal F}_S\left ( s,t_{0}+ \right )= \exp \left ( - u_{\alpha_0} s \right ).
\label{eq:initcondPDEtrfCart}
\eea
Specifically, by the physical nature of energy density, we are interested in solutions that satisfy the Dirichlet boundary conditions 
\bea
f_{U_\alpha}(0+,t) = 0,~~~~f_{U_\alpha}(+\infty,t) = 0
\label{eq:boundcond}
\eea 
in accordance with similar properties of known PDFs for the Cartesian energy density of imperfect quasi-stationary random fields\footnote{The former condition is not satisfied for the negative exponential PDF for the Cartesian energy density of a perfect reverberation field, for which $f_{U_\alpha}(0+,t) = 1$. For an imperfect field, more elaborate models \cite{arnaKdf}, \cite{arnaCE} show that the boundary condition is fulfilled, which underlines the generality of the solution found in this section in all practical cases.} \cite{arnaKdf}, \cite{arnaCE}.
The second condition in (\ref{eq:boundcond}) is the natural boundary condition for all PDFs; the former condition makes the right hand side of (\ref{eq:PDEtrfCart_copy}) vanish (homogeneous PDE). 
The PDF will also satisfy the Neumann condition 
$\partial f_{U_\alpha}(u_\alpha=0+,t) / \partial u_\alpha = 0$. 
However, (\ref{eq:PDEtrfCart_copy}) is generally independent of its specific value $\partial f_{U_\alpha}(u_\alpha=0+,t) / \partial u_\alpha$, as a result of the pre-factor $u_\alpha$ for the term in $\partial^2 f_{U_\alpha} / \partial {u_\alpha}^2$ in (\ref{eq:FPenergystirCart_copy}).

The first-order PDE (\ref{eq:PDEtrfCart_copy}) may be converted to an ODE of same order, through a suitable transformation of coordinates.
Defining
\bea
s^\prime \doteq s
,~~~~~~
t^\prime \doteq t - t_0 + \frac{\tau^2}{2 \sigma^2_{X_\alpha} s}
\label{eq:trfvarPDE}
\eea
so that 
\bea
\frac{\partial {\cal F}_S }{ \partial s }
= \frac{\partial {\cal F}_S }{ \partial s^\prime  }- \frac{\tau^2}{2 \sigma^2_{X_\alpha} s^2} \frac{\partial {\cal F}_S }{ \partial t^\prime }
,~~~~~~
\frac{\partial {\cal F}_S }{ \partial t } = \frac{\partial {\cal F}_S }{ \partial t^\prime},
\eea 
the terms containing $\partial {\cal F}_S(s,t^\prime) / \partial t^\prime$ in (\ref{eq:PDEtrfCart_copy}) then cancel, whence this PDE reduces to 
\bea
&~&\frac{\partial {\cal F}_{S^\prime}\left ( s^\prime,t^\prime \right ) }{\partial s^\prime} 
+ \frac{1}{s^\prime} 
\left ( 1 - \frac{u_{\alpha_0} \tau}{\sigma^2_{X_\alpha}} \right ) \exp \left ( - \frac{ 2 t^\prime }{\tau} + \frac{\tau}{\sigma^2_{X_\alpha} s^\prime} \right ) 
{\cal F}_{S^\prime} \left ( s^\prime,t^\prime \right )\nonumber\\
&~&~~~~~~~~~~~~~~
=
\left [ -1 + \left ( 1- \frac{u_{\alpha_0}\tau}{\sigma^2_{X_\alpha}} \right ) \exp \left ( - \frac{ 2 t^\prime }{\tau} + \frac{\tau}{\sigma^2_{X_\alpha}s^\prime} \right ) 
\right ] \frac{f_{U_\alpha}(0+,t)}{{s^\prime}^2}
\label{eq:FPEtoODE}
\eea
with the re-scaled initial condition
\bea
{\cal F}_{S^\prime} \left ( s^\prime,t^\prime = \frac{\tau^2}{2 \sigma^2_{X_\alpha} s} \right ) 
= \exp ( - u_{\alpha_0} s^\prime ) 
= 
\exp \left ( - \frac{\tau^2 u_{\alpha_0} }{2 \sigma^2_{X_\alpha} t^\prime} \right )
\eea
in accordance with (\ref{eq:initcondPDEtrfCart}) and (\ref{eq:trfvarPDE}). This yields the general solution
\bea
\fl
&~&{\cal F}_{S^\prime}(s^\prime,t^\prime)
\exp \left [
\int^{s^\prime}_{\frac{\tau^2}{2\sigma^2_{X_\alpha}t^\prime} }
\frac{1}{s^{\prime\prime}} \left ( 1 - \frac{u_{\alpha_0}\tau}{\sigma^2_{X_\alpha}} \right ) 
\exp \left ( - \frac{2 t^\prime }{\tau} + \frac{\tau}{\sigma^2_{X_\alpha} s^{\prime\prime}} \right ) {\rm d} s^{\prime\prime} 
\right ]
\nonumber\\
\fl
&~&~
=
\exp \left ( - \frac{\tau^2 u_{\alpha_0} }{2 \sigma^2_{X_\alpha} t^\prime} 
\right )
+
\int^{s^\prime}_{\frac{\tau^2}{2\sigma^2_{X_\alpha}t^\prime} }
\left [ -1 + \left ( 1- \frac{u_{\alpha_0}\tau}{\sigma^2_{X_\alpha}} \right ) \exp \left ( - \frac{ 2 t^\prime }{\tau} + \frac{\tau}{\sigma^2_{X_\alpha} s^{\prime\prime}} \right ) 
\right ] \frac{f_{U_\alpha}(0+,t)}{{s^{\prime\prime}}^2}
\nonumber\\
\fl
&~&~~~~~~~~~~~~~~~~~~~~~~~~~~~~~
\times
\exp \left [
\int^{s^{\prime\prime}}_{\frac{\tau^2}{2\sigma^2_{X_\alpha}t^\prime} }
\frac{1}{s^{\prime\prime\prime}} \left ( 1 - \frac{u_{\alpha_0}\tau}{\sigma^2_{X_\alpha}} \right ) 
\exp \left ( - \frac{2 t^\prime }{\tau} + \frac{\tau}{\sigma^2_{X_\alpha} s^{\prime\prime\prime}} \right ) {\rm d} s^{\prime\prime\prime} 
\right ]
{\rm d}s^{\prime\prime}.
\label{eq:solPDEtoODE}
\eea
A further transformation of (\ref{eq:solPDEtoODE}) via $t^{\prime\prime(\prime)} \doteq \tau^2 / (2 \sigma^2_{X_\alpha} s^{\prime\prime(\prime)})$, followed by a transformation back to the original $s$ and $t$, yields the final result given by (\ref{eq:PDEfinalLTCart}).

The integral expression of the solution can be re-expressed with the aid of residue calculus.
The point $s=0$ is an essential isolated singularity in the complex $s$-plane, 
because $s^{-1}$ appears in the argument of exponential functions in ${\cal F}_S(s,t)$. Therefore, calculation of (\ref{eq:FPEgeneralsolutionCartEnergy}) using the residue theorem requires calculation of the Laurent series for ${\cal F}_S(s,t) \exp \left ( u_\alpha s \right )$. 
The asymptotic points $\gamma \pm \rmj \infty$ are also essential isolated singularities, because this series contains terms with positive as well as negative powers, on account of the factor $\exp (u_\alpha s)$. The Bromwich contour has its vertical straight segment located at $\Re[s]=\gamma > 0$ and closure is across the negative (left) half of the $s$-plane. When tracking this contour, the transition across the positive real axis is governed by $\lim_{|s-\gamma|\rightarrow 0} \exp (1/s)= \exp(1/\gamma)$, which is defined provided $\gamma \not = 0$; otherwise $ \lim_{|s|\rightarrow 0} \exp (1/s) = \lim_{n\rightarrow + \infty} \exp (-\rmj n)$ which is indeterminate. 
The integral across the closing arc of the contour can be verified to satisfy the Jordan lemma for sufficiently small $\gamma$, whence its contribution vanishes. The inverse transformation then yields
(\ref{eq:residue})--(\ref{eq:residue_expan}).

\section{~~~~~~~~~~~~~~~Limit PDFs\label{app:limitPDFCartEnergy}}
In this Appendix, we derive the early-time limit PDF (\ref{eq:condprobenergyCart_smallt_largetau}) of the Cartesian energy density governed by the FPE (\ref{eq:FPenergystirCart}), i.e., (\ref{eq:FPenergystirCart_copy}).
For $(t-t_0)/\tau \rightarrow 0$, the exponential time dependence in (\ref{eq:FPenergystirCart_copy}), (\ref{eq:PDEtrfCart_copy}) and (\ref{eq:FPEtoODE}) disappears. 
Upon scaling the time variable to 
\bea
t^\prime \doteq \frac{2 \sigma^2_{X_\alpha} }{\tau^2} \left ( t-t_0 \right ) ,
\label{eq:deftprime}
\eea
the FPE (\ref{eq:FPenergystirCart_copy}) transforms as
\bea
\frac{\partial f_{U_\alpha} \left ( u_\alpha,t^\prime \right ) }{\partial t^\prime} = 
\left ( 1 + \frac{u_{\alpha_0}\tau}{\sigma^2_{X_\alpha}} \right ) \frac{\partial f_{U_\alpha}\left (u_\alpha,t^\prime \right ) }{\partial u_\alpha} 
+
u_\alpha 
\frac{\partial^2 f_{U_\alpha}\left (u_\alpha,t^\prime \right ) }{\partial u_\alpha^2} 
\label{eq:FPE_Ualphasmallt}
\eea
and (\ref{eq:PDEfinalLTCart}) becomes
\bea
{\cal F}_S \left ( s, t^\prime \right )
&=& 
\exp \left ( - \frac{u_{\alpha_0}s}{1+s t^\prime} \right ) 
\exp \left [ - \left ( 1-\frac{u_{\alpha_0}\tau }{\sigma^2_{X_\alpha}} \right ) \int^{s^{-1}+t^\prime}_{s^{-1}} \frac{ {\rm d}t^{\prime\prime} }
                                                    {t^{\prime\prime}} \right ]\nonumber\\
&=&
\exp \left ( - \frac{u_{\alpha_0}s }{1+s t^\prime} \right )
\left ( 1 + s t^\prime \right )^{-1 + \frac{u_{\alpha_0} \tau}{\sigma^2_{X_\alpha}} } .
\label{eq:Laplacetrfspecial}
\eea
We now make use of the identity \cite[eq. (6.643.2)]{grad1}
\bea
\fl \hspace{1cm}
\frac{\Gamma\left ( \mu + \nu + \frac{1}{2} \right )}
     {\Gamma \left ( 2 \nu + 1 \right )} 
\exp \left ( \frac{\beta^2}{2\alpha} \right ) \frac{ \alpha^{-\mu}}{\beta} M_{-\mu, \nu} \left ( \frac{\beta^2}{\alpha} \right )
=
\int^{+\infty}_0 x^{\mu-\frac{1}{2}} \exp \left ( - \alpha x \right ) I_{2\nu} \left ( 2 \beta  \sqrt{ x} \right ) {\rm d} x
\label{eq:GRAD}
\eea
where $M_{-\mu,\nu}(\cdot)$ and $I_{2\nu}(\cdot)$ are Whittaker and modified Bessel functions of the first kind and order $2\nu$, respectively,
in which we define the parameters as
$\alpha \doteq {t^\prime}^{-1}+s$, 
$\beta \doteq \sqrt{u_{\alpha_0}}/{t^\prime}$, and
\bea
\nu \doteq \mu - \frac{1}{2} \doteq -\frac{u_{\alpha_0}\tau}{2\sigma^2_{X_{\alpha}}} \leq 0 .
\label{eq:defnu}
\eea 
With this choice, we have that
$
M_{-\mu,\nu}(z) = z^{\mu} \exp (z/2)
$ 
with which the left member of (\ref{eq:GRAD}) can be rewritten as $ (u_{\alpha_0})^\nu t^\prime \exp \{ u_{\alpha_0} / [t^\prime (1 + s t^\prime)] \} / (1+s t^\prime)^{2\nu+1} $. 
Hence (\ref{eq:GRAD}) becomes
\bea
\fl \hspace{1cm}
\frac{\exp \left [ - \frac{u_{\alpha_0} s}
                          {1+s t^\prime} 
           \right ]}
     {\left ( 1+s t^\prime \right )^{2\nu+1}}
=
\frac{\exp \left ( - \frac{u_{\alpha_0}}
                          {t^\prime} 
           \right )}
     { (u_{\alpha_0})^\nu ~ t^\prime  } 
\int^{+\infty}_0 
\left ( u_\alpha \right )^{\nu}
\exp \left [ - \left ( {t^\prime}^{-1} + s \right ) u_{\alpha} \right ] 
I_{2\nu} \left ( \frac{2\sqrt{u_\alpha u_{\alpha_0}}}{t^\prime} \right ) {\rm d}u_\alpha
\label{eq:Laplacetrfinalbis}
\eea
which corresponds to (\ref{eq:Laplacetrfspecial}) and yields the inverse transform of ${\cal F}_S(s,t^\prime)$.
After a final transformation back to the original time variable, the PDF of $U_\alpha$ for $(t-t_0)/\tau \rightarrow 0$ is obtained as (\ref{eq:condprobenergyCart_smallt_largetau}).

\section{~~~~~~~~~~~~~~~PDF for energy density of vector field\label{app:solvingFPEVecEnergy}}
In this Appendix, we derive the kernel (\ref{eq:PDEfinalLTTot}) for the general solution of the FPE-based boundary value problem (\ref{eq:FPenergystirTot})--(\ref{eq:BC_FPenergystirTot}) for the energy density $U_\rmt(t)$ of a vector field $Y_\rmt(t)$.
Compared to the scalar case (\ref{eq:PDEtrfCart_copy}), the transformed FPE is now 
\bea
&~&\frac{\partial {\cal F}_S(s,t)}{\partial s} 
+ 
\frac{3\tau^2}{2 \sigma^2_{X_\rmt} s^2} \frac{\partial {\cal F}_S(s,t)}{\partial t} 
+
\frac{3}{s} 
\left ( 1 - \frac{u_{\rmt_0} \tau}{\sigma^2_{X_\rmt}} \right ) \exp \left [ - \frac{2 \left ( t - t_0 \right )}{\tau} \right ] 
{\cal F}_S(s,t) \nonumber\\
&~&~~~~~~~~~~~~~~~~~~~~~~~~~~~ 
= \left \{ - 1 + \left ( 1 - \frac{u_{\rmt_0}\tau}{\sigma^2_{X_\rmt}} \right ) 
\exp \left [ - \frac{2 \left ( t - t_0 \right )}{\tau} \right ] 
\right \}
\frac{f_{U_\rmt}(0+,t) }{s^2} 
\label{eq:PDEtrfTot_copy}
\eea
with transformed initial condition and Dirichlet boundary conditions
\bea
{\cal F}_S\left ( s,t_{0}+ \right )= \exp \left ( - u_{\rmt_0} s \right ) ,
\label{eq:initcondPDEtrfTot}
\eea
\bea
f_{U_\rmt}(0+,t) = 0,~~~f_{U_\rmt}(+\infty,t) = 0 .
\label{eq:boundcondPDEtrfTot}
\eea 
The pertinent transformation of variables is
\bea
\left \{
\begin{array}{l}
s^\prime \doteq s\\
t^\prime \doteq t - t_0 + \frac{3\tau^2}{2 \sigma^2_{X_\rmt} s}
\end{array}
\right .
\label{eq:trfvarPDETot}
\eea
with which (\ref{eq:PDEtrfTot_copy})--(\ref{eq:boundcondPDEtrfTot}) transform to
\bea
&~&\frac{\partial {\cal F}_{S^\prime}\left ( s^\prime,t^\prime \right ) }{\partial s^\prime} 
+ \frac{3}{s^\prime} 
\left ( 1 - \frac{u_{\rmt_0} \tau}{\sigma^2_{X_\rmt}} \right ) \exp \left ( - \frac{ 2 t^\prime }{\tau} + \frac{3\tau}{\sigma^2_{X_\rmt} s^\prime} \right ) 
{\cal F}_{S^\prime} \left ( s^\prime,t^\prime \right )\nonumber\\
&~&~~~~~~~~~~~~~~
=
\left [ -1 + 3 \left ( 1- \frac{u_{\rmt_0}\tau}{\sigma^2_{X_\rmt}} \right ) \exp \left ( - \frac{ 2 t^\prime }{\tau} + \frac{3\tau}{\sigma^2_{X_\rmt} s^\prime} \right ) 
\right ] \frac{f_{U_\rmt}(0+,t)}{{s^\prime}^2}
\label{eq:FPEtoODETot}
\eea
with 
\bea
{\cal F}_{S^\prime} \left ( s^\prime,t^\prime = \frac{3\tau^2}{2 \sigma^2_{X_\rmt} s} \right ) 
= \exp ( - u_{\rmt_0} s^\prime ) 
= 
\exp \left ( - \frac{3\tau^2 u_{\rmt_0} }{2 \sigma^2_{X_\rmt} t^\prime} \right ) .
\eea
Similar to the procedure in \ref{app:solvingFPECart}, the transformation $t^{\prime\prime(\prime)} \doteq 3 \tau^2 / ( 2 \sigma^2_{X_\rmt} s^{\prime\prime(\prime)})$, followed by a transformation of ${\cal F}_{S^\prime} (s^\prime, t^\prime)$ back to a form in function of the original $s$ and $t$, yields (\ref{eq:PDEfinalLTTot}).

\section{~~~~~~~~~~~~~~~Mellin integral for asymptotic PDF of energy density \label{sec:Mellin}}
In this Appendix, we seek to calculate the inverse Laplace transform of 
\bea
F(s)
&=
\exp \left ( - \frac{u_0 s}{1+a s} \right ) 
\left ( 1+ a s \right )^{-\left ( b - \frac{c}{s} \right )} 
\label{eq:MellinF0}
\eea
via the Mellin integral 
\bea
f_{U} \left ( u \right ) = \left ( \rmj 2 \pi \right )^{-1} \int^{\gamma+\rmj\infty}_{\gamma-\rmj\infty}
F(s) \exp \left ( u s \right ) {\rm d}s,~~~\gamma = \Re[s] > \gamma_0
\label{eq:Mellin}
\eea
where $\gamma_0$ is the abscissa of convergence, in connection with the early-time asymptotic PDF (\ref{eq:finalsolMellinFCart}) for the scalar or Cartesian field, and (\ref{eq:finalsolMellinFTot}) for the vector field.
For the case of scalar or Cartesian energy density,
\bea
a &\doteq \frac{2 (t-t_0) \sigma^2_{X_\alpha}}{\tau^2} \label{eq:defaCart} \\
b &\doteq \left ( 1 - \frac{u_{\alpha_0}\tau}{\sigma^2_{X_\alpha}} \right ) \left [ 1 - \frac{2 (t-t_0)}{\tau} \right ]\label{eq:defbCart}  \\
c &\doteq \left ( 1 - \frac{u_{\alpha_0}\tau}{\sigma^2_{X_\alpha}} \right ) \frac{\tau}{\sigma^2_{X_\alpha}} \label{eq:defcCart} \\
u_0 &\doteq {u_{\alpha_0}}
\eea
with $0 \leq (t-t_0)/\tau \ll1$, $u_{\alpha_0} \geq 0$ and $0 < u_{\alpha_0}\tau / {\sigma^2_{X_\alpha}} \ll 1$, whence $a\geq 0$, $0 < b \sim 1$, $0 < c \ll 1/u_{\alpha_0}$, and $0 < c/b \ll 1/u_{\alpha_0}$. For the energy density of a vector field,
\bea
a &\doteq \frac{2 (t-t_0) \sigma^2_{X_\rmt}}{3 \tau^2} \label{eq:defaTot} \\
b &\doteq 3 \left ( 1 - \frac{3u_{\rmt_0}\tau}{\sigma^2_{X_\rmt}} \right ) \left [ 1 - \frac{2 (t-t_0)}{\tau} \right ] \label{eq:defbTot} \\
c &\doteq 3 \left ( 1 - \frac{3u_{\rmt_0}\tau}{\sigma^2_{X_\rmt}} \right ) \frac{\tau}{\sigma^2_{X_\rmt}} \label{eq:defcTot} \\
u_0 &\doteq {u_{\rmt_0}}.
\eea
with corresponding conditions holding.

With regard to the general case (\ref{eq:MellinF0}),
the issue centers around the integration along the incisions $C_{1,\pm}$ which surround and exclude the singularity $-1/a$ (see below). On account of the Casorati--Weierstrass theorem, $\exp [ - u_0 s / (1+as)]$ takes on any given nonzero complex value in the neighbourhood of this singularity upon integrating. As part of a {product} of exponentials in $F(s)\exp (u_\alpha s)$, however, this factor does not cancel when considering mirrorred locations on pairs of corresponding segments or arcs as part of the overall contour of integration.

The special case for which $u_0=0$, as well as being of considerable practical interest, is easier to solve, i.e.,
\bea
F(s)
=
\left ( 1+ a s \right )^{-\left ( b - \frac{c}{s} \right )} 
\equiv
\exp \left [ - \left ( b - \frac{c}{s} \right ) {\rm ln} \left ( 1+ a s \right ) \right ],~~~a,b,c > 0 \label{eq:MellinF}
\eea
which we shall consider first.
Since $\Re [ b - c/s ]$ may take values smaller than one, the order condition  
\bea
F(s) = \Or (s^{-k}),~~~|s| \rightarrow +\infty,~~~ k > 1
\eea
that would guarantee the existence of (\ref{eq:Mellin}), is not satisfied. Since this condition is merely {\em sufficient} for the existence of (\ref{eq:Mellin}), an a posteriori check ${\cal L} [ f_{U_\alpha}(u_\alpha) ] (s) = F(s)$ for $f_{U_\alpha}(u_\alpha) $ will therefore be required.

\begin{figure}[t] \begin{center} 
\begin{tabular}{c} \ 
\epsfxsize=12cm 
\epsfbox{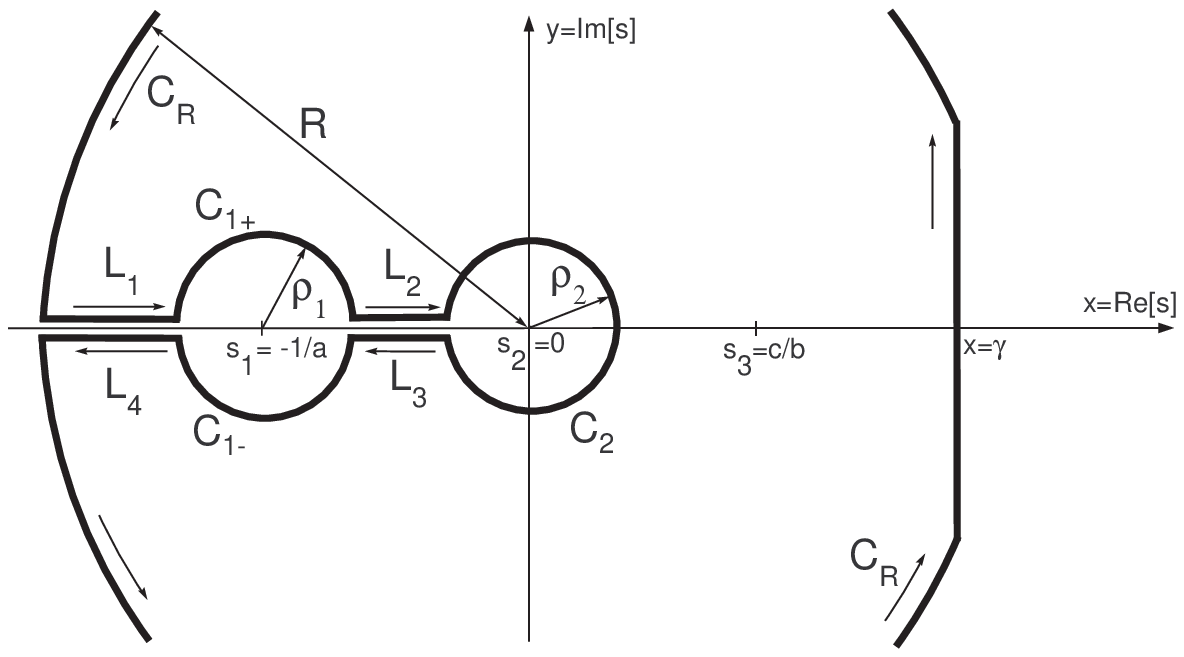}\ \\ 
\\
\end{tabular} \end{center}
\caption{\label{fig:contourMellin} \small
Contour of integration $C$ in complex $s$-plane.}
\end{figure}

To invert (\ref{eq:MellinF}),
we apply contour integration in the complex $s$-plane (figure \ref{fig:contourMellin}). To this end, we extend $F(s)$ analytically\footnote{For notational simplicity, we shall denote this extension by the same symbol $F(s)$.} across the left half of the plane.
Since $(b-c/s)$ is in general (complex) fractional, we apply a branch cut to maintain a single-valued function. We choose this cut along the negative real axis, with the phase reference $\theta = \arg(s) = 0$ along the positive real axis:
\bea
s \in {\cal C}_c \doteq {\cal C} \setminus \{ s \thinspace | \thinspace s=x<0 \},~~~{\rm arg} (s=x>0) = 0, ~~~ - \pi \leq {\rm arg} (s) < \pi.
\eea
Define the following special points from (\ref{eq:MellinF}): 
\bea
s_1 \doteq - a^{-1},~~~ s_2 \doteq 0,~~~ s_3 \doteq \frac{c}{b}
\eea 
where $s_1$ is the critical point of (\ref{eq:MellinF}), $s_2$ is the branch point, and $s_3$ is a regular auxiliary point introduced to assist in the further analysis.
Rewriting (\ref{eq:MellinF}) as 
\bea
F(s) &=& \left [ \left | 1+as \right | \exp \left ( \rmj \theta_1 \right ) \right ]^{\left | b- \frac{c}{s} \right | \exp \left ( \rmj \theta_0 \right )}\\
       &=& \left | 1+as \right |                                ^{\left | b- \frac{c}{s} \right | \exp \left ( \rmj \theta_0 \right )} 
           \exp \left [ \rmj \theta_1 \left | b- \frac{c}{s} \right | \exp \left ( \rmj \theta_0 \right ) \right ]
\eea
where, on account of $a, b>0$,
\bea
\theta_1 &\doteq& {\rm arg} (1+ as) ~\equiv~ \arg \left ( s-s_1 \right )\\
\theta_0 &\doteq& \pi + {\rm arg} \left ( b - \frac{c}{s} \right )  \equiv \pi - {\rm arg} (s-s_2) + {\rm arg} (s-s_3) ,
\eea
Table \ref{tbl:arg} shows the phase progressions when moving along the contour $C$ around the branch cut, in the positive direction indicated in the figure.
\begin{table}
\begin{center}
\begin{tabular}{||l||c|c|c|c||}
\hline\hline
~ & $L_1$ & $L_2$ & $L_3$ & $L_4$ \\
\hline\hline
$\theta_1=\arg (s-s_1)$ & $\pi$ & $0$       & $0$    & $-\pi$\\ \hline\hline
$\theta_2=\arg (s-s_2)$     & $\pi$ & $\pi$     & $-\pi$ & $-\pi$\\ \hline
$\theta_3=\arg (s-s_3)$ & $\pi$ & $\pi$ & $-\pi$ & $-\pi$\\
\hline\hline
$\theta_0=\pi-\theta_2+\theta_3$ & $\pi$ & $\pi$ & $\pi$ & $\pi$\\
\hline\hline
\end{tabular}
\caption{\label{tbl:arg} \small Argument of $s\in C$ with respect to critical points $s_i$ when moving along the branch cut.}
\end{center}
\end{table}

The joint contribution by the segments $L_1$ and $L_4$ is
\bea
\fl
\left ( \int_{L_1} + \int_{L_4} \right ) F(s) \exp \left ( u_\alpha s \right ) {\rm d}s
&=&
\lim_{\rho_1\rightarrow 0,R\rightarrow \infty}
\left \{
\int^{-(1/a)-\rho_1}_{-R}
\left [ \left | 1 + a x \right | \exp \left ( \rmj \pi \right ) \right ]^{\left | b-\frac{c}{x} \right | 
\exp \left (\rmj \pi \right )} 
\exp \left ( u_\alpha x \right ) 
{\rm d}x\nonumber
\right. \\
\fl
&~&~~~~~~~~~~~~~
\left.
+
\int_{-(1/a)-\rho_1}^{-R}
\left [ \left | 1 + a x \right | \exp \left ( - \rmj \pi \right ) \right ]^{\left | b-\frac{c}{x} \right | 
\exp \left (\rmj \pi \right )} 
\exp \left ( u_\alpha x \right ) 
{\rm d}x\nonumber
\right \}
\fl
\\
&=& 
\lim_{\rho_1\rightarrow 0,R\rightarrow \infty}
\left \{
\rmj 2
\int^{-(1/a)-\rho_1}_{-R}
\left | 1 + a x \right |^{-\left | b-\frac{c}{x} \right | }
\sin \left ( - \pi \left | b- \frac{c}{x} \right | \right ) 
\exp \left ( u_\alpha x \right )
{\rm d} x
\right \}
\nonumber\\
\fl
&=&
\rmj (-1)^{1-b+\frac{c}{s}} 
2 
\int^{-1/a}_{-\infty}
\left (1 + a x \right )^{- \left ( b-\frac{c}{x} \right ) }
\sin \left [ \pi \left ( b- \frac{c}{x} \right ) \right ]
\exp \left ( u_\alpha x \right )
{\rm d} x .\label{eq:contourinttemp}
\eea
A similar calculation for the other pair of segments $L_2$, $L_3$ can be shown, with the aid of table \ref{tbl:arg}, to yield a pairwise vanishing contribution to the Mellin integral.

Upon application of the Jordan lemma, the contribution across the arc $C_R$ vanishes in the limit $R\rightarrow +\infty$, on account of 
\bea
\lim_{R \rightarrow \infty} \left ( {\rm sup}_{C_R} |F(s)| \right ) 
&=& 
\lim_{R \rightarrow \infty} \left \{ {\rm sup}_{C_R} \left | \left [ 1 + a R \exp \left ( \rmj \theta \right ) \right ]^{- b + \frac{ c}{R \exp \left ( \rmj \theta \right )}} \right | \right \}
\nonumber\\
&=&
\lim_{R \rightarrow \infty} \left \{ \left | \left [ 1 + a R \exp \left ( \rmj \theta \right ) \right ]^{-b} \right |\left | \exp \left [ \frac{c}{R\exp(\rmj\theta)} \right ]\right | \right \}\\
&=&
0,
\eea
because $|\exp (\rmj \theta)| \not = 0$ for all $\theta$.

For the contribution along the circle $C_{2}$ with radius $\rho_2$, on substituting $s=s_2 + \rho_2 \exp (\rmj \theta)$, we have
\bea
\fl \hspace{1cm}
\lim_{(s-s_2) \rightarrow 0} \left [ \left ( s - s_2 \right ) F(s) \exp \left (u_\alpha s \right ) \right ] = 0 
~~\Rightarrow~~
 \lim_{\rho_2 \rightarrow 0} \int_{C_2} F [s(\rho_2)] \exp \left [ u_\alpha s(\rho_2) \right ] {\rm d}\rho_2 = 0 ,
\eea 
on account of the integral limit theorem for vanishingly small $\rho_2$ across an arc of given opening angle \cite{garn1}. 
Note that $s_2$ is a regular point [$F(s_2)=\exp(ac)$].
Likewise for the semi-circles $C_{1,\pm}$, using de l'H\^{o}pital's rule on the right member of (\ref{eq:MellinF}), we obtain
\bea
\fl \hspace{1cm}
\lim_{s \rightarrow s_1} \left [ \left ( s-s_1 \right ) F(s) \exp \left ( u_\alpha s \right ) \right ] 
&=
a^{-b-ac} 
\exp \left [ \frac{1 + \rmj \theta}
                  {\cos \theta} 
             \left ( \frac{t-t_0}{\tau} \right ) 
             \exp \left ( \rmj \theta \right ) - \frac{u_\alpha}{a} 
     \right ] \nonumber\\
\fl
&~~~~\times
\lim_{\rho_1 \rightarrow 0} 
\exp \left [ \frac{{\rm ln} \left ( \rho_1 \right )}{\cos \theta} \left ( \frac{t-t_0}{\tau} \right ) \exp \left ( \rmj \theta \right ) + u_\alpha \rho_1 \exp \left ( \rmj \theta \right ) \right ] .
\eea 
Although we have been unable to calculate the latter limit analytically, numerical computation for selected values of $\theta \not = \pm\pi/2$ indicates that this limit is zero because of pairwise cancelling contributions.
Hence the contributions by the arcs $C_{1,\pm}$ vanish as well.

Collating the above results and applying the Cauchy integral theorem on the closed contour $C$, we thus obtain a real integral expression for $f_{U_\alpha} (u_\alpha)$: from (\ref{eq:Mellin}) and (\ref{eq:contourinttemp}),
\bea
f_{U_\alpha} (u_\alpha) 
=
\frac{ (-1)^{1-b-\frac{c}{s}} }{\pi} 
\int^{+\infty}_{1/a} 
\left ( 1 - a x \right )^{ - b - \frac{c}{x} }
\sin \left [ \pi \left ( b + \frac{c}{x} \right ) \right ] 
\exp \left ( - u_\alpha x \right )
{\rm d} x
\label{eq:finalsolMellinF}
\eea
with which (\ref{eq:finalsolMellinFCart}) and (\ref{eq:finalsolMellinFTot}) follow by substituting (\ref{eq:defaCart})--(\ref{eq:defcCart}) or (\ref{eq:defaTot})--(\ref{eq:defcTot}), respectively.

\section{~~~~~~~~~~~~~Mean, variance, and covariance of quasi-stationary cavity field and for a second-order system\label{app:secondordersyst}}

In this Appendix, we derive expressions (\ref{eq:musecondorderRx})--(\ref{eq:sigmasqmixsecondorderRx}) for the first- and second-order statistics for a general second-order system with a quasi-stationary input field, as characterized by (\ref{eq:2ndorderSDE}), incorporating the particular case of a nonstationary cavity field filtered by a first-order system as specified by (\ref{eq:LangevinChamber}).
If we define
\bea
\ul{Y}(t) \doteq 
\left [ \begin{array}{l}
Y(t)\\
\dot{Y}(t)
\end{array} \right ]
&,&~
\dul{a} \doteq 
\left [ \begin{array}{lr}
0 & -1\\
\omega^2_o & 2\zeta
\end{array} \right ],
\nonumber\\
\ul{h}(t) \doteq 
\left [ \begin{array}{l}
0\\
v(t)
\end{array} \right ]
&,&~
\ul{g} \doteq 
\left [ \begin{array}{l}
0\\
{\cal T}^{-2}
\end{array} \right ],
\eea
then (\ref{eq:2ndorderSDE}) can be rewritten as the 1st-order vector ODE
\bea
\ul{\dot{Y}}(t) = - \dul{a} \cdot \ul{Y}(t) + \ul{h}(t)  + \ul{g} \dot{B}(t). 
\label{eq:Langevin2D}
\eea
To integrate (\ref{eq:Langevin2D}), we multiply by $\exp \left ( \dul{a} t \right ) \doteq \sum^{+\infty}_{n=0} \left ( \dul{a}t \right )^n/n!$, viz.,
\bea
\exp \left ( \dul{a} t \right ) \cdot {\rm d}\ul{Y}(t) + \exp \left ( \dul{a} t \right ) \cdot \dul{a} \cdot \ul{Y}(t) {\rm d}t
=
\exp \left ( \dul{a} t \right ) \cdot \left [ \ul{h}(t) {\rm d}t + \ul{g} {\rm d}B(t) \right ].
\label{eq:Langevin2Dtempa}
\eea
Since the left hand side of (\ref{eq:Langevin2Dtempa}) represents ${\rm d}\left [ \exp \left ( \dul{a}t \right ) \cdot \ul{Y}(t) \right ]$, integration of this equation yields
\bea
\exp \left ( \dul{a} t \right ) \cdot \ul{Y}(t) = 
\ul{Y}_0 + \int^t_{t_0} \exp \left ( \dul{a} t^\prime \right ) \cdot \ul{h} \left ( t^\prime \right ) {\rm d}t^\prime
+
\int^t_{t_0} \exp \left ( \dul{a} t^\prime \right ) \cdot \ul{g} {\rm d}B \left ( t^\prime \right ).
\label{eq:Langevin2Dtempb}
\eea
The second integral in (\ref{eq:Langevin2Dtempb}) can be integrated by parts as
\bea
\int^t_{t_0} \exp \left ( \dul{a} t^\prime \right ) \cdot \ul{g} \thinspace {\rm d}B \left ( t^\prime \right )
&=&
\exp \left ( \dul{a} t \right ) \cdot \ul{g} \left [ B(t) - B(t_0) \right ]
-
\int^t_{t_0} B \left ( t^\prime \right ) ~ {\rm d}{\left [ \exp \left ( \dul{a} t^\prime \right )\cdot \ul{g} \right ]}\nonumber\\
&=&
\exp \left ( \dul{a} t \right ) \cdot \ul{g} \left [ B(t) - B(t_0) \right ]
-
\dul{a} \cdot 
\int^t_{t_0} \exp \left ( \dul{a} t^\prime \right ) \cdot \ul{g} B\left ( t^\prime \right ) {\rm d} t^\prime
\eea
whence substitution into (\ref{eq:Langevin2Dtempb}) and taking $B(t_0)=0$ yields 
\bea
\ul{Y}(t) = \exp \left ( - \dul{a} \thinspace t \right ) \cdot
\ul{Y}_0 + \ul{g} B(t) 
+ \int^t_{t_0} \exp \left [ - \dul{a} \thinspace \left ( t - t^\prime \right ) \right ] \cdot \left [ \ul{h} \left ( t^\prime \right ) + \ul{g} B \left ( t^\prime \right ) \right ] {\rm d} t^\prime.
\eea
The expansion of the matrix exponentials (into an infinite series of matrix powers) can be avoided by making use of the matrix Laplace transform ${\cal L}\left [ \exp \left ( - \dul{a} t \right ) \right ](s) 
= \left ( s \dul{I} + \dul{a} \right )^{-1}
$
i.e., 
\bea
\int^{+\infty}_0 \exp \left ( - \dul{a} t \right ) \exp \left ( -s t \right ) {\rm d}t
= 
\left ( s^2 + 2 \zeta s + \omega^2_o \right )^{-1}
\left [
\begin{array}{lr}
s + 2 \zeta & 1\\
-\omega^2_o & s
\end{array}
\right ] .
\eea
Taking the scalar inverse transform, element by element, yields 
\bea
\fl
&~ \exp \left ( - \dul{a} t \right ) =
\left ( \lambda_1 - \lambda_2 \right )^{-1} \nonumber\\
\fl
&~ ~~ \times
\left [
\begin{array}{lr}
- \lambda_1 \exp \left ( \lambda_1 t \right ) + \lambda_2 \exp \left ( \lambda_2 t \right ) 
+
2 \zeta  \left [ \exp \left ( \lambda_1 t \right ) - \exp \left ( \lambda_2 t \right ) \right ]
& \exp \left ( \lambda_1 t \right ) - \exp \left ( \lambda_2 t \right ) \\
-\omega^2_o \left [ \exp \left ( \lambda_1 t \right ) - \exp \left ( \lambda_2 t \right ) \right ]
 & - \lambda_1 \exp \left ( \lambda_1 t \right ) + \lambda_2 \exp \left ( \lambda_2 t \right ) 
\end{array}
\right ]
\eea
where $\lambda_{1,2} = - \zeta \mp \sqrt{\zeta^2-\omega^2_o}$ are the roots of $s^2 + 2 \zeta s + \omega^2_o=0$. 

Alternatively, one may orthogonalize $\ul{Y}$. To this end, we seek an orthogonal transformation $\ul{Y}^\prime = \dul{c} \cdot \ul{Y}$, i.e., such that
\bea
\dul{c} \cdot \dul{a} \cdot \dul{c}^{-1} = 
\left [ \begin{array}{lr} \lambda_1 & 0\\ 0 & \lambda_2 \end{array} \right ].
\eea
Here, a primed quantity signifies a transformed (new) variable.
This transformation is specified by
\bea
\left [
\begin{array}{c}
Y^\prime(t) \\
\dot{Y}^\prime(t) 
\end{array}
\right ]
=
\left ( \lambda_2 - \lambda_1  \right )^{-1}
\left [
\begin{array}{lr}
\lambda_2 & - 1\\
-\lambda_1 & 1
\end{array}
\right ]
\cdot
\left [
\begin{array}{c}
Y(t) \\
\dot{Y}(t) 
\end{array}
\right ] 
,
\eea
in which $\lambda_{1,2}$ as defined above are the solutions of det$(\dul{a}-\lambda\dul{I})=0$.
The associated vector FPE for $\ul{Y}^\prime(t)$ yields a bivariate Gauss normal PDF with a mean value
\bea
\ul{\mu}_{Y^\prime} =
\left [
\begin{array}{c}
y^\prime_0 \exp \left ( \lambda_1 t \right ) \\
\dot{y}^\prime_0 \exp \left ( \lambda_2 t \right ) \\
\end{array}
\right ],~~~{\rm where}~
\left [
\begin{array}{c}
y^\prime_0\\
\dot{y}^\prime_0
\end{array}
\right ]
\doteq
\dul{c} \cdot 
\left [
\begin{array}{c}
y_0\\
\dot{y}_0
\end{array}
\right ]
\label{eq:musecondorder}
\eea
and covariance matrix
\bea
\left [ \left ( {\dul{\sigma}^2_{Y^\prime} } \right )_{ij} \right ]
&=&
\left[ \frac{\exp \left [ \left ( \lambda_i + \lambda_j \right ) t \right ] -1}
     {\lambda_i + \lambda_j}
\right ]
\dul{c} 
\cdot
\dul{D}^{(2)}
\cdot
\dul{c}^{\rm T}\nonumber\\
&=&
\frac{\sigma^2_X }{4 {\cal T}^4 \left ( \omega^2_o - \zeta^2 \right )}
\left [
\begin{array}{lr}
\frac{1 - \exp \left ( 2 \lambda_1 t \right )}{2 \lambda_1} & 
- \frac{1 - \exp \left [ \left ( \lambda_1 + \lambda_2 \right ) t \right ]}{\lambda_1 + \lambda_2} \\
\\
- \frac{1 - \exp \left [ \left ( \lambda_1 + \lambda_2 \right ) t \right ]}{\lambda_1 + \lambda_2} & 
\frac{1 - \exp \left ( 2 \lambda_2 t \right )}{2 \lambda_2} 
\end{array}
\right ],
\label{eq:sigmasecondorder}
\eea
where
\bea
\dul{D}^{(2)}
=
\sigma^2_X \ul{g} \thinspace \ul{g}^{\rm T} 
=
\left [
\begin{array}{lr}
0 & 0\\
0 & \frac{\sigma^2_X}{{\cal T}^4}
\end{array}
\right ].
\eea
In terms of the individual processes $Y(t)$ and $\dot{Y}(t)$, the PDF $f_{\ul{Y}}(\ul{y},t|\ul{y}_0,t_0)$ is also bivariate Gauss normal, with $ \ul{\mu}_Y = \dul{c}^{-1} \cdot \ul{\mu}_{Y^\prime}$ and $\dul{\sigma}_Y = \dul{c}^{-1} \cdot \dul{\sigma}_{Y^\prime} \cdot \left (\dul{c}^{-1} \right )^{\rm T}$. This leads to expressions
(\ref{eq:musecondorderRx})--(\ref{eq:sigmasqmixsecondorderRx}) for the mean, variance and covariance of $Y(t)$ and $\dot{Y}(t)$.


\section*{References}
\begin{thebibliography}{99}
\bibitem{arnaZurich03} Arnaut L R 2003
Nonstationary effects in mode-stirred reverberation 
\it Proc. 15th Z\"{u}rich Int. Symp. \& Techn. Exhib. on Electromagnetic Compatibility \rm (18--20 Feb. 2003, Z\"{u}rich, CH) pp 245--50 
\bibitem{meye1} Meyer E 1930
Ein neues automatisches Verfahren der Nachhallmessung
\it Z. Techn. Physik \rm {\bf 7} 253--9
\bibitem{knud1} Knudsen V O 1933
The absorption of sound in air, in oxygen, and in nitrogen -- Effect of humidity and temperature
\it J. Acoust. Soc. Am. \rm {\bf 5} 112--21
\bibitem{meye2} Meyer E, Kurtze G, Kuttruff H and Tamm K 1960
Ein neuer Hallraum f\"{u}r Schallwellen und elektromagnetische Wellen 
\it Acustica \rm {\bf 1}(supplement) 253--64
\bibitem{bell1} Bello P A 1963
Characterization of randomly time-variant linear channels
\it IEEE Trans. Comm. Syst. \rm {\bf 11} 360--93
\bibitem{gesb1} Gesbert D, B\"{o}lcskei H, Gore D A, and Paulraj A J 2002
Outdoor MIMO wireless channels: models and performance prediction
\it IEEE Trans. Comm. \rm {\bf 50}(12) 1926--34
\bibitem{poon1} Poon A S Y, Brodersen R W and Tse D N C 2005
Degrees of freedom in multiple-antenna channels: a signal space approach
\it IEEE Trans. Inf. Theo. \rm {\bf 51}\rm(2) 523--36
\bibitem{arnaRS3} Arnaut L R 2007
Probability distributions of random electromagnetic fields in the presence of a semi-infinite isotropic medium
\it Radio Sci. \rm {\bf 42}(6) RS3001
\bibitem{nye1} Nye J F and Hajnal J V 1987 
The wave structure of monochromatic electromagnetic radiation 
\it Proc. Roy. Soc. \rm A {\bf 409} 21–-36
\bibitem{reve1} Reverberi A P, Scalas E and Vegli\`{o} F 2002
Numerical solution of moving boundary problems in diffusion processes with attractive and repulsive interactions 
\it \JPA \rm {\bf 35}(7) 1575--88
\bibitem{losk1} Loskutov A, Ryabov A B and Akinshin L G 2000
Properties of some chaotic billiards with time-dependent boundaries 
\it \JPA \rm {\bf 33}(44) 7973--86
\bibitem{koil1} Koiller J, Markarian R, Kamphorst S O and de Carvalho S P 1996
Static and time-dependent perturbations of the classical elliptical billiard
\it J. Stat. Phys. \rm {\bf 83}(1/2) 127--43
\bibitem{murp1} Murphy K A 1992 
Estimation of unknown variable parameters in moving boundary problems
\it SIAM J. Control Optim. \rm {\bf 30}(3) 637--74
\bibitem{wilk1} Wilkinson M 1988
Statistical aspects of dissipation by Landau--Zener transitions
\it \JPA \rm {\bf 21}(21) 4021--37
\bibitem{weis1} Weiss G 1975
Time-reversibility of linear stochastic processes
\it J. Appl. Prob. \rm {\bf 12} 831--36
\bibitem{sill1} Sills J A and Kamen E W 2000
On some classes of nonstationary parametric processes
\it J. Franklin Inst. \rm {\bf 337} 217--49
\bibitem{alt1} Alt H, Gr\"{a}f H-D, Hofferbert R, Rangacharyulu C, Rehfeld H, Richter A, Schardt P and Wirzba A 1996
Chaotic dyanmics in a three-dimensional superconducting microwave billiard 
\it \PR \rm E {\bf 54}(3) 2303--12
\bibitem{diet1} Dietz B, Heine A, Richter A, Bohigas O and Leb{\oe}uf P 2006
Spectral statistics in an open parametric billiard system 
\it \PR \rm E {\bf 73}(3) 035201(R)
\bibitem{kuhl1} Kuhl U, St\"{o}ckmann H-J and Weaver R 2005
Classical wave experiments on chaotic scattering
\it \JPA \rm {\bf 38}(49) 10433--64
\bibitem{fyod1} Fyodorov Y V, Savin D V and Sommers H-J 2005
Scattering, reflection and impedance of waves in chaotic and disordered systems with absorption
\it \JPA \rm {\bf 38}(49) 10731--60
\bibitem{tich1} Tichy J and Baade P K 1974
Effect of rotating diffusers and sampling techniques on sound-pressure averaging in reverberation rooms
\it J. Acoust. Soc. Am. \rm {\bf 56}(1) 137--43
\bibitem{arnaTEMCv47n4} Arnaut L R 2005
On the maximum rate of fluctuation in mode-stirred reverberation
\it IEEE Trans. Electromagn. Compat. \rm {\bf 47}(4) 781--804
\bibitem{petr1} Petrov B M 1972
Spectral characteristics of the scattered field from a rotating impedance cylinder in uniform motion
\it Radiotekhn. Elektron. \rm {\bf 17} 1431--37
\bibitem{vanb1} van Bladel J 1976
Electromagnetic fields in the presence of rotating bodies
\it Proc. IEEE \rm {\bf 64}(3) 301--18
\bibitem{chua1} Chuang C W, Pathak P H and Chuang C C 1982
On wave modulation by a rotating object
\it IEEE Trans. Antennas Propag. \rm {\bf 30}(3) 486--9
\bibitem{lamb1} Lamb W E (Jr) 1946
Theory of a microwave spectroscope
\it \PR \rm {\bf 70}(5/6) 308--17
\bibitem{rich1} Richardson R E (Jr) 1985
Mode-stirred chamber calibration factor, relaxation time, and scaling laws'
\it IEEE Trans. Instrum. Meas. \rm {\bf 34}(4) 573--80
\bibitem{arnaDEMEM012} Arnaut L R 2006
Time-domain measurement and analysis of mechanical step transitions in mode-tuned reverberation
\it National Physical Laboratory Report {\bf DEM-EM 12} \rm http://www.npl.co.uk/publications/
\bibitem{davi1} Davidovich M V 2001 
On the theory of the nonstationary excitation of a cavity
\it Radiotekhn. Elektron. \rm {\bf 46}(10) 1198--205
\bibitem{arnaKdf} Arnaut L R 2003
Limit distributions for imperfect electromagnetic reverberation
\it IEEE Trans. Electromagn. Compat. \rm {\bf 45}(2) 357--77
\bibitem{hillv40n3} Hill D A 1998
Plane wave integral representation for fields in reverberation chambers
\it IEEE Trans. Electromagn. Compat. \rm {\bf 40}(3) 209--17
\bibitem{arnaPRE} Arnaut L R 2006
Spatial correlation functions of inhomogeneous random electromagnetic fields
\it \PR \rm E {\bf 73} 036604
\bibitem{soko1} Sokolov V V and Zelevinsky V G 1989
Dynamics and statics of unstable quantum states
\it \NP \rm A {\bf 504} 562--88
\bibitem{soko2} Sokolov  V V Zelevinsky V G 1992
Collective dynamics of unstable quantum states
\it \APNY \rm {\bf 216} 323--50
\bibitem{arnaCEM11} Arnaut L R and West P D 1998
Evaluation of the {\em NPL\/} untuned stadium reverberation chamber using mechanical and electronic stirring techniques
\it National Physical Laboratory Report {\bf CEM 11} \rm http://www.npl.co.uk/publications/
\bibitem{loev1} Lo\`{e}ve M 1977
\it Probability Theory II \rm 4th ed. (New York, NY: Springer) ch XI sec 37.4
\bibitem{prie1} Priestley M B 1965
Evolutionary spectra and non-stationary processes
\it J. Roy. Stat. Soc. \rm B {\bf 27}(2) 204--37
\bibitem{stra1} Stratonovich R L 1963 \& 1967
\it Topics in the Theory of Random Noise \rm vols 1 \& 2 (New York, NY: Gordon \& Breach)
\bibitem{prie2} Priestley M B 1988 
\it Non-Linear and Non-Stationary Time Series Analysis \rm (New York, NY: Academic) ch 6
\bibitem{PRL} Arnaut L R and Knight D A 2007
Observation of coherent precursors in pulsed reverberation fields
\it \PRL \rm {\bf 98}(5) 053903
\bibitem{risk1} Risken H 1989
\it The Fokker--Planck Equation \rm 2nd ed. (Berlin, DE: Springer)
\bibitem{vank1} van Kampen N G 1992
\it Stochastic Processes in Physics and Chemistry \rm revised \& enlarged ed (Amsterdam, NL: Elsevier) sec. VIII.5
\bibitem{arnaTEMC1} Arnaut L R 2001
Effect of local stir and spatial averaging on the measurement and testing in mode-tuned and mode-stirred reverberation chambers
\it IEEE Trans. Electromagn. Compat. \rm {\bf 43}(3) 305--25
\bibitem{vanm1} Vanmarcke E 1983
\it Random Fields: Analysis and Synthesis \rm (Cambridge, MA: MIT Press)
\bibitem{voge1} Vogel K, Risken H and Schleich W 1989 Noise in a ring-laser gyrospcope -- In: Moss F and McCintock P V E (eds) \it Noise in Nonlinear Dynamical Systems \rm vols. I and II (Cambridge, UK: Cambridge University Press) ch 11 pp 271--92
\bibitem{coff1} Coffey W T, Kalmykov Yu P, and Waldron J T 1996 \it The Langevin Equation \rm 1st ed (Singapore: World Scientific)
\bibitem{arnaCE} Arnaut L R 2002 
Compound exponential distributions for undermoded reverberation chambers
\it IEEE Trans. Electromagn. Compat. \rm {\bf 44}(3) 442--57
\nonum 
Arnaut L R 2003
Corrections
\it IEEE Trans. Electromagn. Compat. \rm {\bf 45}(3) 568--9
\bibitem{grad1} Gradsteyn I S and Ryzhik I M 1994 \it Table of Integrals, Series and Products \rm 5th ed (New York, NY: Academic)
\bibitem{okse1} \O ksendahl B 2000 \it Stochastic Differential Equations \rm (new York, NY: Springer Universitext)
\bibitem{arnaRadioSci} Arnaut L R 2005
Statistical distribution of dissipated power in electronic circuits immersed in a random electromagnetic field
\it Radio Sci. \rm {\bf 40}(6) RS6S06
\bibitem{khol1} Kolmogorov A N 1956 \it Foundations of the Theory of Probability \rm 2nd ed (New York, NY: Chelsea)
\bibitem{gard1} Gardiner C W 1985 \it Handbook of Stochastic Methods \rm 2nd ed (Berlin, DE: Springer)
\bibitem{garn1} Garnir H and Gobert J 1965 \it Fonctions d'une Variable Complexe \rm (Paris, FR: Dunod)
\bibitem{moss1} Moss F and McCintock P V E (eds) 1989 \it Noise in Nonlinear Dynamical Systems \rm vols. I and II (Cambridge, UK: Cambridge University Press)
\bibitem{lead1} Cram\'{e}r H and Leadbetter M R 1967 \it Stationary and Related Stochastic Processes \rm (New York, NY: Wiley) ch 13
\end{thebibliography}
\end{document}